\documentclass[11pt]{report}
\usepackage{amsfonts}
\usepackage{amsmath}
\usepackage{amssymb}
\usepackage{graphicx}

\makeatletter
\@addtoreset{equation}{section} 
\makeatother

\newcommand{\eqn}[1]{(\ref{#1})}
\def\appendix#1
{
\addtocounter{section}{1}\setcounter{equation}{0}
 \renewcommand{\thesection}{\Alph{section}}
 \section*{
 \thesection\protect\indent \parbox[t]{16.715cm}{#1}}
 \addcontentsline{toc}{section}{Appendix \thesection\ \ \ #1}
}
\newcommand{\complex}{{\mathbb{C}}} 
\newcommand{\real}{R} 

\newcommand{\Dirac}{{D\mkern-11.5mu/\,}} 
\newcommand\opname[1]{\mathop{\mathrm{#1}}\nolimits}

\newcommand{\Tr}{\opname{Tr}}

\newcommand{\del}{\partial}
\newcommand{\kM}{$\kappa$-Minkowski}
\newcommand{\kkP}{$\kappa$-Poincar\'e}
\newcommand{\de}{{\mathrm{d}}}
\newcommand{\cM}[1]{[M_j,{#1}_k]=i\epsilon_{jkl}{#1}_l}
\newcommand{\Da}[1]{\Delta(#1)=#1\otimes 1+1\otimes #1}

\newcommand{\e}{\epsilon}
\newcommand{\mA}{\mathcal{A}}
\newcommand{\cplus}{\dot+}

\newcommand{\ft}{\tilde{f}}
\newcommand{\gt}{\tilde{g}}

\newcommand{\al}{\alpha}
\newcommand{\bt}{\beta}

\newcommand{\x}{{\bf x}}
\newcommand{\ts}{\left(}
\newcommand{\td}{\right)}
\newcommand{\qs}{\left[}

\newcommand{\qd}{\right]}
\newcommand{\vq}{\vec{q}}
\newcommand{\vs}{\vec{s}}
\newcommand{\vt}{\vec{t}}
\newcommand{\vk}{\vec{k}}
\newcommand{\vl}{\vec{l}}
\newcommand{\vx}{\vec{x}}
\newcommand{\vp}{\vec{p}}
\newcommand{\trir}{\triangleright}
\newcommand{\D}{{\mathcal{D}}}
\newcommand{\E}{{\mathcal{E}}}
\newcommand{\mN}{\mathcal{N}}
\newcommand{\mM}{\mathcal{M}}
\newcommand{\mP}{{\mathcal{P}}}
\newcommand{\mL}{{\mathcal{L}}}

\newcommand{\uno}{\bar{(1)}}
\newcommand{\due}{\bar{(2)}}
\newcommand{\tril}{\triangleleft}

\newcommand{\Ns}{\mathcal{N}}
\newcommand{\n}{{\bf \mathcal{N}}}
\newcommand{\p}{{\bf p}}
\newcommand{\wb}{\trir\!\!\!\blacktriangleleft}

\newcommand{\mT}{\mathcal{T}}
\newcommand{\q}{{\bf q}}

\def\<#1,#2>{\left\langle#1,#2\right\rangle} 
\def\hil{\mathcal{H}}
\def\uni{\mathcal{U}}

\def\nn{\nonumber}
\def\be{\begin{equation}}
\def\ee{\end{equation}}
\def\bea{\begin{eqnarray}}
\def\eea{\end{eqnarray}}

\begin{document}

\begin{titlepage}
\begin{center}
December 19, 2003\\
\vskip .2in {\large \bf FIELDS AND SYMMETRIES \\
IN\\
 $\kappa$-MINKOWSKI NONCOMMUTATIVE
SPACETIME }\\
(Ph.D Thesis)\footnote{Under the supervision of Dr. G. Amelino-Camelia and Prof. F. Lizzi}\\
 \vskip .2in Alessandra AGOSTINI\footnote{e-mail: agostini@na.infn.it} \vskip .2in
{\it Dipartimento di Scienze Fisiche, Universit\`{a} di Napoli ``Federico II'',\\
and INFN Sez.~Napoli, Via Cintia, 80125 Napoli, Italy} \vskip .5in
{\bf Abstract}
\end{center}

\noindent We have investigated some issues relevant for the
possibility to construct physical theories on the \kM\
noncommutative spacetime. The notion of field in \kM\ has been
introduced by generalizing the Weyl system/map formalism and a
comparative study of the star products arising from this
generalization has been done. A line of analysis of the symmetries
of \kM\ has been proposed that relies on the possibility to find a
``maximally"-symmetric action which is invariant under a
10-generator Poincar\'e-like symmetry algebra. The equation of
motion for scalar particles has been obtained  by a generalized
variational principle. An extension of the Dirac equation for
spin-1/2 particles has been proposed by using a five-dimensional
differential calculus on \kM .
\end{titlepage}

\font\ninerm = cmr9
\def\footnoterule{\kern-3pt \hrule width \hsize \kern2.5pt}
\pagestyle{empty}

\tableofcontents
\addcontentsline{toc}{chapter}{Introduction}

\newcounter{contatore}

\baselineskip 19pt plus .5pt minus .5pt

\pagenumbering{arabic}
\setcounter{page}{0}

\pagestyle{plain}

\chapter*{Introduction}
It is widely expected that our current description of particle
physics would require a profound revision in order to describe
processes with energies of the  order of the Planck energy
($E_P=\sqrt{\frac{\hbar c}{G_N}}\simeq 10^{19}GeV$). At such high
energies both quantum and gravitational effects are important, and
the Standard Model of particle physics appears to be incomplete
since it neglects gravity. Different arguments can be produced to
identify the Planck scale as the special scale at which quantum
and gravitational effect are equally important. For example,
adopting, a ``semi-Newtonian" approximation of gravity, one
 finds that at the Planck scale the gravitational
force between two Planck-energy particles is $F_g\simeq
\frac{E_P^2G_N}{r^2}$, while their electronic attraction is
$F_e=\frac{e^2}{r^2}$. Taking into account that at such high
energy scale the electromagnetic coupling constant is expected to
be of order $1$ ($\frac{e^2}{\hbar c}\sim 1$), the two forces turn
out to be comparable
$F_g/F_e\simeq\frac{E_P^2G_N}{e^2}\simeq\frac{E_P^2G_N}{\hbar c}
=1$.

A large research effort has been devoted to the search of a
``Quantum Gravity" \emph{i.e.} a theory giving  a unified
description of Quantum Mechanics and General Relativity, the two
theories that  respectively govern quantum and gravitational
phenomena.

Quantum Mechanics reigns supremely in low-energy ($E<\!<E_P$)
processes where gravity is negligible. In particular, Quantum
Field Theory, following the unification of Special Relativity with
Quantum Mechanics, successfully describes all experimental data up
to energies currently achievable in the laboratory which are in
the TeV range.
Several characteristic predictions of the Standard Model  of
strong, electromagnetic and weak interactions have been very
successful as in the case of the discovery of the $W$ and $Z$
gauge bosons.

On the other hand, Einstein's General Relativity successfully
describes the motion of  macroscopic bodies where quantum effects
are negligible.

However a unified description of these two theories is necessary
in order to produce predictions for  some interesting situations
in which both are required, for example the ``Big Bang"- the first
few  moments of the Universe, when gravitational interactions were
very strong and the scales involved were all microscopic.

If one simply attempts to quantize General Relativity, in the same
sense that quantum electrodynamics is a quantization of Maxwell's
theory, the result is an inconsistent theory. This is due to the
fact that  Newton's constant is dimensionful and consequently, the
divergences can not be disposed of by the technique of
renormalization. In addition to this ``renormalizability" problem,
great difficulties of a unified description of Quantum Mechanics
and General Relativity originate from their deep
incompatibilities. One of the most evident aspects of this
incompatibility regards  the way in which  the geometry of space
and time is treated. In the Quantum Mechanical picture spacetime
is a fixed arena in which quantum observables (such as position of
a particle) are described. But in General Relativity spacetime can
not be treated as a fixed background since it acquires a
geometrodinamical structure.

The lack of reliable data on the space-time at very small distance
scales (\emph{i.e.} at very high energies) has led to the proposal
of  various models for Quantum Gravity (see for
example~\cite{carloHISTO,ash,smo}). These models are sometimes
very different in the way they approach the technical and
conceptual problems emerging from a Quantum-Gravity theory;
however, they lead to a common Quantum-Gravity intuition: from any
approach to the unification of General Relativity and Quantum
Mechanics emerges the idea of a limitation to the localization of
the space-time point~\cite{Maggiore9301067}. One can even find
evidence of this fact through simple heuristic arguments. In fact,
Heisenberg's uncertainty principle asserts the measurability
resolution of space distances increases with the energy of the
probe. However, increasing the probe energy produces a greater
disturbance of the space-time metric. Consequently, the
measurability uncertainty tends to grow. The competing
contributions from these two uncertainties lead to emergence of a
minimum uncertainty  beyond which the position of the particle can
not be specified. More precisely, on the basis of ``gedanken"
experiments some candidate generalized uncertainty principles have
been proposed~\cite{Maggiore9301067} for the description of a
minimum Planck-length uncertainty in particle position and
distance measurements.


In this scenario it is conceivable that at Planck-length distance
scales geometry can have a form which is quite different from the
classical one with which we are familiar at large scales. The
description of the spacetime as differentiable manifold might need
a revision and a new description of geometry might lead to the
development of a completely new understanding of physics.

In this perspective Noncommutative Geometry acquires an important
role in the search of a unified description of Quantum Mechanics
and General Relativity. Noncommutative Geometry emerges in
different ways in the Quantum Gravity approaches but we can single
out essentially two ways.

In the first case, there is an \emph{a priori} assumption of
spacetime noncommutativity. In these approaches (see for
example~\cite{Snyder,DFR,dsr1,Kow02-NST}) it has been explored the
possibility that a noncommutative geometry might be needed for the
correct fundamental description of spacetime.


In the second case, Noncommutative Geometry turns out to be useful
at the effective theory level in the  description of certain
Quantum Gravity contexts -- for example, in String
Theories~\cite{SW,susskind,douglasnovikov} spacetime
noncommutativity provides an effective theory description of the
physics of strings in presence of a corresponding external
background.

At the fundamental level it has been argued~\cite{DFR, garay,
dsr1} that one can describe algebraically  Quantum Gravity
corrections replacing the traditional (Minkowski) spacetime
coordinates $x_{\mu}$ with Hermitian operators $\x_{\mu}$ that
satisfy commutation relation of the type: $$
[\x_{\mu},\x_{\nu}]=i\theta_{\mu\nu}(\x) $$
 A noncommutative
spacetime of this type embodies an impossibility to fully know the
short distance structure of spacetime, in the same way that in the
phase space of the ordinary Quantum Mechanics there is a limit on
the localization of a particle. This fact agrees with the above
mentioned intuition of a measurability bound in the
Quantum-Gravity framework.

There is a wide literature on the simplest ``canonical"
noncommutativity characterized by a constant value of the
commutators $$ [\x_{\mu},\x_{\nu}]=i\theta_{\mu\nu}, $$ where
$\theta_{\mu\nu}$ is a matrix of dimensionful parameters. Such
spaces have also emerged in the context of string
theory~\cite{witten} where Y-M theories arises from the
compactification of $M$-theory on a torus in the presence of a
constant background field~\cite{CDS}, or as low-energy limit of
open strings in a constant background $B$-field~\cite{LLS, SW}.

In this thesis we consider another much studied noncommutative
spacetime, the \kM\ spacetime, characterized by the commutation
relations: $$ [\x_0,\x_j]=\frac{i}{\kappa}\x_j\;\;\;[\x_j,\x_k]=0,
$$ where $\kappa$ is a dimensionful parameter.
 This type of noncommutativity is an example of Lie-algebra-type
noncommutativity in which commutation relations among space time
coordinates exhibit a linear dependence on the spacetime
coordinate themselves $$ [\x_{\mu},\x_{\nu}]=i
\zeta^\rho_{\mu\nu}\x_\rho, $$ with coordinate-independent
$\zeta_{\mu\nu}^\rho$. \kM\ represents one example of algebras
proposed in the Quantum Group approach for the Planck scale
Physics~\cite{Maj88, MajPhD, majid}, where \kM\ appears as a
natural candidate for a quantized spacetime in the zero-curvature
limit.

Recently \kM\ gained remarkable attention due to the fact that it
provides an example of noncommutative spacetime in which Lorentz
symmetries are
preserved as deformed (quantum) symmetries. This property is not
achieved for example in the much studied canonical
noncommutative-spacetimes where classical Lorentz symmetries are
manifestly broken by the presence of a constant two-tensor
$\theta_{\mu\nu}$~\cite{susskind}. Thus the latter can not be
connected with any quantum symmetry group. The quantum deformation
and even a break down of Lorentz symmetry is not surprising for a
quantum spacetime because of the existence of a minimum spatial
length that is not a Lorentz invariant concept. If ordinary
Lorentz invariance was preserved we could always perform a boost
and squeeze any given length as much as we want and therefore a
minimal length could not exist. If a minimal length really exists
we have to contemplate the possibility that the Lorentz invariance
is lost. The peculiarity of \kM\ spacetime is that the symmetry is
lost as classical symmetry being preserved as ``quantum symmetry".
The presence of a symmetry (even a quantum symmetry) has profound
implications in a physical theory.

The fact that symmetries are deformed in \kM\ has emerged
in~\cite{MajidRuegg} where \kM\ has been connected with a
dimensionful deformation of the Poincar\'e algebra called \kkP .

The analysis of the physical implications of the deformed \kkP\
algebra have led to interesting hypotheses about the possibility
that in \kM\ particles are submitted to modified dispersion
relations~\cite{AmelinoLukNowik}. Since the growing sensitivity
and accuracy of the astrophysical observations renders
experimentally accessible such modified dispersion relations,
there is now strong interest on  a systematic analysis of a field
theory in \kM .

\bigskip
\bigskip
\bigskip

The primary objective of our work is the development of a
systematic construction of a field theory on \kM\ noncommutative
spacetime. In spite of the mentioned strong interest in this
problem, the results obtained so far are only partial. One
encounters many difficulties in constructing a field theory in \kM
, and these difficulties are already present in classical field
theory.

Among the various problems there are two that are the main focus
of our work: the problem of introducing a suitable
\emph{``ordering" prescription} for a field in a Lie-algebra-type
noncommutative spacetime, and the problem of characterizing the
symmetries at the level of the action of the theory.

The study of a classical field theory provides us a natural arena
for these problems, avoiding the further complications that arise
in a quantum field theory, for which some additional technical
tools would be needed.

Moreover, beyond the results that we obtain in the specific case
of \kM\ spacetime, this study allows us to develop some new
techniques that might be used in order to deal with field theory
in other types of noncommutative spacetimes with a non-simple
structure as \kM .

As we discuss in Chapter 1, there are actually two distinct
approaches which have attained some important results in
noncommutative-geometry field theories: Connes
approach~\cite{ConnesLott} and the Quantum-Group
approach~\cite{Woro}. In both methods new mathematical instruments
are introduced  to deal with the noncommutative spacetime
structure. Our approach is strictly speaking none of them but we
are inspired from both using some of the mathematical tools
introduced by them and elaborating an original approach much based
on the physical intuition (following as close as possible the
analogy with the commutative-spacetime case). In this optics, we
attempt to translate some abstract mathematical notions in
physical conditions.

In Chapter 2 we define some fundamental tools (such as the
generalized Weyl map) in order to introduce the concept of
``field" in \kM\ and to manage the ambiguity regarding the
``ordering" due to the noncommutativity. The problem of the
``ordering" consists in the fact that, in a noncommutative
spacetime, a field (that is a function of the spacetime
coordinates) can be written in different forms, which reduce all
to the same function in the limit of commutative spacetime. For
example, in \kM , the simple functions $\x{\bf t}$, with the time
to the right, and ${\bf t}\x=\x{\bf t}+i\lambda \x\neq \x{\bf t}$,
with the time to the left, reduce to the same function $xt=tx$ in
the commutative limit $\lambda\to 0$. However, they are different
and can be distinguished by the ordering of the time and space
variables. Therefore there is an ``ordering" ambiguity in
extending the function $xt=tx$ from the commutative spacetime to a
noncommutative one.

In the canonical noncommutative spacetime, which has a structure
very similar to the Quantum Mechanics phase space, it is used to
introduce fields through the Weyl map~\cite{weyl}.
In~\cite{Monaco} was suggested the idea to introduce fields in
Lie-algebra-type noncommutative spacetimes generalizing the Weyl
map procedure. Following this idea, we introduce a field in \kM\
through a generalized Weyl map based on the notion of generalized
Weyl system. An explicit construction shows that the generalized
Weyl systems related to \kM\ satisfy a non-Abelian group law, and
then they have a much more complex group structure with respect to
the Weyl systems underlying the Weyl map in Quantum Mechanics (or
in canonical noncommutative spacetimes). In particular, using a
Weyl map, one can define a deformed (``star") product and
correspondingly deformed (``star") algebra in which functions are
not multiplied with the commutative (pointwise) product, but with
a new, noncommutative product, the ``star" product. In this way
the noncommutative space is studied as the structure space of a
deformed $*$-algebra. The definition of this $*$-algebra is very
important, as it is the first step toward the use of Connes'
machinery for the construction of physical theories, and the
construction of field theories on noncommutative spaces
(deformation quantization). From the side of star products in
Chapter 2 we have led back to the same origin (of Weyl systems)
different *-products for \kM\ present in literature and we have
derived some new ones through a reduction procedure.

The analysis reported in Chapter 2 also clarifies that the
different choices of Weyl systems in \kM\ (that lead to different
*-products) correspond to different ordering prescription of the
functions of the \kM\ noncommutative coordinates, so we can (in
principle) relate the star product with an ordering prescription.


The Weyl-system description allows to introduce a field in \kM\ as
a generalized Fourier transform that establishes a correspondence
between positions (NC coordinate generators of \kM ) and certain
variables that could have the meaning of momenta. Thus, such a
generalized Fourier transform allows us to rewrite structures
living on NCST as structures living on classical (commutative) but
non-Abelian ``energy-momentum" space. This would lead to an easier
treatment of field theory rewriting Feynman rules in terms of new
momenta, that would coincide with the translation generators of
the \kkP\ quantum group. However the interpretation that the
Quantum Group language gives to ``momenta" as generators of the
translations (i.e. the real physical particle momentum) is based
on the notion of quantum group symmetry. It is puzzling in fact
that in the Quantum Group literature it is stated (see, {\it
e.g.}, Refs.~\cite{lukieAnnPhys,Kow02-NST}) that the symmetries of
\kM\ can be described by any one of a large number of \kkP\ Hopf
algebras. The nature of this claimed symmetry-description
degeneracy remains obscure from a physics perspective, since it is
only supported by an (equally obscure) ``duality"
criterion~\cite{lukieAnnPhys,Kow02-NST}. In particular we are used
to associate energy-momentum with the translation generators and
it is not conceivable that a given operative definition of
energy-momentum could be equivalently described in terms of
different translation generators. The difference would be easily
established by testing, for example, the different dispersion
relations that the different momenta satisfy (a meaningful
physical property, which could, in particular, have observable
consequences in astrophysics~\cite{gacEMNS97,gactp} and in
cosmology~\cite{jurekCOSMO,maguCosmoDispRel}).


In Chapter 3 we propose a new line of analysis of the symmetry of
the \kM\ spacetime in order to make clear the ambiguity concerning
the description of symmetry emerging from the mathematical Quantum
Group approach. Our new line of analysis of the
noncommutative-spacetime symmetries relies on the introduction of
a  Weyl map (connecting a given function in the noncommutative
Minkowski  with a corresponding function in commutative Minkowski)
and of a compatible notion of integration in the noncommutative
spacetime. In this context we have translated in physical language
some mathematical axioms of the Hopf algebra
structures~\cite{aad03}.

We have confirmed and established more robustly (supported by a
more physical approach) the important result that the
commutative-spacetime notion of Lie-algebra symmetries must be
replaced, in the noncommutative-spacetime context, by the one of
Hopf-algebra symmetries. Symmetries are introduced directly at the
level of the action, following very strictly commutative field
theory in which the symmetry of a theory is defined as
transformation of coordinates that leaves invariant the action of
the theory. Firstly we apply our description of symmetry to  a
free scalar theory in \kM\ spacetime. In this way we prove that in
$\kappa$-Minkowski it is possible to construct an action for
scalar theory which is invariant under a Poincar\'{e}-like Hopf
algebra of symmetries with 10 generators, in which the
noncommutativity length scale has the role of relativistic
invariant.

Although our analysis allowed us to reduce the amount of ambiguity
in the description of the rotation/boost symmetries of theories in
these noncommutative spacetimes, we are left with a choice between
different realizations of the concept of translations in the
noncommutative spacetime. We have clarified that such an ambiguity
might have to be expected on the basis of the type of coordinate
noncommutativity here considered, but it remains to be seen
whether by appropriate choice of the action of the theory one can
remove the ambiguity, {\it i.e.} construct a theory which is
invariant under one specific type of translations and not under
any other type.


Chapter 4 is devoted to the construction of the wave equations of
free particles in \kM . We analyze the case of scalar particles
and spin-$1/2$ particles. The wave equation of a free scalar
particle is obtained through a "generalized Action Principle". It
is the generalization of the well-known Action Principle that, in
the Minkowski commutative spacetime, allows us to obtain the
equations of motion for particles starting from the action of the
theory. Thus, we derive the equations of motion for a scalar
particle in \kM\ using the maximally-symmetric action found
Chapter 3. So we obtain a $\kappa$-deformed Klein-Gordon equation
which reduces to the standard Klein-Gordon equation in the
$\kappa\to \infty$ limit.

The wave equation for spin-$1/2$ particles in \kM\ is obtained
using a \kM\ version of the standard procedure introduced by
Dirac~\cite{Dirac} for the construction of spinorial wave
equations. The starting point of this procedure, in the
commutative case, is a linear differential equation for the Dirac
spinor. This equation is written in terms of unknown coefficients
(matrices) that are completely determined imposing that each
single component of the Dirac spinor must satisfy the Klein-Gordon
equation (physical condition). In generalizing this procedure to
\kM\ case we start from the same ingredients: a linear equation in
certain vector fields (that generalize the notion of derivatives
in \kM ) with unknown coefficients (matrices) to be determined
imposing a physical condition analogous to the one imposed in the
commutative case. The choice of the vector fields generalizing the
notion of derivative in \kM\ represents a key point of our line of
analysis. In fact, in the commutative case there is only one
(natural) differential calculus involving the conventional
derivatives, whereas in the \kM\ case (and in general in a
noncommutative spacetime) the introduction of a differential
calculus is a more complex problem and, in particular, it is not
unique. In our analysis we focus on a possible choice of
differential calculus in \kM : the "five-dimensional" differential
calculus introduced in~\cite{Sitarz}. The vector field
corresponding to this differential calculus have in fact special
covariance properties: they transform under \kkP\ in the same way
that the ordinary derivatives (\emph{i.e.} the vector fields
associated to the differential calculus in the commutative
Minkowski space) transform under Poincar\'e. Imposing the
condition that the spinor components must satisfy the
$\kappa$-deformed Klein-Gordon equation (physical condition), we
determine completely the unknown matrices only in the case
"on-shell", in which the energy and momenta of the particle obey
the deformed dispersion relation of \kM . While in order to
determine the Dirac equation "off-shell", we have to impose that
the Dirac equation be invariant under the action of \kkP\ algebra
(this condition ia automatically satisfied by the Dirac equation
in the commutative case found by only imposing the physical
condition). In this way, our study has allowed us to construct a
wave equation for spin-$1/2$ particles in \kM\ space and has
revealed how the five-dimensional differential calculus plays a
fundamental role in this construction.


\chapter{Noncommutative Geometry and \kM\ spacetime}

In this chapter we give a brief overview of Noncommutative
Geometry and we present the main current approaches to it. We
introduce the Hopf-algebras structures which play a fundamental
role in the description of \kM\ noncommutative spacetime and its
quantum \kkP\ symmetry group. A complete treatment of
Noncommutative Geometry would require a large quantity of
mathematical notions and we refer to~\cite{books}\cite{ticos} for
more technical details.

The Quantum Mechanics phase space, \emph{i.e.} the space of the
microscopic states of a quantum particle, provides the first
example of noncommutative space. It is defined replacing canonical
variables of position and momentum of a particle ($q_j,p_j$) with
self-adjoint operators ($\q_j,\p_j$)
satisfying Heisenberg's commutation relations: \be
[\q_j,\p_k]=i\hbar\delta_{jk},\;\;\;j,k=1,2,3 \label{Heisalg} \ee
from which follows the Heisenberg uncertainty principle: \be
\delta \q_j\,\delta\p_k\geq \frac{\hbar \delta_{jk}}{2}. \ee This
principle establishes the existence of an accuracy limitation for
the measurement of the coordinates and the corresponding momenta
of a particle. Consequently, the quantization of phase space can
be viewed as the smearing out of a classical manifold, replacing
the notion of a point with that of a Planck cell.
The idealized classical situation in which one can simultaneously
determine the exact position-momentum measurements is obtained in
the limit $\hbar\to 0$, where the phase space becomes a continuum
manifold.

A very similar idea led to apply noncommutativity to  spacetime itself.

The idea of a new structure of spacetime already came in the late
40's from Snyder~\cite{Snyder} in order to solve the
short-distance singularities of the quantum field theories. He
proposed to consider the possibility that, at small distances,
coordinates
satisfy the following commutation relations rather than commute as
usual
 \be
 [\x_j,\x_k]=i\theta^2\epsilon_{jkl}M_l,\;\;\;[\x_0,\x_j]=i\theta^2N_j\label{snyder}
\ee where $\theta$ a is noncommutativity parameter with the
dimensions of a length,  and $M_j,N_j$ are the infinitesimal
elements of the four-dimensional Lorentz group. This quantized
spacetime results to be a Lorentz invariant spacetime in which
there is a natural unit of length $\theta$. Snyder expected that
such a unit of length represented a sort of cut-off and solved
many of the divergence troubles of quantum field theory.

Later on, the attention was focused on a general noncommutative
spacetime of Lie-algebra type with central extension,
characterized by the commutation relations: \be
[\x_{\mu},\x_{\nu}]=i\theta_{\mu\nu}+i\zeta^{\al}_{\mu\nu}\x_{\al}
\ee with coordinate-independent $\theta_{\mu,\nu}$ and
$\zeta^{\al}_{\mu\nu}$; in particular, the attention was
concentrated on the canonical noncommutative spacetime,
characterized simply by Heisenberg-like commutation relations: \be
 [\x_{\mu},\x_{\nu}]=i\theta_{\mu\nu}. \label{canonical}
\ee As in the case of quantum phase space, this spacetime
description can be viewed as the smearing out of the classical
manifold loosing the  notion of the \emph{point}: in fact, a
Heisenberg-type uncertainty principle implies  that the notion of
the point is  replaced by an analogous of the Planck cell of the
quantum phase space. For this reason, Von Neumann defined the
study of such a quantized  spacetime  as ``pointless geometry''.

A systematic study of this concept, largely due to Connes, has
originated Noncommutative Geometry and the possibility to
construct field theories on noncommutative spaces in the same way
as on the traditional commutative spaces~\cite{books}. In
particular, Connes' idea of Noncommutative Geometry is based on
the re-formulation of the manifold geometry in terms of a  $C^*$
algebras of functions defined over the manifold, with a
generalization of the corresponding results of differential
geometry to the case of a noncommutative algebra of functions.
However the Connes approach is not the only one, in fact,
Noncommutative Geometry emerges also from other approaches. In
particular it is interesting how it emerges in the Quantum Groups
framework. From this point of view it was Woronowicz~\cite{Woro}
who initiated a systematic study of the "noncommutative
differential geometry" built on some "pseudogroups" that are the
generalization of the standard Lie groups related to the
commutative differential geometry.

 In the next sections we give a
brief overview about the Connes and Quantum Group approaches and
we attempt to clarify their relationships and their differences.
We want to stress however that our approach to the study of \kM\
spacetime is essentially inspired from the Quantum Group
perspective.

\section{Noncommutative Geometry in the Connes formulation}
The main concept at the basis of Connes' approach is the
definition of \emph{$C^*$-algebra}. Thus, it is worthwhile first
reminding the definitions of some mathematical tools that are
needed in the $C^*$-algebra definition.

\subsection{$C^*$-algebras}

An (associative and unital) \emph{algebra} $\mathcal{A}$ over a
complex field $\complex$ is a vector space endowed with two linear
maps: a product $m:\mathcal{A}\otimes \mathcal{A}\rightarrow
\mathcal{A}$\footnote{Here the symbol $\otimes$ means the
algebraic tensor product over $\complex$} and a unit $\eta:
\complex \to \mathcal{A}$ that satisfy the following properties
($\forall\, a,b,c \in \mathcal{A}$):
\begin{enumerate} \item $m(a\otimes(b+c))=m(a\otimes b)+
m(a\otimes c)\;$ and $\;m((a+b)\otimes c)=m(a\otimes c)+
m(b\otimes c)\;\;\;$ (distributivity) \item $m(m(a\otimes
b)\otimes c)=m(a\otimes m(b\otimes c))\;\;\;$ (associativity)
\item $m(\mathbf{1}\otimes a)=m(a\otimes \mathbf{1})=a,\;\;
\mathbf{1}=\eta(1)\in\mathcal{A}\;\;\;$ (existence of unity),
\end{enumerate}
where $\mathbf{1}=\eta(1)$ is the neutral element of the algebra
$\mathcal{A}$.
 If the properties $2.,3.$ are missing we have
simply an algebra (not-associative, not-unital), but we will
always refer to associative and unital algebras unless indicated
otherwise. In the following we will indicated the product of two
elements simply by $ab:=m(a\otimes b)$.

$\mathcal{A}$ is called \emph{Banach algebra} if it is complete
with respect to a norm $|\!|\cdot|\!|:A\to R$ with the usual
properties
\begin{enumerate}
\item $|\!|a|\!|\geq 0,\; |\!|a|\!|=0 \;\Leftrightarrow a=0 $
\item $ |\!|\al a |\!|=|\al||\!|a|\!|$
\item $ |\!|a+b |\!|\leq |\!|a |\!|+|\!|b|\!| $
\item $ |\!| ab|\!|\leq |\!|a|\!||\!|b|\!| $
\end{enumerate}
for any $a,b\in \mathcal{A}$ and $\al,\bt\in \complex$.

A Banach algebra $A$ is  a \emph{$C^*$-algebra} if, in addition to
the properties above, a conjugation operation * has been defined
on it, such that:
\begin{enumerate}
\item $a^{**}=a$
\item $(ab)^*=b^*a^*$
\item $(\al a+\bt b)^*=\bar{\al}a^*+\bar{\bt}b^*$
\item $|\!| a^*|\!|=|\!|a|\!|$
\item $|\!| a^*a|\!|=|\!|a|\!|^2$
\be
\label{conjug}
\ee
\end{enumerate}
for any $a,b\in \mathcal{A}$ and $\al,\bt\in \complex$, and
$\bar{\al}$ denoting the usual complex conjugate of the complex
numbers.

It is also useful to define the concept of \emph{ideal} $I$ of an
algebra $\mathcal{A}$. An ideal  $I$ is a subspace of
$\mathcal{A}$ which is closed under multiplication by $A$ from the
left (left ideal) or from the right (right ideal) or from both
side (two-side ideal).
A \emph{maximal ideal} is an ideal that is not contained in any
other ideal (apart from the trivial ideal $\mathcal{A}$ itself).

A simple example of a $C^*$-algebra is the algebra $C(M)$ of the
continuous complex-valued functions over a compact topological
space $M$. This is  a commutative $C^*$-algebra with product \be
(f\cdot g)(x)=f(x)h(x),\;\;f,g\in C(M)\;\;x\in M. \ee
 In this case the norm is given by the maximum value attained by the function on $M$:
\be |\!|f|\!|_{\infty}=\sup_{x\in M}|f(x)| \ee and the unit is
defined by: \be \eta(x)=x\mathbf{1},\;\;
\mathbf{1}(x)=1,\;\;\;\forall x\in M. \ee Ideal in this algebra is
the set of continuous functions vanishing on some subset of $M$
and  maximal ideals are represented by the set of functions
vanishing at a single point $x\in M$.

\subsection{Connes approach}
Connes approach to Noncommutative Geometry comes from the
statement that a topological space (\emph{i.e.} a set of points
with the notion of their open neighborhood) can be recovered from
the algebra of continuous functions on it. Thus one can study the
topology (and the geometrical properties) of a space by not seeing
it as a set of points, but rather by investigating the set of
functions defined on it. At the basis of this study there are a
series of theorems due to Gel'fand and Naimark~\cite{GN} (see
also~\cite{fell}\cite{dix}) stating a complete equivalence between
compact Hausdorff spaces and commutative unital $C^*$-algebras.

Any commutative $C^*$-algebra can be realized as the $C^*$-algebra
of complex valued function $C(M)$ on some compact Hausdorff space.
The correspondence in the other direction is more complex but it
is possible in a constructive way, in fact all the information of
the topology of $M$ are collected into the algebra $C(M)$. This
reconstruction can be realized trough the notion of maximal ideals
of the algebra. The first step is the identification of the points
of the space $M$ under reconstruction; this can be done with the
definition of the maximal ideal $I_x\subset C(M)$ as the function
that vanishes at the point $x$.
After the identification of the points of $M$ we are able to
construct the topology of $M$ \emph{i.e.} the relations between
the points of $M$. There are in principle many ways to give the
topology of a set, one of them is to give the closure of every
set. In $C(M)$ we can define the closure $\bar{W}$ of any subset
$W$ of the set of the maximal ideals. We define the closure
$\bar{W}$ of $W$ as the set of maximal ideals $I_m$ with: \be
\bar{W}=\big\{I_m\subseteq C(M):(\cap_{I\in W})\subseteq I_m
\big\} \ee That is, the point corresponding to the ideal $I_m$
belongs to the closure of $W$ if $I_m$ contains the intersection
of all ideals $I$ of $W$.

To be simple, let us consider the example of the open interval. In
this case $W$ is the ideal of all functions vanishing at some
point in the open interval. The intersection of all these ideals
is the ideal of functions vanishing on the open interval. Thus an
ideal $I_m$ belongs to $\bar{W}$ if it contains all functions that
vanish in the open interval. But the functions we are considering
are continuous, and, therefore, if they vanish in the open
interval they vanish also at the end-points. Thus the functions
which vanish at any point of the closed interval belong to
$\bar{W}$.

 This example is very simple and involves a
commutative algebra. In the case in which one replaces a
commutative algebra with a noncommutative algebra the
reconstruction of the space fails. So, noncommutative C*-algebra
will be thought as the algebra of continuous functions on some
``virtual noncommutative space''. Connes approach to
Noncommutative Geometry, in fact, switches the attention from
spaces, which do not exist concretely, to algebras of functions,
that are well-defined.

However, a physical space has much more structure than just
topology, which is in fact the most basic aspect of it. Connes has
shown that metric and other aspects can be encoded at the level of
algebras. The key property of the Connes construction is another
important result due to Gel'fand stating that any C*-algebra can
be faithfully  represented as a subalgebra of the algebra
$\mathcal{B}(\mathcal{H})$ of bounded operator on a
infinite-dimensional separable Hilbert space $\mathcal{H}$.

In the noncommutative case, the metric structure and the
noncommutative generalization of differential and integral
calculus, are  obtained via an operator which is the
generalization of the usual Dirac operator. From the point of view
of Noncommutative Geometry the Dirac operator is an operator $D$
on $\mathcal{H}$ with the following properties:
\begin{enumerate}
\item $D$ is self-adjoint ($D^{\dag}=D$)
\item The commutator
$[D,a]$ is bounded on a dense subalgebra of $\mathcal{A}$ for all
$a\in \mathcal{A}$ \item $D$ has compact resolvent, \emph{i.e.}
$(D-\alpha)^{-1}$ (for $\al\not\in$ spectrum of $D$) is a compact
operator on $\mathcal{H}$
\end{enumerate}

In the commutative algebras the Dirac operator enables one to give
the topological space $M$ a metric structure via the definition of
the $distance$ between two points of the space $(x,y)\in M$ in the
following way: \be d(x,y)=\sup_{a\in \mathcal{A}}\{ |a(x)-a(y)|:
\; |\!| [D,a]|\!|\leq 1\}. \ee In the case $M$ is a Riemannian
manifold with Euclidean-signature metric $g_{\mu\nu}$ the distance
coincides with the usual definition of geodesic distance and the
Dirac operator is given by $D=i\gamma_{\mu}\partial^{\mu}$ where
the commutators of the $\gamma_{\mu}$ matrices define the space
metric $[\gamma_{\mu},\gamma_{\nu}]=2g_{\mu\nu}$.

The set of data $(\mathcal{A},\mathcal{H},D)$, i.e. a C*-algebra
$\mathcal{A}$ of bounded operators on a Hilbert space
$\mathcal{H}$ and a Dirac operator D on $\mathcal{H}$, is called
\emph{spectral triple}, and it encodes the geometry and topology
of a space.

\section{Noncommutative Geometry from Quantum Groups}
Quantum Groups or Hopf algebras are a generalization of the
ordinary groups (\emph{i.e.} collections of transformations on a
space that are invertible). They have a rich mathematical
structure and numerous roles in physical situations where ordinary
groups are not adequate (an example of this fact is given in
Chapter 3 where we discuss the symmetries of \kM ). Quantum Groups
allow us to generalize many ``classical" physical ideas in a
completely self-consistent way. This generalization  is realized
through a ``deformation" induced by the presence of one or more
parameters. The classical case is recovered by setting these
parameters to some fixed values. A very similar case of
quantization is represented by Quantum Mechanics, in which the
deformation is introduced by the Planck constant $\hbar$, and the
classical case is recovered in the limit $\hbar\to 0$. As we will
show below Quantum Groups have structures, such as the coproduct
or the antipode, that generalize some properties of ordinary
groups, such as the representation on a vector-space tensor
product or the existence of an inverse. These properties are at
the basis of the origins and the applications of Quantum Groups in
a wide physical domain, from Statistical Physics to Quantum
Gravity.

Quantum Groups arose in the framework of the quantum integrable
field theories in the beginning of 80's. In~\cite{kulres} was
shown that the linear problem of the quantum sine-Gordon equation
was not associated with the Lie algebra $sl_2$ as in the classical
case, but with a deformation of this algebra. Subsequently,
in~\cite{Skly} was shown that deformations of Lie algebraic
structures were not special to the quantum sine-Gordon equation
but they were part of a more general theory. In this context some
works~\cite{Drinfeld1,Drinfeld2,Drinfeld3} of V.I. Drinfel'd had a
crucial role: he showed that a suitable algebraic quantization of
so called Poisson Lie groups reproduce exactly the deformed
algebraic structures encountered in the theory of the quantum
inverse scattering (for example the KdV equation). In a slightly
different way M. Jimbo~\cite{Jimbo} reached the same results.
Following the direction of Drinfel'd and Jimbo the theory of
Quantum Groups was largely investigated. Their importance is
mainly due to the relation with the quantum Yang-Baxter
equation~\cite{Baxter} which plays a role in different physical
problems such as the knot theory~\cite{restur}, solvable lattice
models and quantum integrable systems \cite{FRT}.

A different approach to Quantum Group, based on the $C^*$-algebra
theory, was introduced by Woronowicz \cite{Woro}. As we have
explained in Section 1.1, the Gel'fand-Naimark theorem states that
any commutative $C^*$-algebra is equivalent to an algebra of
continuous functions on some compact topological manifold.
Replacing the commutative $C^*$-algebra with a noncommutative one
corresponds to describe an underlying ``pseudospace" whose
interpretation in terms of manifold is lost, but the theory can
still be described in terms of the $C^*$-noncommutative algebra.
The topological space of the Gel'fand-Naimark theorem can be
represented by a topological group and one can consider the space
of continuous functions on it. The space of functions on a
topological group picks up extra structures and can be promoted to
commutative Hopf algebra. Deforming (\emph{i.e.} making
noncommutative) the Hopf algebra of functions on the group,
corresponds to dealing with ``Quantum Groups'', called
``pseudogroups'' in the Woronowicz language.

A special class of Quantum Groups (called of ``bicrossproduct"
type) was largely investigated by S.~Majid in the approach to
Planck-scale Physics~\cite{MajPhD}. As we will see below, this
line of research represents an important point for our study of
\kM .

 In order to explain these
arguments we provide some preliminary notions about the definition
of  Hopf algebras (or Quantum Groups). A complete and systematic
treatment on quantum groups can be found for example
in~\cite{CP}\cite{majidbook}.

\subsection{Hopf Algebras}
The extra structures that characterize a Hopf algebra with respect
to an algebra turn out to be very useful in order to translate in
the mathematical language some physical properties. For example,
finite and infinitesimal transformations are expressed with the
notions of \emph{action} and \emph{coaction}. In particular some
new structures are introduced as a result of the necessity to
compose representations. For example, a physical particle can be
viewed as a representation of the Poincar\'e group, characterized
by its spin and mass, and for the study of a system of particles
some mathematical structure is needed to describe the composition
of the representations. A representation of an algebra
$\mathcal{A}$ over a vector space $V$ is a set $(V,\rho)$, where
$\rho$ is a linear map from $\mathcal{A}$ to the space of linear
operator in $V$, $Lin(V)$, satisfying $$ \rho(a
b)=\rho(a)\rho(b)\;\;\;a,b\in \mathcal{A}. $$ Suppose that we have
two vector spaces $V_1$ and $V_2$ and the representations of an
algebra $\mathcal{A}$ on them are respectively ($V_1,\rho_1$) and
($V_2, \rho_2$). Can we use the two representations ($V_1,\rho_1$)
and ($V_2, \rho_2$) to determine the representation of
$\mathcal{A}$ on the tensor product of the spaces $(V_1\otimes
V_2,\rho)$?

If we limit us to use the structures of $\mathcal{A}$, we can
construct two possible types of candidates for the representation
on $V_1\otimes V_2$: \bea
1. && \rho(a)(v_1\otimes v_2)=\rho_1(a)v_1\otimes \rho_2(a)v_2 \;\;\;a\in C\nonumber\\
2. &&  \rho(a)(v_1\otimes v_2)=\rho_1(a)v_1\otimes v_2+v_1\otimes
\rho_2(a)v_2\;\;\; a\in C\nonumber \eea but the first is nonlinear
and the second, in general, is not a homomorphism
$(\rho(ab)=\rho(a)\rho(b))$. Thus a new structure is needed that
satisfies linearity and homomorphism property, and reflects the
associativity of the algebra. This structure is the coproduct,
defined as a $linear$ map that spits an algebra element into a sum
of elements belonging to the tensor product of algebras: \be
\Delta: \mathcal{A}\rightarrow \mathcal{A}\otimes \mathcal{A}. \ee
In this way the coproduct is a sum of tensor products and can be
indicated with the symbolic notation
$\Delta(a)=\sum_{j}a^j_{(1)}\otimes a^j_{(2)}$. Alternatively the
Sweedler notation $\Delta(a)=a_{(1)}\otimes a_{(2)}$, where the
sum over the index $j$ is implicit, is often used and we adopt it
in this work.

Using the coproduct the representation of $\mathcal{A}$ is given
by \be \rho(a)(v_1\otimes v_2)=((\rho_1\otimes \rho_2)\cdot
\Delta(a))(v_1\otimes v_2)\;\;a\in \mathcal{A} \ee To ensure the
homomorphism property of $\Delta$ and associativity of the
algebra, $\Delta$ must satisfy these conditions \bea
\Delta(a b)&=&\Delta(a)\Delta(b),\label{homo}\\
(\Delta\otimes id)\Delta &=&(id \otimes\Delta)\Delta \mbox{ $$
(coassociativity)},\label{coaxi1} \eea where $id:\mathcal{A}\to
\mathcal{A}$ denotes the identity map on $\mathcal{A}$.

It is then natural to generalize also the unit in the so called
co-unit, a map $\epsilon$ such that: \bea
\epsilon:&& \mathcal{A}\rightarrow \complex\\
(id \otimes \epsilon)\cdot\Delta&=&(\epsilon\otimes
id)\cdot\Delta=id \mbox{
$$ (counity)}\label{coaxi2} \eea
In this way, we can give the definition of a coalgebra.\\
A $coalgebra$ $C$ is a vector space over a field $\complex$
endowed with a linear coproduct $\Delta:C\rightarrow C\otimes C$
and a linear counit $\epsilon: C\rightarrow \complex$, which
satisfies the \emph{coassociativity} (\ref{coaxi1}) and
\emph{counity} (\ref{coaxi2}) properties.

Let us notice that the homomorphism property (\ref{homo}) is not
required in the mathematical definition of coalgebra. The
homomorphism, in fact, represents a compatibility condition
in a structure that is both algebra and coalgebra, the so called bialgebra.\\
A $bialgebra$ ($B,m,\eta;\Delta,\epsilon$) is a vector space that
is both an algebra and a coalgebra in a compatible way. The
compatibly is given by the following homomorphism properties \bea
\Delta(a b)=\Delta(a)\Delta(b)&&\Delta(1)=1\otimes 1\\
\epsilon(ab)=\epsilon(a)\epsilon(b)&&\epsilon(1)=1
\eea
for all $a,b\in B$. From a bialgebra one can construct a Hopf algebra.

A $Hopf$ $algebra$ ($H,m,\eta;\Delta,\epsilon,S$) is a bialgebra
endowed with a linear antipode map $S:H\rightarrow H$ such that:
\be m(S\otimes id)\Delta =m(id \otimes S)\Delta=\eta \epsilon
\label{antipode} \ee
By this definition it follows that the antipode is unique and
satisfies: \bea
S(a\cdot b)&=&S(b)S(a),\;\;S(1)=1\mbox{$ $ \emph{(algebra antirepresentation)}}\nonumber\\
(S\otimes S)\Delta(a)&=&\tau\Delta S(a) \eea $a,b\in H$ and $\tau$
represent the flip map $\tau: \tau(a\otimes b)=b\otimes a$. In a
Hopf algebra the antipode plays a role that generalizes the
concept of group inversion.

We introduce also the notion of the ``quasitriangular" Hopf
algebras, that plays a fundamental role in one of the approaches
to Quantum Groups (see for example~\cite{Drinfeld1}).

 A \emph{quasitriangular
Hopf algebra} is a Hopf algebra together with an invertible
element $\mathcal{R}=r_\al\otimes r^\al$ (summation implied) in
$\mathcal{A}\otimes \mathcal{A}$ which satisfies the relations
\bea (\Delta\otimes
id)(\mathcal{R})&=&\mathcal{R}_{13}\mathcal{R}_{23}\nn\\
(id \otimes
\Delta)(\mathcal{R})&=&\mathcal{R}_{13}\mathcal{R}_{12}\nn\\
(\tau\circ\Delta )(a)&=&\mathcal{R}\Delta
(a)\mathcal{R}^{-1}\;\;\;a\in \mathcal{A} \eea where $\tau$:
$\mathcal{A}\otimes\mathcal{A}\to \mathcal{A}\otimes\mathcal{A}$
is the flip map $a\otimes b\to b\otimes a$ ($a,b\in \mathcal{A}$),
and \bea \mathcal{R}_{12}&=&r_\al\otimes r^\al\otimes \mathbf{1}=\mathcal{R}\otimes \mathbf{1} \nn\\
\mathcal{R}_{13}&=&r_\al\otimes\mathbf{1}\otimes r^\al\nn\\
\mathcal{R}_{23}&=&\mathbf{1}\otimes r_\al\otimes
r^\al=\mathbf{1}\otimes \mathcal{R}. \eea where $\mathbf{1}$ is
the unit element of $\mathcal{A}$. $\mathcal{R}$ is called the
\emph{universal R-matrix} of $\mathcal{A}$, and, satisfies the
\emph{quantum Yang-Baxter equation} (see~\cite{Drinfeld1}): \be
\mathcal{R}_{12}\mathcal{R}_{13}\mathcal{R}_{23}=\mathcal{R}_{23}\mathcal{R}_{13}\mathcal{R}_{12}.
\ee A simple proof of this can be found writing
$\tau\circ\Delta\equiv\Delta'$ and computing: \bea
(\mathbf{1}\otimes\Delta')(\mathcal{R})&=&r_\al\otimes
\Delta'(r^\al)=r_\al\otimes
(\mathcal{R}\Delta(r^\al)\mathcal{R}^{-1})\nn\\
&=&(\mathbf{1}\otimes \mathcal{R})(r_\al\otimes
\Delta(r^\al))(\mathbf{1} \otimes \mathcal{R}^{-1})
=\mathcal{R}_{23}(id\otimes \Delta)\mathcal{R}\mathcal{R}_{23}^{-1}\nn\\
&=&\mathcal{R}_{23}\mathcal{R}_{13}\mathcal{R}_{12}\mathcal{R}_{23}^{-1}
\eea on the other hand: \bea
(\mathbf{1}\otimes\Delta')(\mathcal{R})&=&(id\otimes \tau)(id
\otimes \Delta)\mathcal{R}=(id\otimes
\tau)\mathcal{R}_{13}\mathcal{R}_{12}\nn\\
&=&\mathcal{R}_{12}\mathcal{R}_{13}. \eea Equating these two
results we get the required result. The name ``quasitriangular" is
due to the fact that an algebra is called \emph{triangular} if the
element $\mathcal{R}$ satisfies the extra relation
$\mathcal{R}_{12}\mathcal{R}_{21}=1$.
 With respect to a generic Hopf algebra, the
quasitriangular Hopf algebras have the property that the
representations on the vector-space tensor products $V_1\otimes
V_2$ and $V_2\otimes V_1$ are isomorphic, the isomorphism being
provided by the element $\mathcal{R}$. In fact if the tensor
product representation $V_1\otimes V_2$ is related to $\Delta$,
the $V_2\otimes V_1$ is related to $\Delta'=\tau\circ\Delta$. So,
$\Delta$ and $\Delta'$ are related by the invertible element
$\mathcal{R}$ (isomorphism property). In an arbitrary Hopf
algebra, instead, $\Delta$ and $\Delta'$ are not related.\\

 Another very useful notion in the framework of Hopf algebras
is the notion of \emph{duality} that we want to introduce here.
Given a vector space $V$ over $\complex$, there is a dual vector
space $V^*=Lin(V)$, such that one can introduce the inner product
map $<\!\cdot, \cdot\!>: V^*\otimes V\to \complex$.

If $C$ is a coalgebra, then the maps $\Delta$ and $\epsilon$
define adjoint maps $m:C^*\otimes C^*\to C^*$ and $\eta:\complex \to C^*$ by
\bea
<ab,c>&=&<a\otimes b,\Delta(c)>,\;\;\;a,b\in C^*,\,c\in C \\
<1_{C^*},c>&=&\epsilon(c) \eea and the coalgebra
axioms(\ref{coaxi1},\ref{coaxi2}) are just the requirements that
make $(C^*,m,\eta)$ into an algebra. On the other hand a coalgebra
$C$ defines equivalently an algebra $C^*$. \emph{Thus the theory
of coalgebras is essentially dual to the theory of algebras.}

The definition of duality can be extended to Hopf algebras. Two
Hopf algebras $H$ and $H^*$ are said to be \emph{dually paired} if
there exists a non degenerate inner product $<,>$ such that the
following axioms are satisfied \bea
<ab,c>&=&<a\otimes b,\Delta(c)>\label{ax1}\\
<1_{H^*},c>&=&\epsilon(c)\\
<\Delta(a),c\otimes d>&=&<a,cd>\\
\epsilon(a)&=&<a,1_{H}>\\
<S(a),c>&=&<a,S(c)>\label{ax5} \eea where $a,b\in H^*$ $c,d\in H$
and $<a\otimes b, c\otimes d>=<a,c><b,d>$. It is easily shown that
all the relevant consistency relations between the various
operations are satisfied.

Note that the relations above may be used constructively,
\emph{i.e.} given a Hopf algebra $H$, one can construct a dually
paired Hopf algebra $H^*$; this method is used to construct the
spacetime coordinate algebra from the Hopf algebra of the
translation generators, as we will show in the following for \kM\
spacetime, obtained by duality from the momenta sector of the
\kkP\ Hopf algebra.

One can show that to each proposition over an algebra corresponds
a dual proposition over the dual structure that is obtained by
substituting each operation over the algebra with the
corresponding operation over the dual structure. In this way one
can establish, for example, some propositions about the
commutativity $m\tau=m$ and cocommutativity $\tau\Delta=\Delta$ of
a Hopf algebra. It is easy to prove that the dual of a
\emph{commutative} Hopf algebra  is \emph{co-commutative}, and
vice-versa. In fact from the commutativity of $H$ ($cd=dc$
$\forall c,d \in H$) it follows
 \bea <a_{(1)},c><a_{(2)},d>&=&<\Delta(a),c\otimes d>=<a,cd>=<a,dc>=<a_{(1)},d><a_{(2)},c>\nn\\
&=&<a_{(2)},c><a_{(1)},d>\nn
\eea
comparing the first and the last members we find that $\tau\Delta=\Delta$.

Let us make two examples of Hopf algebra realizations that are
crucial in order to understand the Quantum Groups theory. They
also provide the example of a dual pair.
\begin{itemize}
\item Let G be a compact topological group. Consider the space of
continuous functions on G denoted by C(G) endowed with the maps:
\bea
(f\cdot h)(g)&=&f(g)h(g),\;\;\;f,h\in C(G),g\in G\\
\Delta(f)(g_1\otimes g_2)&=&f(g_1g_2),\;\;\;\;f\in C(G),g_i\in G\\
\eta(x)&=&x1,\;\; 1(g)=1\;\;\;\forall g\in G\\
\epsilon(f)&=&f(e) \;\;\;\mbox{where $e$ is the neutral element of $G$}\\
S(f)(g)&=&f(g^{-1}) \eea In this way $C(G)$ assumes a
(commutative) Hopf-algebra structure. If $G$ is commutative $C(G)$
is also cocommutative but if $G$ is not, then $C(G)$ is not
cocommutative. Note that we assume $C(G)\otimes C(G)=C(G\times G)$
for the definition of a coproduct\footnote{This is obviously true
if $G$ is finite, but it is true also in the infinite dimensional
case introducing algebraic completions.}. In this case the
coproduct expresses the group multiplication and the counit
expresses the group identity element $e$. As mentioned in the
introduction of this Section, the Gel'fan-Naimark theorem states
the  equivalence between commutative Hoph algebras $C(G)$
(actually endowed with an involution *) and compact topological
groups $G$. Considering Quantum Groups means making the Hopf
algebras $C(G)$ noncommutative.

\item Let $\mathtt{g}$ be a Lie algebra and $U(\mathtt{g})$ its
universal enveloping algebra, then $U(\mathtt{g})$ becomes a Hopf
algebra if we define \bea
\Delta(x)&=&x\otimes 1+1\otimes x, \;\;\;x\in \mathtt{g}\label{coco}\\
\eta(\lambda)&=&\lambda \mathbf{1} \\
\epsilon(x)&=&0\;\;\forall x\neq\mathbf{1},\;\;\;\epsilon(\mathbf{1})=1\\
S(x)&=&-x \eea An element $x\in \mathtt{g}$ verifying this
properties is called \emph{primitive element}. We have defined
$\Delta,\eta,\epsilon,S$ only on the subset $\mathtt{g}$ of the
universal enveloping algebra, but it is easily seen that these
maps can be extended uniquely to all of $U(\mathtt{g})$ such that
the Hopf axioms are satisfied everywhere. In particular the
property (\ref{coco}) makes the algebra $U(\mathtt{g})$
cocommutative. The coproduct here provides the rule by which
actions extend to a tensor product. Thus the cocommutativity of
the  coproduct states that when an algebra element acts on an
algebra it acts as a derivation.
\end{itemize}

It can be shown that $C(G)$ and $U(\mathtt{g})$ are a dual pair,
if $\mathtt{g}$ is the Lie algebra of G (see for
example~\cite{Tjin}).

\subsection{Actions and Coactions}
An algebra can act on other structures. A left action of an
algebra $H$ over an algebra  $\mathcal{A}$ is a linear map $\al:
H\otimes \mathcal{A}\rightarrow \mathcal{A}$ such that: \bea
\al((h\cdot g)\otimes a)&=&\al(h\otimes \al(g\otimes a))\;\;h,g\in H,\;a\in \mathcal{A}\nonumber\\
\al(h\otimes a)&=&\epsilon(h)\,a\label{act} \eea We can use the
short notation $\al(h\otimes a)=h\trir a$, so the (\ref{act}) can
be written as \bea
(hg)\trir(a)&=&h\trir(g\trir a)\\
h\trir 1&=&\epsilon(h)1. \eea Usually in physical applications,
the request of \emph{covariant action} is made in order that the
action of a Hopf algebra preserves the structure of the object on
which it acts. We say that an Hopf algebra $H$ acts $covariantly$
(from the left) over an algebra $\mathcal{A}$ (or equivalently
that $\mathcal{A}$ is a left $H$-module algebra) if $\forall h\in
H$: \bea h\trir(a\cdot b)=(h_{(1)}\trir a)(h_{(2)}\trir b), &&
a,b\in \mathcal{A}\nonumber \eea

The action of $H$ over a coalgebra $C$ ($C$ is a left $H$-module
coalgebra) states that: \bea \Delta(h\trir c)&=&(h_{(1)}\trir c_{(1)})\otimes(h_{(2)}\trir c_{(2)})=(\Delta h)\trir\Delta c\nonumber\\
\epsilon(h\trir c)&=&\epsilon (h) \epsilon (c),\;\;c\in C
\eea

In the same way it can be defined a right action of a Hopf algebra
$H$ on $\mathcal{A}$ (algebra, coalgebra or Hopf algebra):
$$\tril: \mathcal{A}\otimes H\rightarrow \mathcal{A}$$
and a \emph{covariant} right-action should satisfy \bea (a\cdot
b)\tril h=(a\tril h_{(1)})(b\tril h_{(2)}), && h\in H,\;a,b\in
\mathcal{A}\nonumber \eea The duality relations connect the left
action over an algebra $\mathcal{A}$ and the corresponding  right
dual action over the dual coalgebra $\mathcal{A}^*$ in the
following way \be <a,h\trir b>=<a\tril^* h,b>, \;\;\;b\in
\mathcal{A},\,a\in \mathcal{A}^*,\,h\in H \ee

Let us make some examples of left actions such as the regular,
canonical and the adjoint one.
\begin{itemize}
\item The left and right \emph{regular actions} of an algebra on
itself are linear maps $\mathcal{A}\otimes \mathcal{A}\rightarrow
\mathcal{A}$ defined as \be
a\stackrel{reg}{\trir}b=ab=a\stackrel{reg}{\tril}b,\;\;\; a,b\in
\mathcal{A}\;\;\;a,b\in \mathcal{A}
 \ee
the first action is the left regular action of $a$ over $b$
whereas the second one is the right action of $b$ over $a$. This
action is not covariant. If one tries to ``dualize" the left
regular action of the algebra $\mathcal{A}$ one finds that \bea
<a,b\stackrel{reg}{\trir}c>&=&<a\stackrel{reg*}{\tril}b,c>,\;\;\;b,c\in
\mathcal{A},\;a\in \mathcal{A}^*\nn \eea The left side of this
equation is \bea <a,bc>=<\Delta(a),b\otimes
c>&=&<a_{(1)},b><a_{(2)},c>, \eea thus comparing with the right
side, one finds \be a\stackrel{reg*}{\tril}b=<a_{(1)},b>a_{(2)}.
 \ee
 This
action defines the right canonical action
$\stackrel{reg*}{\tril}\equiv \stackrel{can}{\tril}$ and in the
same way one can derive the definition of the left canonical
action. \item The left and right \emph{canonical actions} of an
algebra $\mathcal{A}$ over the dual coalgebra $C\equiv
\mathcal{A}^*$ are defined as: \bea
c\stackrel{can}{\trir}a=c_{(1)}<a,c_{(2)}>, && \nn\\
c\stackrel{can}{\tril}a=<c_{(1)},a>c_{(2)},&& a\in \mathcal{A},c\in C
\eea
These actions are covariant.
\item The adjoint left and right actions of a Hopf algebra $H$ on itself are  linear maps $H\otimes H\rightarrow H$ such that
\bea
a\stackrel{ad}{\trir}b&=&a_{(1)}b S(a_{(2)}),\\
b\stackrel{ad}{\tril}a&=&S(a_{(1)})ba_{(2)},\;\;\;a,b\in H \eea
These actions are covariant as well.
\end{itemize}

Some of the actions introduced above have a role in the framework
of the symmetry transformations of the spacetime. We want to show,
for example, how the notion of canonical action reduces to the
physical notion of translation in the case it is applied to the
standard generators of the Poincar\'e-translations acting on its
dual space (\emph{i.e.} the commutative Minkowski space). The
translation sector $T$ of the Poincar\'e algebra is a Lie algebra
generated by the operators $P_\mu$ that has the following Hopf
algebra structure: \be [P_{\mu},P_{\nu}]=0,\;\;\;\Delta
P_{\mu}=P_{\mu}\otimes 1+1\otimes
P_{\mu},\;\;\;\epsilon(P_\mu)=0,\;\;\;S(P_{\mu})=0.\label{triv}
\ee The duality axioms (\ref{ax1}--\ref{ax5}) allow us to
reconstruct the Hopf-algebra structure of the Minkowski space
$\mathcal{M}$ from the Hopf-algebra structure of it dual space
$T$. In fact, assuming that the duality relations between the
generators $P_{\mu}$ of $T$ and the generators $x_{\mu}$ of its
dual space $\mathcal{M}=T^*$ be\footnote{We will usually use
four-dimensional Greek indexes ($\mu,\nu=0,\ldots,3$) and three
dimensional Latin indexes ($j,k=1,2,3$) and we use a $(+,-,-,-)$
signature.} \be <P_{\mu},x_{\nu}>=-i\eta_{\mu\nu}, \label{dual}\ee
one can easily find that also $\mathcal{M}$ has a Lie-algebra
structure: \be [x_{\mu},x_{\nu}]=0,\;\;\;\Delta
(x_{\mu})=x_{\mu}\otimes1 +1\otimes x_{\mu},\;\;\;S(x_{\mu})=-x.
\ee Using these relations, the canonical action of $P_{\mu}\in T$
on $x_{\mu}\in T^*$ can be obtained: \be
P^{\mu}\stackrel{can}{\trir}
x^{\nu}=x^{\nu}_{(1)}<P^{\mu},x^{\nu}_{(2)}>=<P^{\mu},x^{\nu}>=-i\eta^{\mu\nu}
\ee This is just the usual definition of the Poincar\'e
translation in the  case of commutative Minkowski spacetime in
which the translation generators take the differential form
$P_{\mu}=-i\partial_{\mu}$ and its action over the coordinates is
just $P_{\mu}x_{\mu}=-i\eta_{\mu\nu}$.

The dual concept to the action of an algebra  is the
\emph{coaction} of a coalgebra. The left coaction of a coalgebra
$C$ over an algebra $\mathcal{A}$ is defined as a linear
application $\bt_L:\mathcal{A}\rightarrow C\otimes \mathcal{A}$.
The map $\bt_L$ satisfies: \bea
(id\otimes \bt_L)\circ\bt_L&=&(\Delta\otimes id)\circ\bt_L\\
(\epsilon\otimes id)\circ\bt_L&=&id \eea The coaction gives a
corepresentation of the coalgebra. One can find in the literature
the coaction denoted by $\Delta_L$ (left coaction), $\Delta_R$
(right coaction). A $covariant$ coaction is required to respect
the algebra structure on which it (co)acts. Thus: \be
\bt_L(ab)=\bt_L(a)\bt_L(b),\;\;\; \bt_L(1)=1\otimes 1,\;\;a,b\in
A. \ee We will adopt the following notation for the coaction:
\be \bt(a)=\sum_ia^{\uno}_i\otimes a^{\due}_i=a^{(\bar{1})}\otimes
a^{(\bar{2})}\;\;\;a,a^{(\bar{2})}\in A,\;a^{(\bar{1})}\in C. \ee

\subsection{Bicrossproduct Hopf algebras}
The notions of action and coaction allow us to define a special
class of algebras that can be constructed by the composition of
two Hopf algebras. These algebras are called \emph{bicrossproduct
algebras} and take an important role in our study since \kkP\
algebra has been showed to be of this type~\cite{MajidRuegg}.
Consider two Hopf algebras, $\mathcal{A}$ and $\mathcal{X}$.
Suppose that we know a right action $\tril$ of the algebra
$\mathcal{X}$ over the algebra  $\mathcal{A}$ and a left coaction
of  $\mathcal{A}$ on $\mathcal{X}$: \bea
\tril:&& \mathcal{A}\otimes \mathcal{X}\rightarrow \mathcal{A},\\
\bt_L:&& \mathcal{X}\rightarrow \mathcal{A}\otimes \mathcal{X}.
\eea For the coaction of $\mathcal{A}$ on the element $x\in
\mathcal{X}$ we adopt the notation $\bt_L(x)\equiv x^{\uno}\otimes
x^{\due}$ with $x^{\uno}\in \mathcal{A}$. We require the following
properties for the action: \bea
\epsilon (a\tril x)&=&\epsilon (a)\epsilon(x)\label{bicrosscomp1}\\
\Delta(a\tril x)&=&(a_{(1)}\tril x_{(1)}){x_{(2)}}^{\uno}\otimes
(a_{(2)}\tril {x_{(2)}}^{\due})\label{bicrosscomp2} \eea and for
the coaction: \bea
\bt_L(xy)&=&(x^{\uno}\tril y_{(1)}){y_{(2)}}^{\uno}\otimes x^{\due}{y_{(2)}}^{\due}\label{bicrosscomp3}\\
{x_{(1)}}^{\uno}(a\tril x_{(2)})\otimes {x_{(1)}}^{\due}&=&(a\tril
x_{(1)}){x_{(2)}}^{\uno}\otimes
{x_{(2)}}^{\due}\label{bicrosscomp4} \eea $a\in \mathcal{A}$,
$x,y\in \mathcal{X}$. A bicrossproduct algebra can be constructed
from two algebras $\mathcal{A}$ and $\mathcal{X}$ acting and
coacting in the following way.\\
A \emph{bicrossproduct algebra} (usually indicated with
the symbol $\mathcal{X}{\wb}\mathcal{A}$) is the tensor
product algebra $\mathcal{X}\otimes \mathcal{A}$ endowed with the maps:
$$
\begin{array}{rlll}
(x\otimes a)\cdot (y\otimes b)&=&xy_{(1)}\otimes (a\tril y_{(2)})b &\mbox{(product)}\\
&&&\nn\\
1_{\mathcal{X}{\wb} \mathcal{A}}&=&1_{\mathcal{X}}\otimes 1_\mathcal{A}&\mbox{ (unity)}\\
&&&\nn\\
\Delta(x\otimes a)&=&[x_{(1)}\otimes {x_{(2)}}^{\uno} a_{(1)}]\otimes [{x_{(2)}}^{\due}\otimes a_{(2)}] &\mbox{(coproduct)}\\
&&&\nn\\
\epsilon(x\otimes a)&=&\epsilon (x)\epsilon (a)&\mbox{ (counit)}\\
&&&\nn\\
S(x\otimes a)&=&(1_\mathcal{X}\otimes S(x^{\uno}a))\cdot (S(x^{\due})\otimes 1_\mathcal{A})&\mbox{ (antipode)}
\end{array}
$$
\be \label{bicrosscond} \ee such that the action $\tril$ and the
coaction $\bt_L$ satisfy the compatibility conditions
(\ref{bicrosscomp1})-(\ref{bicrosscomp4}).

One can consider as generators of the bicrossproduct algebra
$\mathcal{X}{\wb} \mathcal{A}$ the elements of the type
$A=1\otimes a$, with $a\in \mathcal{A}$, and the elements of the
type $X=x\otimes 1$, with $x\in \mathcal{X}$. In fact, following
the definitions above, the single element $x\otimes a\in
\mathcal{X}{\wb} \mathcal{A}$ is given by the product $XA$: \be X
A=(x\otimes 1)(1\otimes a)=x\otimes (1\tril 1)a=x\otimes a, \ee
while the other product $AX$ is: \be A X=(1\otimes a) (x\otimes
1)=x_{(1)}\otimes (a\tril x_{(2)}). \ee Thus, the bicrossproduct
algebra $\mathcal{X}{\wb} \mathcal{A}$ can be viewed as the
enveloping algebra generated by $X$ and $A$, modulo the
commutation relations: \be [X,A]=x\otimes a-x_{(1)}\otimes(a\tril
x_{(2)})\label{bicrosscommutator}. \ee The conditions
(\ref{bicrosscomp1})-(\ref{bicrosscomp1}) and the definitions
(\ref{bicrosscond})  allow to construct the bicrossproduct algebra
$\mathcal{X}{\wb} \mathcal{A}$, if the Hopf-algebra structures of
$\mathcal{X}$ and $\mathcal{A}$ are fully known. The
bicrossproduct \kkP\ algebra  $U(so(1,3)){\wb} \mathcal{T}$ is
constructed in this way, choosing $\mathcal{X}=U(so(1,3))$, the
Lie algebra of the Lorentz rotations, and $\mathcal{A}=\mT$, the
algebra of the Poincar\'e translations with a deformed coalgebra
sector ($\mT$ is a non-trivial Hopf algebra). In particular, in
the Majid-Ruegg construction~\cite{MajidRuegg}, the (deformed)
coalgebra of $\mT$ is chosen such that $\mT$ is a dual space to
\kM , \emph{i.e} to the Lie algebra generated by the elements
$\x_{\mu}$ which satisfy the commutation relations
$[\x_0,\x_j]=i\lambda x_j, [\x_j,\x_k]=0$.

\subsection{Quantum Group approach}

As mentioned in the beginning of this Section, Quantum Groups can
be viewed from different sides: they arose as generalized symmetry
groups in the context of quantum integrable
theories~\cite{Drinfeld1,Jimbo}, but they are useful also in very
different contexts, as in some approaches to Quantum
Gravity~\cite{MajPhD}.

\bigskip
\bigskip
\bigskip

We begin by giving a brief overview on the first (and more famous)
class of Quantum Groups, the so called ``quasitriangular" Quantum
Groups, which emerge as generalized symmetries in the field of
exactly solvable lattice models, as mentioned
above~\cite{Drinfeld1}\cite{FRT}.
 In this approach, the notion of classical symmetry group
coming from classical geometry is naturally generalized to a
quantum group symmetry. These quantum groups depend on some
deformation parameter $q$. The deformation induced by $q$ can be
viewed as a quantization of the symmetry. These groups induce
noncommutativity on the objects on which they act.

To be precise the \emph{quasitriangular} Quantum Groups include
the deformations \be U_q(\mathtt{g}) \ee of the enveloping algebra
$U(\mathtt{g})$ of every complex-semisimple Lie algebra
$\mathtt{g}$. They have the same number of generators as the usual
Lie groups but modified relations and additional structures such
as the ``coproduct". Although these quantum groups arise as
generalized symmetries in certain lattice models, they are also
visible in the continuum limit quantum field theory such as the
Wess-Zumino-Novikov-Witten model~\cite{WZNW}. The coordinate
algebra $C_q[G]$ associated to these quantum groups has the
structure of quantum groups as well, deforming the commutative
algebra of coordinate functions on $G$.

A simple example of nontrivial quasitriangular Hopf algebra can be
obtained from $su(2)$. We will obtain a noncommutative and
noncocommutative Hopf algebra denoted by $U_q(su(2))$, that is a
``deformed" universal enveloping algebra of $su(2)$. This algebra
was first introduced by Drinfeld and Jimbo. Let $U(su(2))$ be the
universal enveloping algebra of the three generators $H,X_+,X_-$
modulo the
Jimbo-Drinfel relations: \bea {}[H,X_\pm ]&=&\pm 2 X_\pm \nn\\
 {}[X_+,X_-]&=&q^{1-H}[H]_q \label{suqalg}
 \eea
where the symbol $[x]_q\equiv\frac{q^{2x}-1}{q^2-1}$, and $q\in
\mathbb{R}$. The coproducts, counits and antipodes are given by:
\bea \Delta(H)=H\otimes 1+1\otimes H,
&&\Delta(X_{\pm})=X_{\pm}\otimes q^{H/2}+q^{-H/2}\otimes X_{\pm},\nn\\
\epsilon(H)=\epsilon(X_{\pm})=0&&\nn\\
S(H)=-H&&S(X_{\pm})=-q^{\pm}X_{\pm}. \label{suqcoalg}\eea The
universal $\mathcal{R}$ matrix for this algebra is given by: $$
\mathcal{R}=\sum_{n=0}^{\infty}\frac{(1-q^{-2})^n}{[n]_q!}q^{\frac{1}{2}(H\otimes
H+nH\otimes \mathbf{1}-n \mathbf{1}\otimes H)}X^n_+\otimes X^n_-.
$$ where the symbol $[n]_q!$ is
$$[n]_q!=\left\{
\begin{array}{ll}
1&n=0,\nn\\
\Pi_{m=1}^n[m]_q& n=1,2,...
\end{array}
\right.
$$
In the classical limit $q\to 1$ $U_q(su(2))\to U(su(2))$ and the
relations (\ref{suqalg},\ref{suqcoalg}) describe the familiar
$su(2)$ Lie algebra. The R-matrix is given by:
$\mathcal{R}=\mathbf{1}\otimes \mathbf{1}$ and satisfies trivially
all the appropriate requests of a $triangular$ Hopf algebra. Thus,
the classical limit $U(su(2)$ can be view as a ``trivial"
quasitriangular Hopf algebra.

Associated to these types of Quantum Groups there are
noncommutative homogeneous $q$-spaces (whose these Quantum Groups
are symmetry), in which noncommutativity is controlled by the
parameter $q$ as well. In particular, in \cite{car}
$q$-deformations of the Lorentz algebra are introduced. The
correspondent quantum spaces called $q$-Minkowski are
characterized by relations of the type: $$
[\x_j,\x_0]=0,\;\;\;[\x_k,\x_k]\neq0. $$\\

Quantum Group of bicrossproduct type have been proposed by S.
Majid~\cite{Maj88} as models to unify quantum theory and gravity
at the Planck-scale physics. This point of view is alternative to
the idea of quantization of a theory as the result of a process
applied to an underlying classical space (as for example in the
case of deformation quantization, that is based on the classical
notion of Poisson brackets or the case of String Theory
description of Quantum Gravity where one quantizes strings moving
in a classical spacetime). Models should instead be built guided
by the intrinsic Noncommutative Geometry at the level of
noncommutative algebras. Only at the end one can consider
classical geometry (with Poisson brackets) as classical limits and
not as a starting point, like in \emph{deformation quantization}.
In a quantum world, in fact, phase-space and probably spacetime
should be ``fuzzy" and only approximately described by the
classical geometry.

The \emph{bicrossproduct} Quantum Groups can be represented in the
form \be U({\mathtt{g}})\wb C[M] \ee associated to the
factorization of a Lie group $X$ into two Lie groups, $X=GM$.
$C[M]$ is the commutative coordinate algebra and $U(\mathtt{g})$
is the enveloping algebra of the Lie algebra $\mathtt{g}$ of $G$.

The idea at the basis of the introduction of these Quantum Groups,
in a Quantum Gravity approach~\cite{Maj88}, is a principle of
\emph{self-duality}, that is peculiar in the Hopf algebras and
that should allow to put \emph{quantum mechanics} and
\emph{gravity} on equal (but mutually dual) footing. This line of
thinking led in the mid 80's to the Planck-scale Quantum Group
$C[q]\wb _{\hbar,G_N}C[p]$ generated by $\q,\p$ with the
relations: \be [\q,\p]=i\hbar(1-e^{-\q/G_N}),\;\;\Delta
\q=\q\otimes1+1\otimes \q,\;\;\Delta \p=\p\otimes
1+e^{-\q/G_N}\otimes \p. \label{bicrossqg} \ee where $G_N$ is the
Newton constant. In any situation where $\q$ can effectively be
treated as having values $>0$, \emph{i.e.} for quantum states
where the particle is confined to this region, the quantum flat
space with Heisenberg algebra is recovered in the limit $G_N\to
0$. If we are given $C[q]$, the position coordinate algebra, and
$C[p]$, defined a priori as the natural momentum coordinate
algebra, then all possible quantum phase spaces built from $q,p$
in a way that preserves the symmetry between $q$ and $p$ and
retains the group structure of the classical phase space as a
quantum group, are of this form, labelled by two parameters
$\hbar,G_N$~\cite{maj00}. Notice that the degree of
noncocommutativity in the coproduct on the quantum algebra of the
momenta is controlled by Newton's constant $G_N$. Such a
noncommutativity measures curvature in the momentum space (called
``cogravity" in the Majid language) and it is connected with the
presence of  noncommutativity in the position space, if the
momentum and position space are dual. In order to explain this
concept we make the example of the classical phase space
$(q_j,p_j)\in \real^{2n},j=1,...,n$. In this case we can consider
the group $G$ of elements $W_k=e^{ik_jq_j}$ labelled by the
parameter $k_j\in \real^n$. This group has an abelian composition
law, $W_{k_1}W_{k_2}=W_{k_1+k_2}$. The algebra of the functions of
positions is given by the enveloping algebra $U(\mathtt{g})$
(where $\mathtt{g}$ the Lie algebra of $G$). The algebra of the
momenta can be viewed as the algebra $C[G]$ dual to
$U(\mathtt{g})$, and the generators $p_j$ of this algebra can be
introduced via the relations: \be p_j(W_k)\equiv <p_j,W_k>=k_j \ee
that follow from the duality relation $<p_j,q_l>=-i\delta_{jl}$
(see \ref{dual}). Thus: \be
p_j(W_{k_1}W_{k_2})=p_j(W_{k_1+k_2})=(k_1+k_2)_j.\ee On the other
hand (by duality), \bea
p_j(W_{k_1}W_{k_2})&=&<p_j,W_{k_1}W_{k_2}>=<\Delta(p_j),W_{k_1}\otimes
W_{k_2}> \nn\\
&=&\Delta(p_j)(W_{k_1}\otimes W_{k_2}).\eea Thus the coproduct
turns out to be: $\Delta(p_j)=p_j\otimes 1+1\otimes p_j$. From
this example one can see the relation between the coproduct and
the composition law in the momentum sector. The commutativity of
the positions is connected with the cocommutativity of coproduct
in the space of momenta (then the momentum space is flat, \emph{it
has a abelian composition law}). If the space of positions is
noncommutative, the group law of $G$ will be in general
non-abelian $W_{k_1}W_{k_2}\neq W_{k_1+k_2}$ and the coproduct of
momenta will be noncocommutative. Thus a noncommutative position
space corresponds to a curved momentum space. The existence of
$self-duality$ however states also the opposite: a curved space in
positions corresponds to a noncommutative momentum space. For
example, when the position space is a 3-sphere $S^3$ the natural
momentum is $su(2)$, with relations: \be
[p_j,p_k]=\frac{i}{R}\epsilon_{jkl}p_l\ee where R is proportional
to the radious of curvature of the $S^3$.

The search for a quantum algebra of observables which is a Hopf
algebra translates in a search of a simple model in which quantum
and gravitational effects are unified and in which they are dual
to each other. In this optics one would have something that is a
quantum algebra of observables of a quantum system and at the same
time preserves something of the geometrical structure of phase
space in the quantum case.

As in the case of Quantum Groups of the type $U_q(g)$, also in the
case of bicrossproduct Quantum Groups there are corresponding
homogeneous spaces.  \kM\ noncommutative spacetime \be
[\x_0,\x_j]=\frac{i}{\kappa}\x_j,\;\;\;[\x_j,\x_k]=0\label{kmincomm}\ee
arose just as the homogeneous spacetime of a bicrossproduct
realization of \kkP\ algebra~\cite{MajidRuegg}. This is the main
reason because we are interested in the bicrossproduct Quantum
Group.

Notice that the $1+1$-dimension algebra (\ref{kmincomm}) turns out
to be a particular case of algebra (\ref{bicrossqg}). It can be
obtained making in (\ref{bicrossqg}) the limit
$G\to\infty,\hbar\to\infty$ with $\frac{G}{\hbar}=\kappa$. So,
part of the study concerning the the bicrossproduct Hopf algebra
(\ref{bicrossqg}) proves useful in the study of \kM\
noncommutative spacetime. In fact, some technical notions that we
will employ in this thesis, such as the introduction of a rule of
integration in \kM , originate in the framework of the
bicrossproduct algebras (\ref{bicrossqg}).

\section{Deformation of the Poincar\'e algebra and \kM\ spacetime}

In the framework of the Quantum Groups $U_q(\mathtt{g})$, the
deformation of the Poincar\'e group has attracted much attention
in the early 1990s for the motivations mostly arising from Quantum
Gravity, in which a loss of the classical Lorentz symmetry has
been predicted due to the existence of a \emph{minimum length}.
Different approaches have been attempted in this direction,
but a particular approach, which has found several implications,
has consisted in looking for a deformation of the algebra rather
than the group. A very interesting technique used in this context
is the contraction procedure introduced in~\cite{firenzegroup}.
One first consider the $q$-deformation of the anti-de Sitter
algebra $U(so(3,2))$. This can be done with the standard
Drinfeld-Jimbo method~\cite{Drinfeld1,Jimbo} that introduces a
dimensionless deformation parameter $q$. Then one sends to
infinity the de Sitter radius $R$ while $q\to 1$ in such a way
that $R\log q= \kappa^{-1}$ fixed. Following this procedure one
recovers a quantum deformation  $U_{\kappa}(\mathcal{P}_4)$ of the
Poincar\'{e} algebra $\mathcal{P}_4$  which depends on a
$dimensionful$ parameter $\kappa$. In this way a fundamental
length $\lambda=\kappa^{-1}$ enters the theory\footnote{The
notation with the parameter $\lambda=1/\kappa$ is also common, and
we prefer it since our emphasis is on the structure of the \kM\
spacetime,  thus we find convenient to write formulas in terms of
the length scale $\lambda$ rather then the dimensionful parameter
$\kappa$ usually used in the \kkP\ approach. However the two
deformation parameter $\lambda$ and $\kappa$ are relied by the
simple reaction $\lambda=\frac{\hbar}{c\kappa}=\kappa^{-1}$, since
we work in units such that the speed-of-light and the Planck
constant are $c=\hbar=1$. }. This quantum algebra has been
obtained firstly in~\cite{LNRT}
in the so called \emph{standard basis}, whose characteristic commutation relations are:
\bea
&&{}[P_\mu,P_\nu]=0,\nn\\
&&{} [M_j,P_0]=0, \;\;\;\;\;\;[M_j,P_k]=i\epsilon_{jkl}P_l\nn\\
&&\qs N_j,P_0\qd =iP_j,\;\;\;\;\;\;[N_j,P_k]=i\delta_{jk}\lambda^{-1}\sinh(\lambda P_0), \nn\\
&&{}[M_j,M_k]=i\epsilon_{jkl}M_l,\;\;\;\;\;\;\qs M_j,N_k\qd =i\epsilon_{jkl}N_l,\nn\\
&&{} [N_j,N_k]=-i\epsilon_{jkl}(M_l\cosh(\lambda
P_0)-\frac{\lambda^2}{4}P_l\vec{P}\cdot\vec{M}). \eea where
$P_{\mu}$ are the four-momentum generators, $M_j$ are the spatial
rotation generators and $N_j$ are the boost generators. The
algebra obtained in this way contains  the subalgebra of the
classical rotations $O(3)$.  The cross-relations between the boost
and rotation generators are instead deformed, and consequently the
full Lorentz sector do not form a sub-algebra. The coalgebra
sector of the \kkP\ standard basis is given by: \bea
&&\Delta P_0=P_0\otimes 1 + 1 \otimes P_0,
\;\;\;\Delta P_j=P_j\otimes e^{\lambda P_0/2}+e^{-\lambda P_0/2}\otimes P_j, \label{stanbas}\\
&&\Delta M_j=M_j\otimes 1+1\otimes M_j,\\
&&\Delta N_j=N_j\otimes e^{\frac{\lambda
P_0}{2}}+e^{-\frac{\lambda P_0}{2}}\otimes N_j+\frac{\lambda}{2}
\epsilon_{jkl}(P_k\otimes M_le^{\frac{\lambda
P_0}{2}}+e^{-\frac{\lambda P_0}{2}}M_k\otimes P_l) \eea The mass
Casimir $C_\lambda(P)$, {i.e.} the function that commutes with all
the generators of the algebra, is: \be
C_\lambda(P)=(2\lambda^{-1}\sinh(\frac{\lambda
P_0}{2}))^2-\vec{P}^2\stackrel{\lambda\to 0}{\longrightarrow}
P_0^2-\vec{P}^2 \ee it provides a deformation of the Casimir of
Poincar\'e algebra $C(P)=P_0^2-\vec{P}^2$.

\kM\ noncommutative spacetime, whose coordinates satisfy the
commutation relations $[\x_0,\x_j]=i\lambda
\x_j,\;\;[\x_j,\x_k]=0,$ as we already noted in
Eq.~(\ref{kmincomm}), is shown to be the spacetime associated to
\kkP\ algebra as we discuss below.

In the Quantum Group language it is said that the pair of a Hopf
algebra and its dual determines a generalized phase space,
\emph{i.e.} the space of the generalized momenta and the
corresponding generalized coordinates. The quantum algebra
$U_\kappa(\mathcal{P}_4)$ contains a translation subalgebra, and
it is natural to consider the dual of the enveloping algebra of
translations as  \kM\ space.  This space must necessarily be
noncommutative, because  the duality axioms (see \ref{ax1}) state
that a non-cocommutative algebra in the momenta corresponds  to a
non commutative algebra in the spacetime coordinates. So, the
non-cocommutative relations~\eqn{stanbas} imply that the
generators of the dual space (spacetime coordinates) do not
commute, and their commutation relations are given
by~\eqn{kmincomm}. It is very easy to show that  the dual Hopf
algebra  $\mT^*$ of the momentum sector $\mT$ of the \kkP\
algebra, $\mT\subset U_{\kappa}(\mathcal{P}_4)$, is the Hopf
algebra of the \kM\ generators $\x_{\mu}$. We can assume the
duality relations: \be <\x_{\mu},P_{\nu}>=-i\eta_{\mu\nu} \ee and,
applying the duality axioms, we can determine the Hopf algebra of
$\x_{\mu}$ if we know the Hopf algebra of $P_\mu$. For example,
using the axiom (\ref{ax1}) and the coproducts (\ref{stanbas}),
one finds: \bea
<[\x_0,\x_j],P_k>&=&<\x_0\otimes \x_j,\Delta P_k>-<\x_j\otimes \x_0,\Delta P_k>\nn\\
&=&<\x_0\otimes \x_j,P_k\otimes 1+e^{-\lambda P_0}\otimes P_k>-<\x_j\otimes \x_0,P_k\otimes 1+e^{-\lambda P_0}\otimes P_k>\nn\\
&=&<\x_0,e^{-\lambda P_0}> <\x_j,P_k>-<\x_j,e^{-\lambda P_0}> <\x_0,P_k>\nn\\
&=&<\x_0,e^{-\lambda P_0}><\x_j,P_k>=-\lambda <\x_0,P_0><\x_j,P_k>=<[i\lambda\x_j],P_k>\nn
\eea
from which it follows:
$$
[\x_0,\x_j]=-i\lambda x_j,
$$
that is the non-zero commutator between space and time
``coordinates" of \kM.

However, one expects that the quantum deformation of a group
symmetry (such as $U_\kappa(\mP_4)$) represents, in some sense, a
``quantum symmetry" for the corresponding homogeneous space. In
our case, for example, \kkP\ is expected to act on \kM\ spacetime
in a covariant way, preserving its algebra structure. For this
reason, a new \kkP\ basis has been introduced, in which the
``covariance" of its action on \kM\ is clearly manifest.  This is
the case of the Majid-Ruegg \emph{bicrossproduct basis} introduced
in~\cite{MajidRuegg}.

One has a large freedom in the choice of the generators of the
quantum algebra $U_\kappa(\mP_4)$. One can define a very large
number of bases through nonlinear combinations of the generators.
Then the choice of the generators of $U_{\kappa}(\mathcal{P}_4)$
is not unique, different choice of the  basic generators modify
the form of the \kkP\ Hopf algebra in the algebra sector
(\emph{i.e.} the commutation relations among generators) and in
the coalgebra sector (\emph{i.e.} the form of the coproduct and
the counit). It has been found in~\cite{MajidRuegg} that the
$\kappa$-deformed Poincar\'e algebra, in a particular choice of
generator basis, has a manifest structure of bicrossproduct Hopf
algebra $U(so(1,3))\wb \mT$, \emph{i.e} the semidirect product of
the classical Lorentz group $so(1,3)$ acting in a deformed way on
the momentum sector $\mT$, and in which also the coalgebra is
semidirect with a back-reaction of the momentum sector on the
Lorentz rotations. The following change of variables: \be
\mathcal{P}_0=-P_0,\;\;\mathcal{P}_j=-P_je^{-\frac{\lambda P_0}{2}},\;\;\;\mathcal{N}_j=N_je^{-\frac{\lambda P_0}{2}}-\frac{\lambda}{2}\epsilon_{jkl}M_kP_le^{-\frac{\lambda P_0}{2}}
\ee leads to the \kkP\ algebra in the so called \emph{Majid-Ruegg
bicrossproduct basis} in which the Lorentz sector is not deformed.
The deformation occurs only in the cross-relations between the
Lorentz and translational sectors: \bea
{}[\mP_\mu,\mP_\nu]&=&0\nn\\ {}[M_j,\mP_k]&=&i\epsilon_{jkl}\mP_l
\nn\\{}[M_j,P_0]&=&0\nn\\
{}[\mathcal{N}_j,\mP_k]&=&i\delta_{jk}\left(\frac{1}{2\lambda}(1-e^{-2\lambda
P_0})+\frac{\lambda}{2} \mP^2\right)-i\lambda \mP_j\mP_k\nn\\
\qs \mathcal{N}_j,P_0\qd &=&i\mP_j \label{kappapoincare}
\eea
and the Lorentz subalgebra remains classical:
\bea
[M_j,M_k]&=&i\epsilon_{jkl}M_l\nn\\
\qs M_j,\mathcal{N}_k\qd &=&i\epsilon_{jkl}\mathcal{N}_l\nn\\
\qs \mathcal{N}_j,\mathcal{N}_k\qd &=&-i\epsilon_{jkl}M_l
\eea
All these commutation relations becomes the standard ones for
$\lambda\to 0$.
The coproducts are given by
\bea
\Delta P_0&=& P_0\otimes 1 + 1 \otimes P_0\nn\\
 \Delta M_j&=& M_j\otimes 1 +1 \otimes M_j\nn\\
 \Delta \mP_j&=&\mP_j\otimes 1+e^{-\lambda P_0}\otimes \mP_j\nn\\
  \Delta \mathcal{N}_j
&=&  \mathcal{N}_j\otimes 1+e^{-\lambda P_0}\otimes
\mathcal{N}_j-\lambda\varepsilon_{jkl}\mP_k\otimes M_l
\label{coproducts} \eea and the antipodes are: $$
S(\mathcal{P}_j)=-e^{\lambda
P_0}\mathcal{P}_j,\;\;S(\mathcal{P}_j)=-\mathcal{P}_j,\;\;\;S(M_j)=-M_j,\;\;\;S(\mathcal{N}_j)=-e^{\lambda
P_0}+\lambda\epsilon_{jkl}{\mP}_k\otimes M_l.
$$
 The mass
Casimir of this algebra, \emph{i.e.} the function that commute
with all the generators of the algebra is given by: \be
C_\lambda(\mP)=\frac{e^{\lambda P_0}+e^{-\lambda
P_0}-2}{\lambda^2}-\vec{\mP}^2e^{\lambda
P_0}\;\;\;\stackrel{\lambda=0}{\longrightarrow}\;\;\;P_0^2-\vec{P^2}
\label{majruecasimir} \ee this deformation of the Poincar\'e
Casimir has led to many discussions about the phenomenological
implications of a deformed group symmetry. This is essentially due
to the connections of \kkP\ with \kM\ spacetime, in which the
relation (\ref{majruecasimir}) is considered to have the
interpretation of deformed dispersion relation for particles.

Expressing the \kkP\ generators in this basis it is possible to
see that $\kappa$-Poincar\'{e} acts covariantly as a Hopf algebra
on \kM\ spacetime. In this way the commutation
relations~\eqn{kmincomm} that characterize \kM\ remain unchanged
under the action of \kkP\ algebra, this is consistent with the
notion of quantum group symmetry that should preserve a covariant
action over the associated homogeneous space. In the limit
$\kappa\to\infty$ ($\lambda\to 0$) one recovers the standard
Minkowski space, with the ordinary Poincar\'{e} group.

Let us analyze more deeply the bicrossproduct structure of the
Majid-Ruegg basis and let us show that it acts $covariantly$ on
\kM . In doing this, we refer to the Section (1.2), where the
general theory of bicrossproduct algebras is presented. Consider
the \kkP\ generators written in the following form: \be
\mP_{\mu}=1\otimes p_{\mu},\;\;\;M_j=m_j\otimes
1,\;\;\;\mN_j=n_j\otimes 1 \ee where $p_{\mu},m_j,n_j$ are the
generators of the standard Poincar\'e algebra $U(so(1,3))$. From
(\ref{bicrosscommutator}) one finds that the commutators among the
\kkP\ generators are connected with the action $\tril$ in the
following way: \bea
{}&&[M_j,P_0]=M_jP_0-m_j\otimes p_0-1\otimes (p_0\tril m_j)=-1\otimes (p_0\tril m_j)\\
{}&&[M_j,\mP_k]=M_jP_k-m_j\otimes p_k-1\otimes (p_k\tril m_j)=-1\otimes (p_k\tril m_j)\\
{}&&[\mN_j,P_0]=\mN_jP_0-n_j\otimes p_0-1\otimes (p_0\tril n_j)=-1\otimes (p_0\tril n_j)\\
{}&&[\mN_j,\mP_k]=\mN_j\mP_k-n_j\otimes p_k-1\otimes (p_k\tril
n_j)=-1\otimes (p_k\tril n_j) \eea where we used that
$M_j\mP_{\mu}=m_j\otimes p_{\mu}$ and $\mN_j\mP_{\mu}=n_j\otimes
p_{\mu}$, that follow from the definition of the product in the
bicrossproduct algebra. Then, substituting the expressions of the
commutators (\ref{kappapoincare}) in the previous relations, one
finds: \bea
{}&&p_0\tril m_j=0             \\
{}&& p_k\tril m_j=-i\epsilon_{jkl}p_l\\
&&p_0\tril n_j=-ip_j   \\
&&p_k\tril n_j=-i\delta_{jk}\left(\frac{1}{2\lambda}( 1-
e^{-2\lambda p_0})+\frac{\lambda}{2}p^2\right)+i\lambda p_jp_k
\eea In the same way, using the definition of the coproduct of a
bicrossproduct algebra (\ref{bicrosscond}), one can obtain the
following coactions: \be \bt_L(m_j)=1\otimes
m_j,\;\;\;\bt_L(n_j)=e^{-\lambda p_0}\otimes n_j
+\lambda\epsilon_{jkl}p_k\otimes m_l \ee One can verify that these
expressions for the action $\tril$ and the coaction $\bt_L$
satisfy all the compatibility conditions of the bicrossproduct
algebras.

In the construction of a bicrossproduct algebra $\mathcal{X}\wb
\mathcal{A}$, the action $\tril$ refers to the action of
$\mathcal{X}$ on $\mathcal{A}$, but there are no prescriptions on
the action of the elements of the bicrossproduct algebra itself.
The choice of this action can be dictated by the generalization of
the classical action of the standard Poincar\'e generators. In the
case of the Poincar\'e algebra the action of the Lorentz rotations
over the translation generators (momenta) is represented by the
commutators, and this action coincides with the adjoint action.
Taking into account that the Poincar\'e algebra is generated by
elements $h$ that satisfy $\Delta(h)=1\otimes h+h\otimes 1,
S(h)=-h$ (\emph{i.e.} they are primitive elements), one easily
finds \be p^j\stackrel{Ad}{\tril}h=S(h_{(1)})p^j h_{(2)}=[h,p_j]
\ee This suggests to consider the adjoint action as a good
generalization of the action of the Lorentz rotations over the
translation generators, also in the deformed case. It is
surprising that in the case of the \kkP\ bicrossproduct basis the
adjoint action is still given by commutators. By a straightforward
calculation one finds: \bea
\mP_{\mu}\stackrel{Ad}{\tril}\mN_k&=&S(\mN^k_{(1)})\mP^j\mN^j_{(2)}=[\mN_k,\mP_{\mu}]=-p_{\mu}\tril n_k\label{canonaction1}\\
\mP_{\mu}\stackrel{Ad}{\tril}M_k&=&S(M^k_{(1)})\mP^j
M^j_{(2)}=[M_k,\mP_{\mu}]=-p_{\mu}\tril m_k \label{canonaction2}
\eea Assuming that $M_j\mN_j$ act on $\mT$ via the adjoint action,
we can determine their action on \kM\ generators, as we show
below.

On the other hand, in the subsection (1.2.2), we have seen that
the action of the ordinary translations on the commutative
spacetime coordinates is described by the canonical action. So, it
appears natural to consider the canonical action as the
generalization of the action of the translation generators
$\mP_{\mu}$ on \kM\ as well.

Under these assumptions we can show that \kM\ transforms
covariantly under the action of the \kkP\ generators.

$P_\mu$ acts on its dual space $\mT^*$ (\emph{i.e.} \kM) by
canonical action: \be t\trir x=<x_{(1)},t>x_{(2)},\;\;\;t\in \mT,
x\in \mT^*, \ee from which follows that $\mP_\mu$ acts as a
derivation on the \kM\ generators: \be \mP_{\mu}\trir
\x_\nu=-i\eta_{\mu\nu}. \ee Their extension to products of the
spacetime coordinates is via the covariance condition $t\trir
xy=(t_{(1)}\trir x)(t_{(2)}\trir y)$. Thus \bea
P_0\trir \x_0\x_k=-i\x_k,&& P_0\trir\x_k\x_0=-i\x_k\nn\\
\mP_j\trir\x_0\x_k=(e^{-\lambda P_0}\trir
\x_0)(\mP_j\trir\x_k)=i(\x_0+i\lambda)\delta_{jk},&&\mP_j\trir\x_k\x_0=\mP_j\trir\x_k=i\delta_{jk}\nn
\eea so, the \kM\ commutation relations are invariant under the
\kkP\ translation.

One can derive also the action of $(M_j,\mN_j)$ on $\mT^*$ because
they act from the right on $\mT$ and this action therefore
dualizes to an action from the left on $\mT^*$: \bea
<\mP_\mu\stackrel{Ad}{\tril}M_k,x>&=&-<\mP_\mu, M_k\stackrel{Ad}{\trir}x>,\;\;\;x\in \mT^*\nn\\
<\mP_\mu\stackrel{Ad}{\tril}\mN_k,x>&=&-<\mP_\mu,
\mN_k\stackrel{Ad}{\trir}x> \nn\eea so, in our case, using
(\ref{canonaction1},\ref{canonaction2}) and the commutators
(\ref{kappapoincare}), one finds: \be M_j\trir
\x_j=i\epsilon_{jkl}\x_l,\;\; M_j\trir \x_0=0,\;\;\mN_j\trir
\x_k=-\delta_{jk}\x_0,\;\;\mN_j\trir \x_0=-i\x_i \ee and,
extending these actions via the covariance property of the adjoint
action $h\trir(xy)=(h_{(1)}\trir x)(h_{(2)}\trir
y),\;\;h=M_j\mN_j,\,x,y\in \mT^*$, \bea
M_j\trir \x_0\x_k=i\epsilon_{jkl}\x_0\x_l,&&M_j\trir\x_k\x_0=i\epsilon_{jkl}\x_l\nn\\
\mN_j\trir (\x_j\x_0)&=&-i\delta_{jk}\x_0^2-i\x_j\x_k\nn\\
\mN_j\trir (\x_0\x_j)&=&-i\delta_{jk}\x_0^2-i\x_j\x_k+i\lambda\delta_{jk}\x_0\nn\\
\mN_j(\x_k\x_l)&=&-i\delta_{jk}\x_0\x_l-i\delta_{jl}\x_k\x_l+\lambda(\delta_{jl}\x_k-\delta_{kl}\x_j) \nn
\eea From these expression it easy to see that the \kM\
commutation relations remain unmodified also under
the action of the boost-rotation generators of \kkP .\\

\section{\kkP\ and DSR theories}
\kkP\ algebra has recently attracted much interest due to the
connections with some recent theories that introduce a minimum
length at the level of Relativity postulates. These theories are
called Doubly Special Relativity (DSR) providing two invariant
scales, a length scale (identified with the Planck length) and a
velocity scale (the speed of light). In this section we want to
show the origin of these theories and how they can be connected to
the \kkP\ algebra.

Most of the Quantum Gravity approaches attribute a fundamental
role to the Planck length in the spacetime structure. In some
approaches, for example, the Planck length plays the role of a
minimum length, setting a limit on the localization of events.
However this fundamental role for $L_P$ in the spacetime structure
may be in conflict with the postulates of Special Relativity. One
of the most direct consequences of them is the FitzGerald-Lorentz
length contraction that is incompatible with the existence of an
invariant length. In fact, an event localized with $L_P$ accuracy
in one frame, would be localized with sub-Planckian accuracy in
some other inertial frames. Thus, according to the
FitzGerald-Lorentz length contraction, different inertial
observers would attribute different values to the same physical
length, violating the Relativity Principle that demands ``the laws
of physics to be the same in all inertial frames". This has led
some authors to consider a modification to Lorentz
transformations, which in particular  would lead to a modified
dispersion relation for particles. In this way the Planck length
plays a role in the kinematic structure of the theory. In some
approaches (for example~\cite{gacEMNS97,gampul}) a deformed
dispersion relations is predicted that would involve Planck length
corrections to Einstein's dispersion relation for particles. The
specific structure of the deformation can differ significantly
from model to model but one can write a general relation for
massless particles introducing a model-dependent function $f$ \be
c^2{\bf p}^2=E^2[1+f(\frac{L_{QG}}{c}E)]\label{drgen} \ee where
$L_{QG}$ is an effective quantum-gravity length scale (expected to
be of the order of Planck length $L_P=\sim 10^{-33}$cm).  One can
write the deformed dispersion relation in the following
parametrized form: \be c^2{\bf p}^2=E^2\qs 1+\xi\ts
\frac{L_{QG}}{c}E\td^{\al}+O\ts
\frac{L_{QG}}{c}E\td^{\al+1}\qd\label{dsrseries} \ee in terms of
the power $\al$ and the sign\footnote{any other coefficient of the
term of order $\al$ can be re-scaled with the parameter $L_{QG}$.}
$\xi=\pm1$. Assuming the special-relativistic rules of
transformation of energy and momentum, these dispersion relations
would allow to select a preferred class of inertial frames,
violating Relativity Principle.

DSR theories are formulated so that Relativity postulates do not
lead to inconsistencies in the case that an invariant length scale
exists. In this way, it is possible for every inertial observer to
agree on the physical laws including the law that would attributed
to the Planck length a fundamental role in the spacetime
structure.

The picture of a spacetime whose structure is governed by an
observer-independent length scale, in addition to an absolute
velocity scale, leads to an intuitive revision of the
FitzGerald-Lorentz transformations. This, in particular, requires
a deep modification of the boost transformations.

The first idea of DSR theory was introduced in~\cite{dsr1} and
subsequently different DSR theories have been proposed (see, for
example,~\cite{Mag01-DSR,dariofrancesco,lukiedsr,judesvisser}). In
the first proposal of DSR theory the two observer-independent
scales of velocity ($c$) and length ($L_P$) enter the theory at
the level of postulates; in~\cite{dsr1} the following postulates
are proposed and their consistency was studied:
\begin{enumerate}
\item The laws of physics involve a fundamental velocity $c$ and a
fundamental length scale $L_P$. \item The value of the fundamental
velocity scale $c$ can be measured by each inertial observer as
the $\lambda/L_P\to \infty$ limit of the speed of light of
wavelength $\lambda$. \item Any inertial observer can establish
the value of $L_P$ by determining the dispersion relation for
photons, which takes the form $E^2=c^2p^2+f(E,p;L_P)$, where the
function $f$ has leading $L_P$ dependence given by
$f(E,p;L_P)\simeq L_PcEp^2$.
\end{enumerate}
The Relativity Principle introduced by Galilei states that the
laws of physics take the same form in all inertial frames
(\emph{i.e.} they are the same for all inertial observers). The
introduction of a fundamental velocity or length scale in the
theory is itself a physical law and combined with Galilei's
Relativity Principle has deep implications for geometry and
kinematics. The Galilei/Newton rules of transformation between
inertial observers can be obtained by combining the Relativity
Principle with the assumption that there are no
observer-independent scales of velocity or length. Einstein's
Special Relativity instead comes from the Relativity Principle by
assuming that there is an observer-independent scale for velocity
but there are no fundamental length scales. Thus, Special
Relativity would be formulated in terms of the following two
postulates:
\begin{enumerate}
\item The laws of physics involve a fundamental velocity $c$.
\item The value of the fundamental velocity scale $c$ can be
measured by each inertial observer as the speed of light of
wavelength $\lambda$.
\end{enumerate}
The first postulate involves only a fundamental velocity, $c$, and
it provides no fundamental lengths. The second postulate provides
a physical interpretation of $c$ as the speed of light. In this
postulate there is the implicit assumption that there is no
observer-independent length scale. In fact, experimental
data\footnote{For example, data of the Michelson-Morley
experiments.} available when Special Relativity was formulated,
only concerned light of very long wavelengths ( very long with
respect to the Planck length $L_P$, introduced by Planck a few
years earlier). But a possible wavelength dependence of the speed
of light would induce a preferred class of inertial frames or an
observer-independent length scale. However in the second postulate
no prescriptions are given for the measure of $c$. So, Special
Relativity does not assume the existence of a  fundamental
lengths. In a DSR theory, instead, the existence of two invariant
scales requires to specify how each inertial observer obtains the
same result for their measurement. The precise role that $L_P$
should play in a DSR theory can only be established
experimentally. When Special Relativity stated the presence of the
fundamental velocity scale $c$, there was already robust data
suggesting a physical interpretation of $c$, whereas we have many
physical arguments and theoretical models predicting some
fundamental role for the Planck length but none of these has yet
experimental support\footnote{Actually, there is some tentative
encouragement from experiments for the idea that $L_P$ is a
spacetime invariant. Observations of multi-TeV photons from Mk501
blazer and of ultra-high energy cosmic rays leads to consider the
existence of a new kinematical length scale. We will make more
clear the phenomenological aspects of DSR theories in the next
section.}. For this reason the postulate $3$ provides just an
example of a possible postulate that agrees with current quantum
gravity intuitions and that leads to a logically consistent
theory. One can notice, in fact, that the choice for the
dispersion relation corresponds to the choice of
$\al=\xi=1,L_{QG}=L_P$ in the expansion of the most general
dispersion relation (\ref{dsrseries}) presented above. One can
also understand that the lack of experimental data for the role of
the Planck constant is in fact at the origin of the existence of
many different proposals for DSR theories.

The postulates presented above lead to a modification of the boost
transformations that at the first order in $L_p$ are: \be
{\mN}_j=icp_j\partial_E+i\qs \frac{E}{c}+L_P\ts\frac{E}{c}
\td^2+\frac{L_P\vec{p}^2}{2} \qd\partial_{p_j}-iL_P
p_jp_l\partial_{p_l}.\label{firstordN} \ee This form of the boost
operators assure the logical consistency of the observations
between two inertial observers that both agree with the law of
propagation of a photon at the first order in $L_P$ given by the
third postulate $E^2-c^2p^2+L_Pcp^2E=0$. Considering $c=1$ and
$L_P=\lambda$, we can see from (\ref{firstordN}) that the cross
relations between boost and momentum generators $(E,p_j)$ are
given by: \be [{\mN}_j,E]=ip_j,\;\;\;[{\mN}_j,p_k]=i\delta_{jk}\qs
E+\lambda\ts E \td^2+\frac{\lambda^2 p^2}{2} \qd-i\lambda p_jp_k
\ee that is just the first order expansion in $\lambda$ of the
commutators of \kkP\ algebra in the bicrossproduct
basis~\ref{kappapoincare}.

The result obtained in~\cite{dsr1} at the leading order in the
length observer-independent scale is generalized to all orders
in~\cite{jurekmax}\cite{bruno}\cite{qgpGAC}. A proposal for the
deformed dispersion relation $E^2=c^2p^2+f(E,p;L_P)$ including the
Planck length is provided by the mass Casimir of \kkP\
(\ref{majruecasimir}), with the identification of the deformation
parameter $\lambda$ with the physical Planck length $L_P$. This
dispersion relation and the corresponding \kkP\ boost
transformations is consistent with the DSR postulates. One can
start from \kM\ spacetime and reconstruct its algebra
symmetry~\cite{tesifra} following exactly the opposite direction
of the line presented above, in which  \kM\ commutation relations
are derived from \kkP\ bicrossproduct basis. On the other hand
\kkP\ transformations, with the deformation parameter coinciding
with the noncommutativity parameter $\lambda$, agree with DSR if
$\lambda$ is identified with the Planck length. Thus \kM\
represents a spacetime in which the noncommutativity length scale
has the role of relativistic invariant.

\section{Phenomenology  of \kkP/\kM}
Predictions of \kM /\kkP\ give rise to a testable scenario. This
fact provides motivation for our study that is not merely of
academic interest.

Recent improvements in the sensitivity and accuracy of
astrophysical observations, which involve particles of energies
much higher with respect to the ones achievable in laboratory,
allow to obtain useful data to test the new quantum gravity
phenomenology emerging at the Planck scale.

In particular, one of the most promising proposals, in the
framework of a large program of \emph{Quantum Gravity
Phenomenology}~\cite{Gac9910089}, is the \emph{test of in-vacuo
dispersion relation}, using observations of electromagnetic
radiation from distant astrophysical sources. This type of test
has a large interest in the quantum gravity studies: in fact, as
discussed in the previous section, several approach to
quantum-gravity leads to the prediction of a minimum length with a
possible fundamental role in spacetime structure. Some of these
approaches predict that this fundamental role for $L_P$ would
require a consequent correction to the Lorentz transformations,
that naturally implies deformed dispersion relations.
The specific structure of the deformation can differ significantly
from model to model, and it is conceivable that in the near future
the results of the experiments might also furnish informations to
discriminate among the various models. As already mentioned, a
general deformed dispersion relation (\ref{drgen}) can be written
in terms of a model-dependent function $f$, and the following
series expansion\footnote{from here we will assume $c=\hbar=1$ }
can be obtained, at small energies $E\ll L_{QG}^{-1}$: \be {\bf
p}^2=E^2\qs 1+\xi\ts L_{QG}E\td^{\al}+O\ts L_{QG} E\td^{\al+1}\qd
\label{serie} \ee in terms of the power $\al$ and the sign
$\xi=\pm1$.

Because $L_{QG}$ is expected to be a very short length
scale\footnote{{i.e.} the energy $E_P=L_P^{-1}$ associated to
$L_P$ is very higher with respect to the energies of Tev-order
currently achievable in laboratory}, it was traditionally believed
that this effect could not be significantly tested experimentally
(for $L_{QG}$ near to $L_P$ experiments would only be sensitive to
values of $\al$ much smaller than 1), but, as we will illustrate
in the following, the recent progress in the phenomenology of
$\gamma$ Ray Bursts and other astrophysical phenomena should soon
allow us to probe values of $L_{QG}$ of the order of $L_P$ for
values of $\al$ as large as $1$.

The dispersion relation for massless spin-0 particles coming from
the \kM /\kkP\ description
\be \lambda^{-2}\ts e^{\lambda E}+e^{-\lambda E}-2 \td-{\bf p}^2
e^{\lambda E}=M^2=0, \label{disprel} \ee can be put in the form
(\ref{serie}) if the deformation length parameter
 $\lambda$ is identified with the length scale introduced above $L_{QG}=\lambda$
\be {\bf p}^2=\lambda^{-2}\ts e^{\lambda E}+e^{-\lambda E}-2 \td
e^{-\lambda E}=E^2\ts1-\lambda E+O(\lambda^2 E^2)\td.
\label{seriekkp} \ee In this way, the \kkP\ dispersion relation
furnishes, at low energies ($E\ll
\frac{1}{\lambda}=\frac{1}{L_{QG}}$), a deformation that is
linearly suppressed by $\lambda$. Thus, for this model the
parameter $\al$ is just equal to $1$ ($\al=-\xi=1$). It is quite
significant that other approaches (different from the one based on
Noncommutative Geometry and Quantum
Groups)~\cite{gampul}\cite{aemn1} have provided evidence for the
possibility that the phenomenon of deformed dispersion relation
goes linearly with the Planck length ($L_p\sim 1/E_P$).

The observation of some astrophysical phenomena provides ideal
features for testing dispersion relations of the form
(\ref{seriekkp}).

The first example of these experiments is based on a possible
energy-dependence of the speed of
light~\cite{gacEMNS97}\cite{gacEMNS98}\cite{Gac9910089}\cite{schaefer99}\cite{biller}.
In fact, the most direct consequence of the deformed dispersion
relations predicted in the \kM/ \kkP\ framework is the emergence
of energy-dependent velocities for particles propagating \emph{in
vacuo}. From the natural assumption that the propagation velocity
of a particle is given by the usual formula $v=\frac{dE}{dp}$, one
can simply derive, using (\ref{disprel}), the following expression
for the velocity \be v=\frac{2\lambda p}{1-e^{-2\lambda
E}-\lambda^2p^2}=\frac{4e^{-\frac{\lambda
E}{2}}\sqrt{\sinh^2(\frac{\lambda
E}{2})-\frac{\lambda^2M^2}{4}}}{1-e^{-2\lambda E}-4 e^{-\lambda
E}(\sinh^2(\frac{\lambda E}{2})-\frac{\lambda^2M^2}{4})}, \ee and
the velocity for massless particles turn out to be \be v(\lambda
E;0)=e^{\lambda E}. \label{lightspeed} \ee According to this
formula,
 two signals respectively of energy $E$ and $E+\Delta E$, emitted {\it simultaneously} from the same astrophysical source at distance $L$, should reach our detectors with a relative time delay $\delta t$ given by:
\be |\delta t|= L|(\frac{1}{v(E)}-\frac{1}{v(E+\Delta E)})|\approx
\lambda L |\Delta E|. \label{dt} \ee The effect that we are
considering might appear very small for the presence of
$\lambda\sim L_P$. Even so, $\delta t$ can be rather significant
for even moderate-energy signals, if they travel over very long
distances. This is the case of astrophysical phenomena of
extraordinary energy, the $\gamma$ Ray Bursts (GRBs),  that
represent an important opportunity for these velocity-based tests.
In fact they typically come from a distance of order $10^{10}$
light-years and their emission are in the range $0.1-100$ Mev.

Actually, the simultaneous emission of signals is only a
schematisation of the phenomenon; in fact the observed photons
could be emitted with a time spread $\Delta T$. In the case that
the time definition $\Delta t$ is mach smaller that $\delta t$,
the relation (\ref{dt}) would allow to determine $\lambda\simeq
\frac{|\delta t|}{L|\Delta E|}$ from $\delta t, L, \Delta E$. The
observed GRBs are known to be emitted with a time spread $\Delta
t$ which in the most favorable conditions is of the order
$10^{-3}$s~\cite{FishmanRAA33(95)}. Unfortunately at present we
have only few GRB observations available for which the distance
$L$ has been determined. In the future the number of observed GRBs
with associated distance measurement should rapidly increase. A
new generation of orbiting spectrometers, e.g. AMS and
GLAST~\cite{glast}, are being developed, whose potential
sensitivities are very impressive. Thanks to cosmological
distances combined with the short time structure, the GRBs will
represent very soon a real ``laboratory" for fundamental physics,
but at present the best GRB based bounds are either ``conditional"
or ``not very robust". We have, at the moment, only an upper bound
for $\lambda$ that turns out to be:
\be
\lambda \leq 500 \times 10^{-35}m=300 L_P.
\ee

Another class of astrophysical experiments concerns the threshold
energy for some collision processes involving Cosmic Radiation and
unsolved related paradoxes such as the Cosmic-ray paradox (or GZK
paradox). These paradoxes concern the kinematic rules for particle
production in a continuum classical spacetime obtained from the
Special Relativity laws. Cosmic-rays  are essentially
ultra-high-energy protons coming from cosmological sources that
should loose energy interacting with the Cosmic Microwave
Background Radiation (CMBR) by producing pions: \be p+\gamma\to
p+\pi^0 \ee Taking into account the typical energy of CMBR
photons, and assuming the validity of the conventional
relativistic  kinematical rules for the production of particles,
these interactions should lead to an  upper limit $E<5\cdot
10^{19}eV$ (``GZK limit''~\cite{GZK}), on the energy of observed
cosmological protons. Instead, some observations of cosmic-rays
with energy above the GZK limit have been reported by the AGASA
observatory~\cite{takeda}.

The deformation of the dispersion relation that we find in \kM\,
in which $\lambda$ is identified with the Planck length $L_P$,
provides a possible solution of the paradoxes. Using the
dispersion relation (\ref{disprel}) and the $standard$
energy-momentum conservation laws one finds~\cite{amelino0107086}
that a collision between a proton of energy $E$ and a CMBR photon
of (much smaller) energy $\epsilon$ can produce a pion (and a
proton) only if \be
E>\frac{2m_pm_\pi+m_\pi^2}{4\epsilon}+\lambda\frac{m^3_\pi(2m_p+m_\pi)^3}{256
\epsilon^4}\ts 1-\frac{m_p^2+m_\pi^2}{(m_p+m_\pi)^2}\td
,\label{soglia} \ee where $m_p$ is the proton mass and $m_\pi$ is
the pion mass. In this formula, the smallness of $\lambda\sim L_P$
is compensated by the huge ratios $m_p/E_\gamma$ and
$m_\pi/E_\gamma$, so that the corrections are enough to allow much
more high energy thresholds. According to (\ref{soglia}), even at
$E\sim 3\cdot 10^{20}$ photo-pion production on CMBR is still not
possible, consistently with the observations reported
in~\cite{takeda}.

We want to underline that the threshold energy (\ref{soglia}) has
been obtained combining the deformed dispersion relation
(\ref{disprel}), that would characterize the propagation of  a
particle in \kM\ spacetime, with the usual equations for the
conservation of energy and momentum that, for a collision process
$1+2\to 3+4$, are: \be E_1+E_2\to
E_3+E_4,\;\;\;\vec{p}_1+\vec{p}_1\to
\vec{p}_3+\vec{p}_4.\label{laws} \ee However, as we will explain
in the next chapters, the study on the propagation of waves in the
\kM\ spacetime leads to a modification of the laws (\ref{laws}).
Combining the dispersion relation (\ref{soglia}) with the modified
conservation laws predicted in \kM\ we are not able to explain the
GZK paradox. This fact has produced further interest in the study
of \kM\ and in particular in the description of particle processes
in such a space.



\chapter{Proposal of a unified Weyl-map-based construction of fields in \kM}

The physical implications of the \kkP\ framework, discussed  in
the previous chapter, motivate the recent strong interest in the
construction of a field theory on \kM\ spacetime. In literature
there are many attempts in this direction  (see e.g.
\cite{KMLS,AmelinoMajid,KLM00,Monaco,AmelinoArzano}), but the
results obtained are still partial, especially in comparison with
the results obtained on the canonical noncommutative spacetime.
In the case of canonical noncommutativity, thanks to the constant
commutation relations, it is possible to develop an analysis of
field theoretical models quite similar to the one adopted in
ordinary commutative spaces. This analysis has revealed some
surprising properties of such models, including the so called
IR/UV mixing \cite{MRS,MR}, which consists in the interdependence
of the small-distance (IR) behavior on large distance (UV) sector,
and the appearance of noncommutative solitons \cite{GMS,BL}. In
general, one of the most fruitful ways to deal with canonical
noncommutative spacetime is via the definition of a deformation of
the usual product among functions. This deformed product, called
``star product", replaces the commutative (pointwise) product, and
allows to map a noncommutative theory into a theory with
commutative functions but with a deformed product among them. In
this way the noncommutative space is studied as the structure
space of a deformed $*$-algebra\footnote{The ``$*$''  here refers
to the presence of an hermitian (complex) conjugation, and has
nothing to do with the deformed products.}.

Given the success of this deformation technique in the case of
canonical noncommutative spaces, we explore the possibility to
introduce the star product description in the approach to \kM\ as
well. The idea was already proposed in \cite{Monaco}, where a
construction of star products for a generic noncommutative
spacetime was performed generalizing the Moyal-Weyl procedure with
which the star product of canonical spacetime is obtained. In
\cite{KLM00}, following this procedure, two types of star product
have been presented in the case of \kM\ spacetime. As we will show
in the following, their origin is connected with two different
ordering prescriptions in \kM . Considering other ordering
conventions one can also obtain other star products. In this way,
different star products can be introduced that both describe the
\kM\ algebra. So, the description of \kM\ is not uniquely defined
in terms of star products. In the canonical noncommutative
spacetime a degeneracy of star products is also present. However,
the various star products differ from each other by a
multiplicative constant (overall a phase). Whereas, in the case of
\kM , their difference is much more substantial because it is
given by a function of the coordinates. Motivated by the necessity
to clarify this degeneracy we attempt, in this chapter, a more
systematic study of the star-product construction.

In this chapter we give a brief review on the origin of the star
product in Quantum Mechanics where the concepts of Weyl system and
Weyl map were firstly introduced. Then we show that a
generalization of these concepts is also possible in spacetimes
with a more complex structure with respect to the quantum
mechanical phase space. We give a general procedure, based on the
generalization of the  Weyl systems and Weyl maps, that allows to
construct star products on a generic noncommutative spacetime of
Lie-algebra type with central extension. Finally we apply such a
technique to \kM , showing that the star products already present
in literature \cite{KLM00} can be obtained with our technique and
can be seen as originating from the same family of generalized
Weyl systems.

A short version of the results of this Chapter can be found
in~\cite{alz02}.


\section{Star product and Deformation quantization in Noncommutative Geometry}

\subsection{Moyal-Weyl star products}
The first example of deformed product is represented by the
Moyal-Weyl product, that arose in studying the noncommutative
structure of Quantum Mechanical phase space
given by  the relations $[\q_j,\p_k]=i\hbar\delta_{jk}$.
Here the deformation parameter is $\hbar$. This product originates
from the Weyl map for quantization. The Weyl quantization
rule~\cite{weyl}, defined as a linear map $\Omega$ from the
classical phase-space functions $f(q,p)$ to functions of quantum
operators $F(\q,\p)$,  was first introduced in this context.
The Weyl map associates to the monomials $q^m_jp^n_l$ the
hermitian operator given by~\cite{Caste}: \be \label{weyl1}
q^m_jp^n_l\to \Omega(q^m_jp^n_l)=\frac{1}{2^n}\sum_{k=0}^n \ts
\begin{array}{c}
n\\
k
\end{array}
\td \p^{n-k}_l\q^m_j\p^k_l \ee This rule can be restated in a
differential form \be \Omega(q^m_jp^n_l)=:e^{-\frac{i}{2}\hbar
\frac{\partial^2}{\partial q_j\partial
p_l}}q^m_jp^n_l:|_{q=\q,p=\p}\label{diffweyl} \ee with the
\emph{normal ordering} prescription (denoted by $:\;:$ ) that
consists in putting in each monomial the variable $q$ to the left
and then to make the substitution $q\to \q,p\to \p$. It easy to
check this formula by a series expansion of the differential
operator $e^{-\frac{i}{2}\hbar \frac{\partial^2}{\partial
q\partial p}}$. For example, if we consider the two dimensional
function $q^2p^2$ the differential expression (\ref{diffweyl})
takes the form: \bea
\Omega(q^2p^2)&=&:e^{-\frac{i}{2}\hbar \frac{\partial^2}
{\partial q\partial p}}q^2p^2:|_{q=\q,p=\p}=:(1-\frac{i}{2}
\hbar \frac{\partial^2}{\partial q\partial p}-\frac{1}{8}\hbar^2\frac{\partial^2}
{\partial^2 q\partial^2 p})q^2p^2:|_{q=\q,p=\p}\nn\\
&=&\q^2\p^2-2i\hbar\q\p-\frac{1}{2}\hbar^2. \eea On the other
hand, the expression obtained from the Weyl prescription
(\ref{weyl1}) can be reordered putting the variable $\q$ at the
right, and one finds $$ \Omega(q^2p^2)=\frac{1}{4}\ts
\p^2\q^2+2\p\q^2\p+\q^2\p^2\td=\q^2\p^2-2i\hbar\p\q-\frac{1}{2}\hbar^2,
$$ that is exactly the expression obtained from the differential
form (\ref{diffweyl}).

The Weyl map (\ref{diffweyl}) can be easily extended to any
function $f(q,p)$ on phase space $$
\Omega(f(q,p))=:e^{-\frac{i}{2}\hbar \frac{\partial^2}{\partial
q\partial p}}f(q,p):|_{q=\q,p=\p} $$ and, introducing the inverse
Fourier transform $\;\tilde{f}(\al,\bt)=(2\pi)^{-1}\int dx dp
f(q,p) e^{-i(\al q+\bt p)}$ of $f$, one can obtain the following
integral expression of the map $\Omega$: \be
\Omega(f(q,p))=\frac{1}{(2\pi)}\int d\al d\bt
\tilde{f}(\al,\bt)e^{i\frac{\hbar}{2}\al\bt}:e^{\al \q+\bt
\p}:\;=\frac{1}{(2\pi)}\int d\al d\bt \tilde{f}(\al,\bt)e^{\al
\q+\bt \p}\label{weylmap} \ee where, in the last expression, we
have absorbed the function $e^{i\frac{\hbar}{2}\al\bt}$ (it is
easy in fact to check through the expansion in $\al,\bt$ that this
factor can be used to take out the normal ordering prescription).
The formula (\ref{weylmap}) can be easily generalized for a
function $f(q,p)$ defined over a $2n$-dimensional phase space: \be
\Omega(f(q,p))=\frac{1}{(2\pi)^{n}}\int d^{n}\al d^{n}\bt
\tilde{f}(\al,\bt)e^{\sum_{j=1}^n(\al_j\q_j+\bt_j
\p_j)}.\label{weylmap2n} \ee The inverse of the map $\Omega^{-1}$,
also called Wigner map, exists and allows to establish a one to
one correspondence between quantum operators and functions on the
commutative phase space. In fact, the product of the Weyl map
$\Omega(f)$ and the Weyl map $\Omega(g)$, is still a Weyl map of
another function, that we call $f\star g$: \be
\Omega(f)\Omega(g)=\Omega(f\star g).\label{arnold} \ee Applying
the Wigner map $\Omega^{-1}$ the $\star$-product is obtained: $$
(f\star g)=\Omega^{-1}(\Omega(f)\Omega(g)). $$
 Using the formula
(\ref{weylmap2n}) and introducing the short notation $x=(q_j,p_j)$
and ${\bf x}=(\q_j,\p_j)$ with $j=1,...,n$, the product between
the two $\Omega$ maps can be written in the following way:
\be \Omega(f)\Omega(g)=\frac{1}{(2\pi)^{2n}}\int d^{2n}s d^{2n}t
\tilde{f}(s) \tilde{g}(t) e^{is{\bf x}}e^{it{\bf x}} \label{int}
\ee where $s=(\al_j,\bt_j)$, $t=(\al'_j,\bt'_j)$ and $s{\bf
x}=\sum_j(\al_j\q_j+\bt_j\p_j)$, $t{\bf
x}=\sum_j(\al'_j\q_j+\bt'_j\p_j)$. We can write the product of
exponentials $e^{is{\bf x}}e^{it{\bf x}}$ with the help of the
Baker-Campbell-Hausdorff formula, in the following way: \be
e^{is{\bf x}}e^{it{\bf x}}=e^{i(s+t){\bf x}-\frac{1}{2}[s{\bf
x},t{\bf x}]}=e^{i(s+t){\bf
x}-\frac{i\hbar}{2}(\al_j\bt'_j-\al'_j\bt_j)}=e^{i(s+t){\bf
x}-\frac{i\hbar}{2}(sJt)}, \label{BCH}
\ee where we have introduced the $2n\times 2n$ matrix $J=\ts
\begin{array}{cc}
0&I_n\\
-I_n&0\\
\end{array}
\td $, with $I_n$ denoting the identity matrix, and the notation
$sJt$ is used as the short notation for the product
$s_kJ_{kl}t_l$, with $k,l=1,...,2n$. Then, introducing (\ref{BCH})
in the integral (\ref{int}), we obtain: \bea
\Omega(f)\Omega(g)&=&\frac{1}{(2\pi)^{2n}}\int d^{2n}s d^{2n}t \;e^{i(s+t){\bf x}}e^{-\frac{i\hbar}{2}(sJt)}\tilde{f}(s)\tilde{g}(t)\nn\\
&=&\frac{1}{(2\pi)^{2n}}\int d^{2n}sd^{2n}t \;e^{is{\bf
x}}e^{-\frac{i\hbar}{2}(sJt)}\tilde{f}(s-t)\tilde{g}(t).\label{1line}
\eea On the other hand, from (\ref{arnold}), the product
$\Omega(f)\Omega(g)$ can be written in terms of the star product
$f\star g$: \bea \Omega(f)\Omega(g)&=&\Omega(f\star
g)=\frac{1}{(2\pi)^{n/2}}\int d^ns \;e^{is{\bf
x}}(\widetilde{f*g})(s).\label{2line} \eea Comparing the equations
(\ref{1line}) and (\ref{2line}) we find: $$ \widetilde{f\star
g}(s)=\frac{1}{(2\pi)^{n/2}}\int d^nt
e^{-\frac{i\hbar}{2}(sJt)}\tilde{f}(s-t)\tilde{g}(t). $$
 Passing
to the Fourier transform, we arrive at the expression of the
deformed star product: \bea (f\star g)(x)
&=&\frac{1}{(2\pi)^{n}}\int d^nsd^nt e^{isx}
e^{-\frac{i\hbar}{2}(sJt)}\tilde{f}(s-t)\tilde{g}(t)
\label{eq:Moyal} \eea This product can be rewritten in a
differential way. Introducing the notation
$\partial_x=(\partial_q,\partial_p)$, we can write: \bea
(f\star g)(x)&=&\frac{1}{(2\pi)^{n}}\int d^nsd^nt e^{i(s+t)x} e^{\frac{-i\hbar}{2}(sJt)}\tilde{f}(s)\tilde{g}(t)=\frac{1}{(2\pi)^{n}}\int d^nt e^{itx}\tilde{g}(t) e^{-\frac{\hbar}{2}(\partial_x J t)}\int d^ns e^{isx}\tilde{f}(s)\nn\\
&=&\frac{1}{(2\pi)^{n/2}}e^{\frac{i\hbar}{2}(\partial_x\,J\,\partial_{x'})}\int
d^nt e^{itx'}\tilde{g}(t)f(x)|_{x'\to
x}=e^{\frac{i\hbar}{2}(\partial_x\,J\,\partial_{x'})}g(x')f(x)|_{x'\to
x}\nn \eea So we have obtained the well-known result: \be (f\star
g)(q,p)=f(q,p)e^{i\frac{\hbar}{2}\triangle}g(q,p)\label{listen}
\ee where $\triangle$ is the bidifferential operator defining the
Poisson bracket $\{\}_{P_B}$: $$ f\triangle g\equiv
\{f,g\}_{P_B}=\partial_qf\partial_pg-\partial_pf\partial_qg.
$$
Such as deformed product was at the basis of the Moyal formalism
in the approach to Quantum Mechanics, and is called ``Moyal
product"~\cite{GroenewoldMoyal}.

Quantum Mechanics noncommutativity is formally identical to
canonical noncommutativity (\ref{canonical}) and we can generalize
straightforwardly the integral form of the Moyal product
(\ref{eq:Moyal}) to the case of noncommutative canonical
spacetime. It is sufficient to make the substitutions: \bea
x&\to& x_\mu,\;\;\;s\to s_\mu,\;\;\;t\to t_\mu \;\;\;\;\nn\\
s{\bf x}&\to& -s_{\mu}\x^\mu \nn\\
sJt=[s{\bf x},t{\bf x}]&\to&
s^{\mu}[x_\mu,x_\nu]t^\nu=is^\mu\theta_{\mu\nu}t^\nu \label{subst}
\eea with $\mu=0,...,3$ and the signature
$\eta_{\mu\nu}=(1,-1,-1,-1)$. In this way the Heisenberg
commutation relations $[\q_j,\p_k]=i\hbar\delta_{jk}$ are mapped
into the canonical commutation relations
$[\x_\mu,\x_\nu]=i\theta_{\mu\nu}$.

Substituting (\ref{subst}) in the the Moyal integral
(\ref{eq:Moyal}), the extension of the Moyal product to the
canonical spacetime will be given by: \be (f\star g)(x) :=
\frac{1}{(2\pi)^{4}}\int\,d^{4}s\,d^{4}t e^{-is_\mu
x^\mu}e^{-\frac{i}{2}s_\mu\theta^{\mu\nu}t_\nu}\,f(s-t)g(t)
\label{eq:Moyalcan} \ee or the more familiar asymptotic expansion:
\be (f\star g)(x)\equiv\left.e^{\frac
i2\theta_{\mu\nu}\partial^{y_\mu}\partial^{z_\nu}}
f(y)g(z)\right|_{y=z=x} \label{defstar} \ee
In this way, the canonical commutators
$[\x_\mu,\x_\nu]=i\theta_{\mu\nu}$ are mapped in the star
commutators: \be [x_\mu,x_\nu]_\star\equiv x_\mu \star x_\nu-
x_\nu\star x_\mu=i\theta_{\mu\nu}. \label{canoncomm} \ee In terms
of this product the exponentiated version of~\eqn{canoncomm}
reads:
 \be
 e^{ik^\mu x_\mu} \star e^{il^\nu x_\nu}=e^{\frac i2k^\mu\theta_{\mu\nu}l^\nu}e^{i(k+l)^\mu x_\mu}\label{expcancomm}
 \ee

\subsection{Deformation quantization}

We have obtained the deformed star product (\ref{defstar}) of the
canonical noncommutative spacetime by a straightforward extension
of the Weyl formalism. However there exists a deformation
technique more general than the one introduced by Weyl. This
technique, called ``deformation quantization" (see for example
\cite{stern}), is based on the possibility to deform the usual
product between commutative functions, introducing some structures
(Poisson brackets) that can be defined on the manifold on which
these functions are defined.

Generalized deformed $*$-products were originally introduced
in~\cite{Bayenetal} as a first attempt to develop a quantization
of classical dynamics on phase space. Let us notice, in fact, that
the Moyal-star commutator between two functions $f(x)$ and $g(x)$
can be written in the following way, using  (\ref{listen}) \bea
[f,g]_{\star}&=&f(x,p)e^{i\frac{\hbar}{2}\triangle}g(x,p)-g(x,p)e^{i\frac{\hbar}{2}\triangle}f(x,p)\nn\\
&=& i\frac{\hbar}{2}(f\triangle g-g\triangle f)+O(\hbar^2)=i\hbar
\{f,g\}_{P_B}+O(\hbar^2).\label{sympl} \eea It has the property
that the $*$-commutator of two functions, which are defined on the
manifold  $S$ ( in this case the manifold is $S=R^2=(x,p)$),
reduces to the Poisson bracket at the first order in the
deformation parameter.

A Poisson bracket is a bilinear, antisymmetric map between the
algebra $\mathcal{F}(M)$ of real functions on $M$: \be
\{\cdot,\cdot\}: \mathcal{F}(M)\otimes \mathcal{F}(M)\to F(M) \ee
that satisfies for all $f,g\in \mathcal{F}(M)$:
\begin{enumerate}
\item the Jacoby identity
\be
\{\{f,g\},h\}+\{\{g,h\},f\}+\{\{h,f\},g\}=0
\ee
\item the Leibnitz rule:
\be
\{fg,h\}=f\{g,h\}+\{f,h\}g
\ee
\end{enumerate}
If one introduces a local reference coordinate system $x_a$,  one
can associate to the Poisson bracket an antisymmetric tensor
$\Lambda$, called Poisson tensor: \be \Lambda
=\{x_a,x_b\}\partial_{x_a}\otimes
\partial_{x_b}=\Lambda_{ab}(x)\partial_{x_a}\otimes
\partial_{x_b}\to \Lambda(df,dg)=\{f,g\} \ee where the sum over
the repeated indices is implicit. If the inverse of
$\Lambda_{ab}(x)$ exists, a two-form $\omega$, called symplectic
form, can  be introduced: \be \omega=\omega_{ab}dx_a\otimes dx_b
\ee such that the function $\omega_{ab}$ turns out to be the
inverse of $\Lambda_{ab}$: \be
\Lambda_{ac}\omega_{cb}=-\delta_{ab}. \ee A manifold on which a
Poisson bracket has been defined is called a Poisson Manifold.

In this way, the phase space of a point particle without
constraints represents the simplest example of Poisson manifold:
the flat  $R^{2N}$ in which the Poisson bracket between classical
variables $\{q_j,p_k\}=\delta_{jk}$ is introduced and, if we use
the notation $x_a=(q_1,...q_n,p_1,...,p_n)$, the symplectic form
is $\omega=\sum_{j=1}^n dq_j\wedge dp_j$. So that the deformed
product (\ref{sympl}) can be defined on a vector space $S$
equipped with a very simple constant symplectic structure.

In the case of canonical noncommutative spacetime, we can imagine
that the underlying Poisson-manifold structure be characterized by
a constant Poisson bracket
$\{x_{\mu},x_{\nu}\}=\theta_{\mu\nu}/\theta$ (choosing in this
case $\xi_a=x_{\mu}$); so, following the idea of the deformation
quantization, we can write: \be [f(x),g(x)]_{\star}=i\theta
\{f,g\}_{P_B}+O(\theta^3) \ee and the deformed commutation
relations among the generators $\x_\mu$ assume exactly the form
(\ref{canoncomm}) \be [x_{\mu},x_{\nu}]_\star=i\theta_{\mu\nu} \ee
This is a special case of Poisson manifold characterized by a
constant symplectic structure.

Not all Poisson manifolds are symplectic however. The general
problem of finding a deformed product for a general Poisson
manifold has been solved by Kontsevich~\cite{Maxim}, at least at
the level of a formal series, that is, without establishing the
convergence of the series. It involves very elaborate
constructions, both conceptually and computationally, and makes an
essential use of ideas coming from string theory.

In particular, \kM, the space we are interested in, is naturally a
Poisson manifold, with bracket: \be
\{f,g\}=\sum_jx_j(\frac{\partial}{\partial
x_0}f\frac{\partial}{\partial x_j} g-\frac{\partial}{\partial x_j}
f \frac{\partial}{\partial x_0} g). \label{poissonkmin} \ee This
allows us to introduce star products of the type \be
[f,g]_{\star}=i\lambda \{f,g\}+O(\lambda^2) \label{starPB} \ee
that lead to the commutators $[x_0,x_i]_{\star}=i\lambda x_i$
between generators $x_{\mu}$.

In the following we will find different forms of star product that
at the first order in $\lambda$ reduce to the same term, given by
the Poisson bracket, according to (\ref{starPB}), while they are
different at order $O(\lambda^2)$. A key aspect of these
relations, as we show below, is that they define a solvable Lie
algebra on the generators, and this makes them particular cases of
more general products.


\section{Weyl System and maps and their Generalization}

In this section we introduce the notion of Weyl system and we show
that it is at the basis of the construction of the Weyl map,
introduced in the previous section. We provide a generalization of
the Weyl system and the correspondent Weyl map.
These generalizations can be used as a tool to construct
$*$-algebras whose generators satisfy some commutation relations.
Generalized Weyl maps and systems enable us  to obtain a wide
class of commutation relations among coordinates. These
commutation relations reproduce Lie algebra structures with the
possibility of considering central extension as well. In
particular, \kM\ commutation relations are included in this class.
This allow us to compare the different products present in the
literature that reproduce the \kM\ commutation relations, and to
show that they can be seen as generalizations of the Moyal
star-product.

\subsection{Standard Weyl Systems and Maps}

The concept of what can now be referred to as a ``standard Weyl
system" was introduced by H.~Weyl~\cite{weyl}. The main motivation
at the time was to avoid that the quantum mechanical formalism was
based on the notion
of unbounded operators.
In fact, starting from the commutation relations between the
quantum mechanical observables $\q,\p$ of a particle in
one-dimension: $$ [\q,\p]=i\hbar $$ the Wintner theorem states
that at least one of them has to be unbounded. In fact consider
the commutator $$ [\q^n,\p]=i\hbar n \q^{n-1}. $$ If both
operators $\q,\p$ were bounded, one could obtain the following
relation between the norms $$ \|\q^n\p-\q\p^n\|=i\hbar n
\|\q^{n-1}\|=i\hbar n \|\q\|^{n-1}. $$ On the other hand from the
triangular inequality $$
\|\q^n\p-\p\q^n\|\leq\|\q^n\p\|+\|\p\q^n\|\leq 2\|\q\|^n\|\p\| $$
one would then have, for any $n$: $$ \|\q\|\, \|\p\|\geq
\frac{n\hbar}{2} $$ in conflict with the initial hypothesis that
$\q$ and $\p$ are bounded. Therefore at least one of them has to
be unbounded.

The introduction of the standard Weyl systems  allows to overcome
this difficulty. Now we give a brief description of the Weyl
system. A more detailed treatment can be found
in~\cite{BaezSegal}.

Given a real, finite-dimensional, symplectic vector space $S$, (a
real vector space on which a constant symplectic structure
$\omega$ is introduced), a \emph{Weyl system} is a map between
this space and the set of unitary operators on a suitable Hilbert
space: \be W\,:\,S\,\mapsto \,\uni\left(\hil\right) \ee with the
properties:
\begin{enumerate}
\item  $W$ strongly continuous \be \lim_{k\to
k_0}||W(k)-W(k_0)||=0 \;\;\;\forall\; k,k_0\in S \label{W1}\ee
\item the map over a sum can be written as the product of the maps
over each element \be
W(k+k^{\prime})=e^{-\frac{i}{2}\omega\left(k,k^{\prime}\right)}
W(k)W\left(k^{\prime}\right) \label{stweylsys} \ee for all $k\in
S$, with $\omega$ the symplectic, translationally invariant, form
on $S$.
\end{enumerate}
 On each one-dimensional subspace of $S$, the formula (\ref{stweylsys})
reduces to ($\alpha$ and $\beta$ real scalars):
$$
W\left(\left(\alpha+\beta\right)k\right)=W\left(\alpha k\right)
 W\left(\beta k\right)=W\left(\beta k\right)W\left(\alpha k\right)
$$

This means that, for each $k$, $W\left(\alpha k\right)$ is a one
parameter group of unitary operators. According to Stone's theorem
$W$ is the exponential of a hermitean operator on the Hibert space
$\hil$: \be W\left(\alpha k\right)=e^{i\alpha X\left(k\right)}
\label{defgenstan} \ee and the vector space structure implies that
$$ X\left(\alpha k\right)=\alpha X\left(k\right)$$ The
relation~\eqn{stweylsys} can be cast in the form \be
W(k)W\left(k^{\prime}\right)= e^{i\omega\left(k,k^{\prime}\right)}
W\left(k^{\prime}\right)W(k).\label{expo} \ee This can be
considered the exponentiated version of the commutation relations,
thus satisfying the original Weyl motivation. The usual form of
the commutation relations between generators can be recovered with
a series expansion of the (\ref{expo}) $$
\left[X\left(k\right),X\left(k^{\prime}\right)\right]
=-i\omega\left(k,k^{\prime}\right) $$

In the usual identification of $S$ with $\real^{2n}$, the
cotangent bundle of $\real^{n}$, with canonical coordinates
$k=\left(\al_{j},\bt_{j}\right),\;j\in1,...,n$ and the symplectic
two-form $\omega=\sum_jd\al_{j}\wedge d\bt_{j}$, this construction
can be given an explicit realization  \be
W\left(\al,\bt\right)=e^{i\sum_j\left(\al_{j}\q_{j}+\bt_{j}\p_{j}\right)}
\label{darbouxstanweylsys} \ee
 where $\q_{j}$ and $\p_{j}$ are the
usual operators that represents the position and momentum
observables for a system of particles, whose dynamics is
classically described on the phase space $\real^{2n}$. In this way
$X(\al,\bt)=\al_j\q_j+\bt_j\p_j$ and
$X(\al',\bt')=\al'_j\q_j+\bt'_j\p_j$, and
 the symplectic structure turns out to be:
 \bea
\omega\left(\al,\bt;\al'\bt'\right)&=&i\left[X\left(\al,\bt\right),X\left(\al',\bt'\right)\right]\nn\\
&=&-i(\al_j\bt'_l-\al'_j\bt_l)[\p_j,\q_l]=-\hbar(\al_j\bt'_j-\al'_j\bt_j)\nn
 \eea

This form of the operator $W$ suggests how to relate an operator
on $\hil$ to a function defined on $S$. It reminds the integral
kernel used to define the Fourier transform. It can be intuitively
seen as a sort of ``plane wave basis" in a set of
operators\footnote{Our considerations are valid for rapid descent
Schwarzian functions, and in the following we will not pay
particular attention to the domain of definition of the product,
discussed at length in~\cite{graciavarillyjmp}.}. Given a function
$f$ on the phase space, whose coordinates we collectively indicate
with $x$, with Fourier transform \be \tilde f(k)=\frac
1{(2\pi)^n}\int d^{2n}x f(x) e^{-ikx} \ee we define the operator
$\Omega(f)$ in the Hilbert space $L^2(\real^{2n})$ via the Weyl
map~\cite{weyl} \be
 \Omega(f)=\frac 1{(2\pi)^n}\int
d^{2n}k \tilde{f}\left(k\right)W\left(k\right)
\label{standardweyl}
\ee
where by $kx$ we mean $\sum_{j=1}^nk_jx_j$ and the integral is taken in the weak operator topology.
$\Omega\left(f\right)$ is the operator that, in the $W(k)$ basis,
has coefficients given by the Fourier transform of $f$.
In the case of the phase space, in which $k=(\al_j,\bt_j)=\real^{2n}$ and $\omega=\sum_jd\al_j\wedge d\bt_j$, using the explicit form of the Weyl system (\ref{darbouxstanweylsys}), the map $\Omega$ reduces to the expression (\ref{weylmap2n})
$$
 \Omega(f)=\frac 1{(2\pi)^n}\int
d^{n}\al d^{n}\bt
\tilde{f}\left(\al,\bt\right)e^{i(\al_j\x_j+\bt_j\p^j)} $$ but the
introduction of the Weyl map, in terms of the Weyl system gives
the possibility to generalize the map to a  noncommutative space
of Lie-algebra type.

The Wigner map, \emph{i.e.} the inverse of $\Omega$, maps an
operator $F$ into a function, whose Fourier transform is: \be
\Omega^{-1}(F)(k)= \Tr F \Omega^\dagger(k)
\ee
 The bijection $\Omega$ can now be used to translate the
composition law in the set of operators on $\hil$ into an
associative
composition law in the space of function defined on $\real^{2n}$.
For two functions on $\real^{2n}$ we define the $\star$ product
(Moyal product) as \be (f\star
g)=\Omega^{-1}\left(\Omega(f)\Omega(g)\right)
\label{stardefweylmap} \ee In this way, we find that for a Weyl
system defined by~\eqn{stweylsys}, with $\omega$ such that
$\omega\left(k,k^{\prime}\right)=k^{\mu}\theta_{\mu\nu}k^{\prime\nu}$,
this reduces to the product defined in~\eqn{eq:Moyalcan}
or~\eqn{defstar}.

The deformed algebra defined by this product is a $*$-algebra
with norm given by:
\be
\parallel f \parallel=\sup_{g\neq
0}\frac{\|f\star g\|_2}{\|g\|_2} \label{defnorm} \ee with
$\|{\cdot}\|_2$ the $L^2$ norm defined as \be \|f\|^2\equiv \int
d^{2n}k |\tilde f(k)|^2 \ee The hermitian conjugation is the usual
complex conjugation. Note that one finds $$
\Omega(f^*)=\Omega(f)^\dagger .$$ These two ingredients enable us
to give this set of functions a very important structure in the
context of noncommutative geometry formalism. In fact they allow
us to construct the $C^*-algebras$ that are at the basis of
Connes's machinery.

\subsection{Generalized Weyl Systems\label{genweyl}}

In the previous section we have shown how the Moyal product arises
via an  explicit realization  (\ref{darbouxstanweylsys}) of the
Weyl systems $W$. It is nothing but a realization, in the space of
functions defined in $\real^{2n}$, of the composition law of the
operators $W$. Following this idea, we show, in this section, how
it is possible to define a class of ``deformed" products in a set
of function defined on $\real^{n}$, without using an explicit
realization of a Weyl system, thus opening the possibility for a
generalization of this concept.

We have previously considered a standard Weyl system simply as a
map (with properties (\ref{W1},\ref{stweylsys})), however it can
be viewed as a unitary projective representation of the
translations group in an even dimensional real vector space. In
fact, the condition (\ref{stweylsys}) of the standard Weyl system,
give the space $S$ the structure of group, with a commutative
composition law $+$ between points. Thus, the most natural
generalization is to consider the manifold $\real^{n}$, and a
non-abelian (associative) composition law $\oplus$ between the
points of the manifold. In this way,  we define a
\emph{generalized Weyl system} as the map in the set of operators
$W:\real^n\to U(\mathcal{H})$, with the following composition rule
\be
 W(k)W(k')=e^{\frac{i}2\omega(k,k')}W(k\oplus k')
\label{genweylrel} \ee where $\oplus$ is a generic (non-abelian)
associative composition law between two points $k,k'\in \real^n$,
and $\omega$ is a function that must satisfies some relation. In
fact, in order to ensure that the algebra of the operators $W$ be
associative, it must be: \bea
[W(k)W(k')]W(k'')&=&W(k)[W(k')W(k'')]\nn \eea According to
(\ref{genweylrel}) the left side is given by: \bea
e^{\frac{i}2\omega(k,k')}W(k\oplus
k')W(k'')&=&e^{\frac{i}2\omega(k,k')}e^{\frac{i}2\omega(k\oplus
k',k'')}W((k\oplus k')\oplus k''),\nn \eea while the right side is
given by: \bea e^{\frac{i}2\omega(k',k'')}W(k)W(k'\oplus
k'')&=&e^{\frac{i}2\omega(k',k'')}e^{\frac{i}2\omega(k,k'\oplus
k'')}W(k\oplus (k'\oplus k'')).\nn \eea Equating the two
expressions, and taking into account that $\oplus$ is associative,
one finds that $\omega(k,k')$ must satisfy the following relation:
\be \omega\left(k,k'\oplus k''\right)+\omega\left(k',k''\right)
=\omega\left(k,k'\right)+\omega\left(k\oplus k',k''\right).
\label{cocycle} \ee It is easy to see that if
$\omega\left(k,k^{\prime}\right)$ is  a two-form (as in the Moyal
case) it necessarily  satisfies the relation (\ref{cocycle}).

Since we are looking for deformations of the algebra driven by a
parameter $\lambda$, we also require that
\bea
\lim_{\lambda\to 0}\, k\oplus k' &=& k+k'\nn\\
\lim_{\lambda\to 0}\, \omega(k,k') &=& 0 \nn\eea

Notice that $\real^n$ endowed with the composition law $\oplus$
acquires a general Lie group structure.  The inverse of an element
$k\in \real^n$ with respect to the composition law $\oplus$ will
be defined as: $$ \bar k \ : \ k\oplus \bar k = \bar k \oplus k=0
$$ where $0$ denotes the neutral element of $(\real^{n},\oplus)$,
and where we are assuming that the right and the left inverse are
the same.

In analogy with the Moyal case we define a map $\Omega$ from
ordinary functions to formal elements of a noncommutative algebra
as
\be \Omega(f)\equiv\frac{1}{(2\pi)^{n/2}} \int d^{n}k \ft(k)
W(k)\label{defOmega}. \ee where $\ft(k)$ is the standard Fourier
transform of $f(x)$. This definition enables us to write
$\Omega(e^{ikx})=W(k)$, and $$ \Omega^{-1}(W(k))=e^{ikx} $$ which
gives the inverse of the map~\eqn{defOmega} for all operators that
can be ``expanded'' in the ``plane wave'' basis given by the
$W$'s.

We show that a $*$-product can be obtained from such a generalized
Weyl map, so that $$ \Omega(f)\Omega(g)=\Omega(f*g) $$ where the
generalized star product has been denoted with the symbol $*$, in
order to distinguish it from the usual Moyal product (denoted by
the symbol $\star$).
 The product $\Omega(f)\Omega(g)$ is given by:
$$ \Omega(f)\Omega(g)=\frac{1}{(2\pi)^{n}}\int d^nkd^nk'
\ft(k)\gt(k')W(k)W(k') $$ and, using the definition of generalized
Weyl systems (\ref{genweylrel}), we obtain \be
\Omega(f)\Omega(g)=\frac{1}{(2\pi)^{n}}\int d^nkd^nk'
\ft(k)\gt(k')e^{\frac{i}{2}\omega(k,k')}W(k\oplus k')
\label{FGfourier} \ee It is useful to write the product as a
twisted convolution. In order to do this, we perform the following
change of variables in eq.~\eqn{FGfourier}: $$
 k\oplus k'=\xi
$$ then, using the associativity of $\oplus$, we have: \be
k=\xi\oplus\bar{k'}\equiv\al(\xi,\bar{k'}), \label{trasfkxi} \ee
and equation~\eqn{FGfourier} becomes \bea
\Omega(f)\Omega(g)&=&\frac{1}{(2\pi)^{n}}\int d^nk'd^n\xi
J(\xi,k')
\ft(\xi\oplus\bar{k'})\gt(k')e^{\frac{i}{2}\omega(\xi\oplus\bar{k'},k')}W(\xi)\nn\\
&=&\frac{1}{(2\pi)^{n}}\int d^n\xi
d^nk'J(\xi,k')\ft(\al(\xi,\bar{k'}))\gt(k')e^{\frac{i}{2}
\omega(\al,k')}W(\xi) \label{twistedconv} \eea where
$J(\xi,k')=|\partial_{\xi}\al(\xi,\bar{k'})|$ is the Jacobian of
the transformation~\eqn{trasfkxi}. The last equation can be cast
in a more suggestive form $$
\Omega(f)\Omega(g)=\frac{1}{\left(2\pi\right)^{n/2}}\int d\xi^n
\left(\frac{1}{\left(2\pi\right)^{n/2}}\int dk'^n
J(\xi,k')\ft(\al(\xi,\bar{k'}))\gt(k')e^{\frac{i}{2}
\omega(\al,k')}\right)W(\xi) $$ Comparison with~\eqn{defOmega}
suggests to define the Fourier transform of the deformed product
of $f$ and $g$ as: $$
\widetilde{\left({f*g}\right)}\left(\xi\right)=
\frac{1}{\left(2\pi\right)^{n/2}}\int dk^{\prime}
J(\xi,k')\ft(\al(\xi,\bar{k'}))\gt(k')e^{\frac{i}{2}
\omega(\al,k')}.$$ A detailed calculation~\cite{alz02} shows that
\be (f* g)(x)=\frac{1}{(2\pi)^{n}}\int d^nkd^nk'
\ft(k)\gt(k')e^{\frac{i}{2}\omega(k,k')}e^{i(k\oplus k')x}.
\label{nota}\ee As we claimed, it has been possible to define a
product simply considering $W$ as a formal device. Now we have
given the set of function on $\real^{n}$ a structure of algebra.
It is easy to check that the neutral element of the product is the
unit element $W(0)$. As is usual for noncompact geometries, it
does not belong to the algebra, which is composed of functions
vanishing at infinity; while associativity is a consequence of the
associativity of $\oplus$.

Now we want to show that the algebra that we have defined can be
promoted to a $C^*$-algebra (see section 1.1). In order to do this
we have to introduce a hermitian conjugation satisfying the
relations (\ref{conjug}). A hermitian conjugation can be defined
using the fact that for the undeformed algebra \be
f(x)^\dagger=f(x)^*=\frac {1}{(2\pi)^{n/2}}\int d^nk \ft^*(k)
e^{-ikx}=\frac {1}{(2\pi)^{n/2}}\int d^nk \ft^*(-k)
e^{ikx}\label{conjuga} \ee and define the hermitian conjugate of
$\Omega(f)$, defined as in eq.~\eqn{defOmega} as $$
\Omega(f)^{\dag}=\frac{1}{(2\pi)^{n/2}}\int d^nk \ft^*(\bar
k)W(k).
$$ This means that we assume, in the set of functions, the
following definition $$
f^{\dagger}\left(x\right)=\frac{1}{\left(2\pi\right)^{n}}\int
d^nk\,d^na\,f^{*}\left(a\right)e^{ikx}e^{i\bar{k}a}. $$

The norm is defined as in~\eqn{defnorm}. The compatibility of the
hermitian conjugation with the product~\eqn{FGfourier},
 $$
(\Omega(f)\Omega(g))^\dagger=\Omega(g)^\dagger \Omega(f)^\dagger,
$$
 imposes further restrictions
on $\omega$ and $\oplus$. Using the definition of hermitian
conjugate we obtain: \be
(\Omega(f)\Omega(g))^{\dag}=\frac{1}{(2\pi)^{n}}\int d^n\xi
d^nk'J^*(\xi',k')|_{\xi' =\bar{\xi}}\ft^*(\al(\bar{\xi},\bar{k'}))
\gt^*(k')e^{-\frac{i}{2}\omega^*(\al(\bar{\xi},\bar{k'}),k')}W(\xi).\label{1}
\ee On the other hand if we compute \bea
\Omega(g)^{\dag}\Omega(f)^{\dag}&=&\frac{1}{(2\pi)^{n}}\int d^nk
d^np
\gt^*(\bar{k})\ft^*(\bar{p})W(k)W(p)\nn\\
&=&\frac{1}{(2\pi)^{n}}\int d^nk' d^n\xi
J'(\xi,k')\gt^*(k')\ft^*(\overline{(k'\oplus\xi)})e^{\frac{i}{2}
\omega(\bar{k'},k'\oplus\xi)}W(\xi)\nn\\
&=&\frac{1}{(2\pi)^{n}}\int d^nk' d^n\xi J'(\xi,k')
\gt^*(k')\ft^*(\al(\bar{\xi},\bar{k'}))e^{\frac{i}{2}\omega(\bar{k'},\al(k',\xi))}W(\xi)
\label{2} \eea where $k'=\bar {k}$ and  $k\oplus p=\xi$, then $$
p=\bar{k}\oplus\xi=k'\oplus\xi\Rightarrow
\bar{p}=\bar{\xi}\oplus\bar{k'} $$ if and only if \be
\overline{(k\oplus k')}=\bar{k'}\oplus\bar{k} \label{ominus} \ee
This is a further requirement on $\oplus$, as group law. The
Jacobian becomes $$
J'(\xi,k')=|\partial_{k'}\bar{k'}\,\partial_{\xi}\al(k',\xi)| $$
Comparing the two equations~(\ref{1}) and~(\ref{2}), we obtain
sufficient conditions for compatibility \bea
J'(\xi,k')\equiv|\partial_{k'}\bar{k'}\partial_{\xi}\al(k',\xi)|
&=&|\partial_{\xi'}\al^*(\xi',\bar{k'})|_{\xi'=\bar{\xi}}\equiv J^*(\xi',k')|_{\xi'=\bar{\xi}}\label{3} \\
\omega^*(\al(\bar{\xi},\bar{k'}),k')&=&-\omega(\bar{k'},\al(k',
\xi))\label{4} \eea The standard Weyl--Moyal system described in
the previous section is an example of this construction, with
$\oplus$ the usual sum.

As we said, the $\oplus$ has given a group structure to
``momentum'' space, and of course there will be a Lie algebra
associated to the group. We will now argue that this Lie algebra
structure corresponds to the noncommutativity of the $x$'s on the
deformed space. First we define of all the generators $x_i$'s $$
X_{\alpha}=\frac{1}{\left(2\pi\right)^{n}}\int
d^{n}x\,d^{n}k\,x_{\alpha}e^{-ikx}W\left(k\right). $$
The product
between them is (performing the integral in a distributional
sense, with suitable boundary conditions) \bea {}
x_{\alpha}*x_{\beta}&=&\frac{1}{\left(2\pi\right)^{2n}}\int
d^{n}z\,d^{n}y\,d^{n}k\,d^{n}l\,z_{\alpha}y_{\beta}e^{-i\left(kz+ly\right)}
e^{\frac{i}{2}\omega\left(k,l\right)}e^{ix\left(k\oplus l\right)}
\nonumber\\
&=&x_{\alpha}x_{\beta}-\frac{i}{2}\left(\frac{\del^{2}\omega\left(k,l\right)}{\del
k_{\alpha}\del l_{\beta}}\right)\Bigg|_{k=l=0}
-ix_{\mu}\left(\frac{\del^{2}}{\del k_{\alpha}\del
l_{\beta}}\left(k\oplus l\right)^{\mu}\right)
\Bigg|_{k=l=0}\nn\eea The commutator is given by the antisymmetric
combination of this product. It is possible to prove~\cite{alz02}
that the second term on the r.h.s. of this relation gives the
structure constants of the Lie algebra defined by the Lie group
$\left(\real^{n},\oplus\right)$: $$ \left(\frac{\del^{2}}{\del
k_{\alpha}\del l_{\beta}}\left(k\oplus
l\right)^{\mu}\right)-\left(\frac{\del^{2}}{\del k_{\beta}\del
l_{\alpha}}\left(k\oplus l\right)^{\mu}\right)
\Bigg|_{k=l=0}=-\zeta^{\mu}_{\alpha\beta} $$ while the cocycle
term gives a central extension that can be cast in the usual form:
$$ \frac{1}{2}\left(\frac{\del^{2}\omega\left(k,l\right)}{\del
k_{\alpha}\del
l_{\beta}}-\frac{\del^{2}\omega\left(k,l\right)}{\del
k_{\beta}\del
l_{\alpha}}\right)\Bigg|_{k=l=0}=-\theta_{\alpha\beta}. $$ Finally
one obtains: $$ [x_\alpha,x_\beta]_*=i\zeta_{\alpha\beta}^\mu
x_\mu+i\theta_{\alpha\beta} $$

In the following we will consider \kM\ without central extensions, and therefore set $\theta=0$.

\section{$*$-products on \kM  }
So far we have abstractly defined a generalization of a Weyl
system. In this section we will show that these systems are a good
description of some $*$-products considered in the study of the
$\kappa$-Minkowski space.

The first two star products, that we describe, come from the
Quantum Group approach \cite{AmelinoMajid,KLM00}. Actually, in the
Quantum Group approach one is used to work directly with the
noncommutative coordinate-generators  $\x_{\mu}$ and their
functions. In particular every function is written in terms of the
exponential noncommutative functions $e^{ik\x}$ with an ordering
prescription. For example in \cite{AmelinoMajid} the prescription
with the time-to-the-right is used, and the exponential functions
are written as $e^{-ik\x}e^{ik_0\x_0}$. We show that there is a
correspondence between the choice of an ordering prescription, in
the space of the noncommutative functions of \kM , and the choice
of a star product.  Therefore they are two equivalent ways to
describe the algebra of functions in \kM.

The third star product is  constructed  through a procedure very
different from the one adopted in the  construction of  the first
two. As we discuss below, this procedure is based on a ``reduction
technique'' that can be applied to any noncommutative space for
which an underlying Poisson manifold is defined. As we have seen
in the section 2.1, \kM\ has a Poisson manifold structure and one
can apply the reduction procedure in order to construct a star
product on it.

\subsection{The CBH product}
The first product that we present is a simple application of the
well known Campbell--Baker--Hausdorff (CBH) formula for the
product of the exponential of noncommuting quantities. Let us
present this product in a more general way for a set of operators
satisfying a Lie algebra condition: $$
[\x_\mu,\x_\nu]=ic_{\mu\nu}^\rho\x_\rho.$$ The CBH formula is \be
e^{ik^\mu\x_\mu}e^{ip^\nu\x_\nu}=e^{i\gamma^\nu(k,p)\x_\nu}
\label{CBH} \ee with: $$
\gamma(k,p)^{\mu}=k^{\mu}+p^{\mu}+\frac{1}{2}c^{\mu}_{\delta\nu}k^{\delta}p^{\nu}+...
$$ We notice that the CBH formula is written in terms of the
exponential functions $e^{ik\x}$, which represent the
straightforward extension of the commutative exponential $e^{ikx}$
to the noncommutative space.

The relation (\ref{CBH}) leads to the  associative star
multiplication introduced by \cite{KLM00}, assuming that the Weyl
map of a function $f(x)$ can be written in terms of the
exponentials $e^{ik\x}$: $$
\Omega_{1}(f)=\frac{1}{(2\pi)^{n/2}}\int\tilde f(k) e^{ik\x} d^nk.
$$ Let us notice that with this definition the commutative
exponential function $e^{ikx}$ is mapped into the noncommutative
exponential: \be \Omega_1(e^{ikx})=\int\tilde \delta(k) e^{ik\x}
d^nk=e^{ik\x}.\label{exp1} \ee

From the definition~\eqn{stardefweylmap} we can obtain the star
product $*_1$ which corresponds to the map $\Omega_1$: \be (f*_1
g)(x)=\Omega_1^{-1}\left(\Omega_1(f)\Omega_1(g)\right).
\label{stardef1} \ee Since this product has been obtained by a
straightforward application of the BCH formula, we call it ``CBH"
star product.

In the case the operators $\hat x_\mu$ are the generators of \kM\
spacetime ~\eqn{kmincomm},  the function $\gamma(\al,\bt)$ can be
easily computed~\cite{KLM2} and its explicit expression is: \bea
\gamma^0(k,p)&=&k^0+p^0\nn\\
\gamma(k,p)^i&=&\frac{\phi(k^0)e^{\lambda
p^0}k^i+\phi(p^0)p^i}{\phi(k^0+p^0)}\nn \eea where the function
$\phi(a)$ is defined by \bea \phi(a)&=&\frac{1}{a\lambda
}(1-e^{-a\lambda})\label{CBHf} \eea In this way, putting in the
(\ref{stardef1}) $f=e^{ikx}$ and $g=e^{ipx}$, and using
(\ref{exp1}), we obtain: \bea
e^{ik^\mu x_\mu}*_1e^{ip^\nu x_\nu}&=&\Omega_1^{-1}\left(\Omega_1(e^{ikx})\Omega_1(e^{ipx})\right)\nn\\
&=&\Omega_1^{-1}\left(e^{ik\x}e^{ip\x}\right)=\Omega_1^{-1}\left(e^{i\gamma^\mu(k,p)\x_\mu}\right)\nn\\
&=&e^{i\gamma^\mu(k,p) x_\mu} \label{CBHstar}
\eea
The product among the generators is obtained differentiating
twice~\eqn{CBHstar} and setting $k=p=0$:
\bea
x_0*_1x_0&=&x_0^2\nn\\
x_0*_1x_i&=&x_0x_i+\frac{i\lambda}{2}x_i\nn\\
x_i*_1x_0&=&x_0x_i-\frac{i\lambda}{2}x_i\nn\\
x_i*_1x_i&=&x_i^2. \label{xstar1} \eea Of course these reproduce
the commutation rules for $\kappa$-Minkowski spacetime.

\subsection{The Time-to-the-Right  Ordered Product \label{sectstar2}}
The time-to-the-right ordered product is a modification of the
previous one and has its roots in the Majid-Ruegg-bicrossproduct
structure of $\kappa$-Poincar\'{e}. It has been first proposed
in~\cite{MajidRuegg} and subsequently investigated, for example,
in~\cite{MO,KMLS,KLM2}. One defines the ``time-to-the-right"
ordering the ordering  for which, in the expansion of a function,
all powers of $\x_0$ appear to the right of the $\x_i$'s. For
example the right-ordered exponential is
$e^{ik_i\x^i}e^{i\al_0\x_0}$. The relation for a time to the right
ordered exponential is: \be
e^{i\alpha^\mu\x_\mu}=e^{i\phi(\alpha_0)\alpha^i\x_i}e^{i\alpha^0\x_0}
\label{timeord}
\ee
where $\phi$ has been defined in˜(\ref{CBHf}).
Introducing the map $\Omega_2$ expressed in terms of the time to the right ordered exponentials:
\be
\Omega_2(f)=\frac{1}{(2\pi)^{n/2}}\int d^{n}k\tilde f(k) e^{ik_0\x_0}e^{-ik_j\x_j}
\ee
the relation (\ref{timeord}) leads to another associative product $*_2$, given by:
\be
(f*_2 g)(x)=\Omega_2^{-1}\left(\Omega_2(f)\Omega_2(g)\right)
\label{stardef2}
\ee
Following the same procedure with which we have obtains the $*_1$ product between two exponentials, we find the following expression for $*_2$:
\be
e^{ik^\mu x_\mu}*_2e^{ip^\nu
x_\nu}=e^{i(k^0+p^0)x_0+i\left(k^i+e^{-\lambda
k^0}p^i\right)x_i} \label{expstar2}
\ee
and the product among the generators:
\bea
x_0*_2x_0&=&x_0^2\nn\\
x_0*_2x_i&=&x_0x_i\nn\\
x_i*_2x_0&=&x_0x_i-i\lambda x_i\nn\\
x_i*_2x_i&=&x_i^2 \label{xstar2}
\eea
These relations are different from the relations (\ref{xstar1}), but they equally reproduce the commutation relations for \kM .

\subsection{The Reduced Moyal Product}

This product is a particular case of a general class of
$*$-products for three-dimensional Lie algebras, introduced
in~\cite{selene}, which for the \kM\ case can be easily
generalized to an arbitrary number of dimensions. The idea is to
obtain a product in an $n$ dimensional space by considering it as
a subspace of an higher dimensional symplectic phase space,
equipped with the usual Poisson bracket, and a Moyal $\star$
product (with deformation parameter $\lambda$). The product is
then defined by first lifting the functions from the smaller space
to the phase space, multiplying them in the higher space, and then
projecting back to the smaller space. The form of the lift (a
generalized Jordan-Schwinger map) and the canonical structure of
the Moyal product ensure that this procedure defines a good $*$
product in the smaller dimensions.

We indicate the coordinate on the phase space $\real^6$ with the
notation: $\real^6 \ni u=(q_1,q_2,q_3;p_1,p_2,p_3)$. We need
to define a map $\pi$:
\be
\pi: \real^6\rightarrow \real^4
\ee
or equivalently the map $\pi^*$ which pulls smooth functions on
$\real^4$ to smooth functions on $\real^6$. The map $\pi$ is
explicitly realized with four functions of the $p$'s and $q$'s,
which we call $x_\mu$. The requirement is that the six dimensional
Poisson bracket of $x$'s reproduce the ``classical'' \kM\
algebra\footnote{This algebra is an extension of the three-dimensional Lie algebra $sb(2,C)$ of $2{\times}2$ triangular
complex matrices with zero trace treated in~\cite{selene}.}:
\be
\{x_0,x_i\}=  x_i,\;\;\;\{x_i,x_j\}=0\;\;\; i,j=1,2,3
\ee
The $*_3$ product is then defined by
\be
\pi^*(f*_3g)=\pi^*f\star\pi^*g
\ee
The fact that, after performing the (nonlocal) product in six
dimensions, we are left with a function still defined using only
the four dimensional coordinates is ensured by the existence of
two local functions, $H_1$ and $H_2$, with the property that
\be
L_{H_i}\pi^*f(x)=0
\ee
and
\be
L_{H_i}\pi^*\left(f(x)\star g(x)\right)=0
\ee
In other words, from the six dimensional point of view, the $x$'s
commute with the $H$'s, and this commutation is stable under the
$\star$-product, thus ensuring that if we multiply a function only
of the $x$'s, the product does not depend on the extra
coordinates. Upon the identification of the parameter $\theta$
with $\lambda$ we obtain a \kM\\ product on $\real^4$.

One of the possible realization of $\pi$ is:
\bea
x_0&=&- \sum_iq_ip_i\nn\\
x_i&=&q_i\label{jsmap}
\eea
Two independent commuting functions are:
\bea
H_1&=&\arctan\frac{q_2}{q_3}\nn\\&~&\nn\\
H_2&=&\arctan\frac{q_3}{q_1}
\eea
Note however that the representation ~\eqn{jsmap} is not unique.
For example the following choice works as well:
\bea
x'_0&=&- \sum_ip_i\nn\\
x'_i&=&e^{q_i} \label{seleneprod'}
\eea
with the commuting functions
\bea
H_1'&=&e^{q_2-q_3}\nn\\
H_2'&=&e^{q_3-q_1}
\eea
connected by a singular canonical transformation to the previous
one.

We will denote the products with the choices~\eqn{jsmap}
and~\eqn{seleneprod'} as~$*_3$ and $*_4$ respectively. The
explicit products between the generators are in~\cite{selene}:
\bea
x_0*_3x_0&=&x_0^2+\frac{3}{4}\lambda^2\nn\\
x_0*_3x_i&=&x_0x_i+i\frac\lambda2x_i\nn\\
x_i*_3x_0&=&x_0x_i-i\frac\lambda2x_i\nn\\ x_i*_3x_i&=&x_i^2
\eea
and
\bea
x_0*_4x_0&=&x_0^2\nn\\
x_0*_4x_i&=&x_0x_i+i\frac\lambda2x_i\nn\\
x_i*_4x_0&=&x_0x_i-i\frac\lambda2x_i\nn\\
x_i*_4x_i&=&x_i^2
\eea
It may be noticed that these relations are the same as the one for $*_1$
in~\eqn{xstar1}. The two products however, although similar, do not
coincide for generic functions.

\section{Weyl Systems for \kM }
In this section we will show how the various products presented
earlier are particular instances of generalized Weyl systems. The
starting point for the construction of the product is the
identification of the $W(k)$'s, which enables the calculation of
the particular relation~\eqn{genweylrel} for the various cases. In
other words we will give an explicit realization of the group
composition law $\oplus$. All Weyl systems presented here have
$\omega=0$, and in all cases a straightforward calculation
verifies relation~(\ref{3}).
\subsection{Weyl System for the CBH product}
In this case the relation is given by the CBH product, which for
\kM\\ has been given in eqs.~(\ref{CBH}-\ref{CBHf}). The
composition $\oplus_1$ can be calculated to give: \bea (k\oplus_1
p)^0&=&k^0+{p}^0\nn\\ (k\oplus_1
p)^i&=&\frac{\phi(k^0)k^i+e^{\lambda
k^0}\phi({p}^0){p}^i}{\phi(k^0+{p}^0)} \eea where $\phi$ has been
defined in eq.~\eqn{CBHf}. We have that \be
W_1(k)=\Omega(e^{ik^\mu x_\mu})=e^{ik^\mu \x_\mu} \ee There are
one check which has to be performed to ensure that $\oplus_1$
defines a group. The requirement~\eqn{ominus}, $\bar{(k\oplus
p)}=\bar{p}\oplus\bar{k}$, is verified with the definition of the
following inverse $\bar{k}$ \be \bar{k}=(-k^0,-k^i) \ee in fact
\be \overline{(k\oplus_1 p)}=\bar{p}\oplus\bar{k} \ee

\subsection{Weyl System for the Time Ordered Product}
This case is similar to the previous one, and a direct calculation
using the CBH relations for the time ordered exponentials give:
\bea
(k\oplus_2p)^0&=&k^0+{p}^0\nn\\
(k\oplus_2p)^i&=&k^i+e^{-\lambda k^0}{p}^i \label{Roplus}
\eea with
\be \bar{k}=(-k^0,-e^{\lambda k^0}k^i) \ee This momenta
composition reflects the coproduct in the Majid-Ruegg
bicrossproduct basis, where the time ordered exponential has been
considered a natural basis for the space of functions. In the
time-to the right case, the Weyl system is given by: \be
W_2(k)=\Omega_2(e^{ikx})=e^{ik^i\x_i}e^{ik^0\x_0}\label{Rmap} \ee
and from the (\ref{timeord}) one can see that it is related with
the $\Omega_1$ map through the relation: \be
W_2(k)=e^{i(k_0\x_0+ik^i\x_i/\phi(k_0))} \ee that, using the
relations (\ref{stardef1}-\ref{CBHstar}), can be cast in the form:
\be W_2=\Omega_1(e^{ik^ix_i}*_1e^{ik^0x_0}) \ee So the product
$*_2$ can be expressed by the choice of a different $W$ \emph{of
the $*_1$ product}.

We want to emphasize that this time-to-the-right Weyl map is equivalent to a corresponding  ordering convention.
In fact, the $\Omega_1$ map associates to each commutative function the
element obtained ordering the time $x_0$ to the right and replacing the $x$
variables with the elements ${\x}$. For example
\begin{eqnarray*}
\Omega_R(x_0x_j) = \Omega_R(x_jx_0)={\x}_j{\x}_0
\end{eqnarray*}
and notice that ${\x}_j{\x}_0 \neq {\x}_0{\x}_j = {\x}_j({\x}_0-i\lambda)
=\Omega_R(x_j x_0-i\lambda x_j)$.
Then naturally one can refer to this choice as
\emph{time--to--the--right ordering} or \emph{right ordering}
convention.

\subsection{Weyl Systems for the Reduced Moyal products}

The path to the definition of the reduced products is
intrinsically different from the first two. These are defined
using straightforwardly the CBH formula with a specific ordering.
The reduced product instead comes from a four dimensional
reduction of a six dimensional product. There is therefore no
warranty that it is possible to obtain them form a \emph{four}-dimensional
Weyl system. This is nevertheless possible.

Notice that if we were to define $W$ as the $*_3$ or $*_4$
exponential of $ik^\mu x_\mu$ we would find the CBH product. This
is because the $*$ commutator of the $x$'s are all the same. We
must therefore use another quantity, and we could use the ordinary
exponential. Care must be taken however because this is not an
unitary operator (with the hermitean conjugation defined in
eq.~\eqn{conjuga}). So that it is necessary to normalize it. The
calculations of the product of two exponentials are in Appendix A.
We define $W_3$ as \be W_3=|a(k^0,-k^0)|^{3/2}\Omega_3(e^{ikx})
\label{W3def} \ee with \bea
a(k^0,p^0)&=&1+\frac{\lambda^2}{4}k^0p^0\nn\\
b(k^0)&=&1-\frac{\lambda}{2}k^0 \label{defab} \eea {}From
equation~\eqn{new} we can read
\be (k\oplus_3p)^\mu=k^\mu+{p^\mu} +\frac{\lambda}{2a}( {p}^0 b(k^0)
k^\mu-{p}^0 b(-{p}^0) {p}^\mu)
\ee
with $\bar k=-k$. Unlike all
other composition laws, $\oplus_3$ is the only one in which the
``time-like'' coordinate has a non-abelian structure: $(k\oplus_3p)^0\neq(p\oplus_3 k)^0$.
Moreover $\oplus_3$ is not well defined for all $(k,p)$, so that it does not define a group structure. In spite of this the integral (\ref{nota})
is well defined and the star product can be always defined.

For the $*_4$ case, the exponential is unitary, and it is possible
to define $W_2(k)=e^{ikx}$. In Appendix A  we calculate the
composition rule $\oplus_4$ which results:
\be
k\oplus_4 p=(k^0+{p}^0,e^{\frac{\lambda}{2}
{p}^0}\vk+e^{-\frac{\lambda}{2} k^0}\vec{p})\label{oplus4}
\ee
with $\bar k=-k$.

Notice that, while the composition law $\oplus_2$, related with
the time-to-the-right star product, is connected to the
coproduct~\eqn{coproducts} in the bicrossproduct basis, the
$\oplus_4$ is related to the coproduct in the standard basis, that
has been discussed in section 1.3. The coproduct of the \kkP\
translation generators in the standard basis is given by
(\ref{stanbas}): \bea
\Delta P_0&=& P_0\otimes 1 + 1 \otimes P_0\nn\\
 \Delta P_i &=& P_i\otimes e^{\frac{\lambda P_0}{2}}+ e^{-\frac{\lambda P_0}{2}}\otimes P_i \label{coproductssta}
\eea Let us note that in this basis the  coproduct for $P_i$
assume a ``symmetric'' form. The similarity goes beyond it, as the
$*_4$ product can be written as as ``symmetrically-ordered''
product with respect to the $\x_0$ \kM\ coordinate defining: \be
\ddag e^{ikx}\ddag = e^{\frac{ik^0\x_0}{2}}e^{-i\vec
k\vec\x}e^{\frac{ik^0\x_0}{2}} \ee so that \be \ddag e^{ik\x}\ddag
\,\ddag e^{ip\x}\ddag = \ddag e^{i(k\oplus_2p)\x}\ddag \ee and it
is possible to repeat the calculations in Sect.~\ref{sectstar2} to
obtain the analog of relation~\eqn{expstar2}. In this way, we can
write the Weyl system for the symmetrically-ordered star product,
in the following form: \be
W_4(k)=\Omega_4(e^{ikx})=e^{\frac{ik^0\x_0}{2}}e^{-i\vec k\vec
\x}e^{\frac{ik^0\x_0}{2}}  \label{Smap} \ee that is related  with
the $\Omega_1$ in the following way: \be
W_4(k)=\Omega_1(e^{i\frac{k_0}{2}x_0}*_1e^{-ik_ix_i}*_1e^{i\frac{k_0}{2}x_0})
\ee As in the case of the time-to-the right-map also in this case
the map corresponds to a clear ordering convention for the
functions on \kM . The $\Omega_4$ map, indeed, associates to each
commutative function a time-symmetrized element, for example
\begin{displaymath}
\Omega_4(x_0x_j)=\Omega_4(x_jx_0)=\frac{1}{2}({\x}_0{\x}_j+{\x}_j{\x}_0)
\end{displaymath}
and could be described
as \emph{time-symmetrized ordering} or \emph{symmetric ordering}.

Since the maps $\Omega_1$ and $\Omega_4$ correspond to two clear ordering prescriptions, we will use them in order to analyse the implications of the non uniqueness of the Weyl map in the study of \kM\ and in particular in the characterization of its symmetries.

\section{Plane waves in \kM }

One can give a physical interpretation to the formulas obtained so far.

The extension of the Weyl formalism that we have analyzed in the
previous sections allows us to introduce a field in \kM\ through
the generalized Weyl map, in the following way: \be
\Phi(\x)=\Omega(\phi(x))=\frac{1}{(2\pi)^2}\int d^4p\,
\tilde{\phi}(p)\Omega(e^{ipx}) \label{kmfield} \ee where $\phi(x)$
is the commutative field to which $\Phi$ reduces in the limit
$\lambda\to 0$, and $\tilde {\phi}(k)$ is the standard Fourier
transform of $\phi(x)$.

In the case of Minkowski commutative spacetime, a field $\phi(x)$
is written in terms of plane waves as a Fourier integral: \be
\phi(x)=\frac{1}{(2\pi)^2}\int d^4p
\,\tilde{\phi}(p)e^{ipx}\label{mfield} \ee in which the variables
$(p_0,\vec{p})$ have the meaning of ``four-momenta'' of particle,
\emph{i.e} its energy and three spatial momenta.

Following the analogy with the commutative case we might give the
variables ($p_0,\vec{p}$) appearing inside the Weyl map
(\ref{kmfield}) the interpretation of energy and momenta of a
particle as well. Consequently, the expression (\ref{kmfield}) can
be viewed as an expansion in "plane-waves basis" given by the
generalized Weyl system $W(k)$. Thus, in our deformed Minkowski
space a natural definition of plane waves associated to a point
$p=(E,\vec{p})$ in the momentum space is provided by the
generalization of the same notion in the commutative case: \be
\psi_{\vec{p},E}=\Omega (e^{ipx})=W(p) \ee From this physical
interpretation of the Weyl system it follows that the composition
law in the momentum group is the non-abelian sum $\oplus$. In
fact, if we multiply two \kM\ plane waves of momenta
$p_{\mu}=(E,\vec{p})$ and $p'_{\mu}=(E',\vec{p'})$, we obtain: \be
\psi_{\vec{p},E}\psi_{\vec{p'},E'}=W(p)W(p')=W(p\oplus
p')=\psi_{p\oplus p'}\label{comp} \ee in which we have used the
property of the generalized Weyl System (\ref{genweylrel}) for \kM
, where $\omega=0$.

In summary, our analysis suggests that in \kM\ there is a natural
plane-wave notion given by the Weyl system, and a natural
interpretation of the Fourier variables $p_\mu$ as four-momenta
associated to plane wave. The momenta of two plane waves compose
with a non-abelian law given by $\oplus$. This composition law
depends on the choice of the Weyl system (or Weyl map) which has
been showed to be non unique. From this fact it is clear that an
ambiguity concerns the composition of momenta. The momenta have
the meaning of generators of the translations over a spacetime,
and their composition is related to their coproduct. Thus, this
ambiguity concerning the momenta composition represents an
ambiguity concerning the Hopf-algebra symmetry of \kM .

We clarify this fact in the following way. Remember that \kM\
represent the enveloping algebra $U(g)$ of the classical Lie
algebra $g$ whose generators satisfy the relations: \be
[\x_0,\x_j]=i\lambda \x_j\;\;\;[\x_j,\x_k]=0 \ee Let us consider
the elements of \kM\ given by the time-to-the-right Weyl map
(\ref{Rmap}), that we call now $\Omega_R$. They are characterized
by the composition rule (\ref{comp}): \bea
\Omega_R(e^{ipx})\Omega(e^{ip'x})&=&\Omega(e^{i(p \oplus_R p')})
\label{oplusR} \eea where $ p \oplus_R p'=(p_0+p'_0,
p_j+e^{-\lambda p_0} p'_j) $. In this way, as we have seen in the
previous sections, the elements $W_R(k)=\Omega_R(e^{ikx})$ form a
compact unitary group that we call $G$, with the group law
$\oplus$, the unity $W(0)$ and the inverse $W(\bar{k})$. \be
G=\{W_R(k)\in \mathcal{M}_k :
W_RW^*_R=W^*_RW_R=1,\;\;\;W_R(k)W_R(k')=W_R(k\oplus_R k')\} \ee
where $k$ are real variables that parametrize  the group elements.
In this way $G$ represents the unitary group of $U(g)$. In fact,
each unitary element of \kM\ that is written in a different
ordering can be rewritten in terms of the time-to-the right
ordering through a redefinition of the Fourier variable $p_\mu$.
For example a simply calculation show that the time-symmetrized
exponential (\ref{Smap}) $W_S(k)=\Omega_S(e^{ikx})$ can be
rewritten in terms of $W_R$ in the following way: \be
W_S(k)=\Omega_S(e^{ikx})=e^{ik_0\x_0/2}e^{-ik\x}e^{ik_0\x_0/2}=e^{-ie^{\frac{-\lambda
k_0}{2}}k\x}e^{ik_0\x_0}=W_R(k_0,e^{\frac{\lambda}{2}k_0}\vec{k})
\ee Thus, the elements $W_S$ are in a one-to-one correspondence
with the elements $W_R$ and they produce the same unitary group
$G$.

Let us consider now the algebra of the functions $C(G)$ over the
group $G$. In the section 1.2 we have shown that the algebra
$C(G)$ and $U(g)$ are dual. The duality relation between them can
be defined in the following way: \be <P_\mu, \Omega_R(e^{ik_\nu
x^\nu})>=k_\mu  \label{luca} \ee where $P_\mu$ are functions of
the algebra $C[G]$. In particular, $P_\mu$ are the generators of
the translations in \kM\ as we can see deriving the relation
(\ref{luca})and then putting $k_\mu=0$ \bea
\partial^{k}_\rho <P_\mu,\Omega_R(e^{ik_\nu x^\nu}) >|_{k=0}&=&<P_\mu,\Omega_R(i x_\rho e^{ik_\nu x^\nu})>|_{k=0}=i<P_\mu,\Omega_R(x_\rho>=i<P_\mu,\x_\rho>\nn\\
&=&\partial_\rho k_\mu=\eta_{\rho\mu} \eea from which we obtain
the duality relation between generators of the translations and
coordinates: \be <P_\mu,\x_\rho>=-i\eta_{\rho \mu} \ee

The duality laws between Hopf algebras allow us to determine the
Hopf algebra structure of $P_\mu$ using (\ref{luca}). For example
\bea
<\Delta P_\mu, \Omega_R(e^{ik_\nu x^\nu})\otimes \Omega_R(e^{ik'_\nu x^\nu})>&=&<P_\mu, \Omega_R(e^{ik_\nu x^\nu})\Omega_R(e^{ik'_\nu x^\nu})>=<P_\mu, \Omega_R(e^{i(k_\nu\oplus k')x^\nu})>\nn\\
&=&k_\mu\oplus k'_\mu=<[P_\mu\otimes 1+e^{\lambda P_0(1-\delta_{\mu 0})}P_\mu],\Omega_R(e^{ik_\nu x^\nu})\otimes \Omega_R(e^{ik'_\nu x^\nu})>\nn
\eea
From which follows that
\be
\Delta(P_0)=P_0\otimes 1+1\otimes P_0,\;\;\;\Delta(P_j)=P_j\otimes 1+e^{-\lambda P_0}\otimes P_j
\ee
and in the same way, one determine the other Hopf algebra structures:
\bea
\epsilon(P_\mu)&=&0\nn\\
S(P_0)&=&-P_0,\;\;\;S(P_j)=-e^{\lambda P_0}P_j \label{antipMR}
\eea This relations define the generators of the translations of
\kkP\ in the Majid-Ruegg bicrossproduc basis.

This analysis shows that the time-to-the right Weyl systems are
connected with the Majid-Ruegg basis. If one repeat the analysis
in the case of other choices of Weyl systems, one finds the
translation generators in other  \kkP\ bases. In particular in the
case of time-symmetrized Weyl systems, the duality relation \be
<P'_\mu,W_s(k)>=k_\mu \ee states that $P'_\mu\neq P_\mu$ are the
translations generators in an other bicrossproduct basis of \kkP .

These examples show that an ambiguity concerns the definition of
the symmetry generators and that the definition of translation
generators depends on the duality relations. Duality relations as
(\ref{luca}) however have no clear physical interpretations. In
the following chapter we analyse more deeply the symmetries of
\kM\ in order to clarify this degeneracy problem.



\chapter{Description~of~noncommutative-spacetime~symmetries}

This chapter is devoted to the study of the symmetry in \kM\
spacetime. In the previous chapter we have shown that there is a
large ambiguity concerning the characterization of the symmetries
of \kM\ spacetime. As discussed in section 2.5, a degeneracy in
the description of the symmetries of \kM\  has been obtained from
a description based on the assumption of certain duality relations
that do not have a clear physical interpretation. Therefore some
studies, supported by these duality criteria, state (see, {\it
e.g.}, Refs.~\cite{lukieAnnPhys,Kow02-NST}) that the symmetries of
\kM\ can be described by any one of a large number of realizations
of the \kkP\ Hopf algebra. But the nature of this claimed
symmetry-description degeneracy remains obscure from a physics
perspective.

In order to clarify the origin of this ambiguity, we give here an
alternative characterization of the symmetries of \kM\ avoiding
the abstract criteria of duality. We hope that such a new
characterization of the symmetries, based on more physical
arguments, allows us to reduce the degeneracy as well. Our
approach is based on a generalization of the notion of symmetries
adopted in the classical (commutative) case. In the classical case
the symmetry algebra of Minkowski spacetime is the
10-generator-Poincar\'e algebra and we can write ``maximally
symmetric" actions, \emph{i.e.} actions which have all the
Poincar\'e symmetries. The existence of maximally symmetric
actions is the most relevant fact from the physical point of view
because the actions describe physical theories whose laws can be
tested experimentally. So, in our approach we focus on the
possibility to introduce symmetries as action symmetries. In \kM\
the form of the symmetry algebra will be a generalization of the
Poincar\'e algebra that reduces to it in the commutative limit
$\lambda=0$. As we show below, we can start from a natural
generalization of the definition of the classical generators and
then find an action invariant under all them. However in our
noncommutative spacetime we make a further request on the
transformation generator in order to guarantee that their action
preserve the ``algebra structure'' (see later) of \kM . The
additional request is that the symmetry algebra be a Hopf algebra.
In this way we define \emph{symmetry operators} for a theory on
\kM\ the operators that:
\begin{itemize}
\item all leave invariant the action of the theory,
\item close a compatible Hopf algebra structure.
\end{itemize}
As we discuss in the following, this definition reduces to the
standard definition of symmetry in the classical case, in which
the Poincar\'e-symmetry algebra is a Lie algebra (\emph{i.e.} a
Hopf algebra generated by primitive elements) and the second
request is naturally satisfied.

We apply our definition of symmetry in the analysis of the
simplest case of an action for a free scalar field.

\section{Construction of an action in \kM }

We have emphasized that our analysis of the symmetries relies on
the possibility to define an action of a theory. In order to
introduce an action on \kM ,  we use two fundamental tools: a Weyl
map (connecting a given function in the noncommutative Minkowski
with a corresponding function in commutative Minkowski) and of a
compatible rule  of integration in this noncommutative spacetime.

As we have shown in the previous chapter one can introduce a
\emph{generalized Weyl map} which establishes a
correspondence between elements of \kM\ and analytical functions
of four commuting variables $x_\mu$. We have shown also that this
correspondence is not unique and the analysis of the symmetries in
\kM\  might depend on the choice of the Weyl map.
In order to study the possible dependence of the symmetry analysis
on the choice of the Weyl map it is convenient for us to focus on
two specific choices of Weyl maps: the time-to-the-right map,
which we denote by $\Omega_R$, and the time-symmetrized map, which
we denote by $\Omega_S$, following the notation introduced in
section 2.5. Let us remember that these two choices are equivalent
to two corresponding ordering conventions, that we have shortly
called \emph{right ordering} and \emph{symmetric ordering},
referring the adjective right (symmetric)  to the position of the
time variable with respect to the spatial variables. One can see
this correspondence on the simple function $f(x)=xx_0=x_0x$, where
the two maps are: \bea
\Omega_R(f)&=&{\x}{\x_0}\;\;\;\mbox{(right ordering)} \\
\Omega_S(f)&=&\frac{1}{2}({\x}{\x_0}+\x_0{\x})\;\;\,\mbox{(symmetric ordering)}
\eea

It is sufficient to specify the Weyl map on the complex exponentials
and extend it to the generic function $f(x)$, whose Fourier transform is
\mbox{$\tilde{f}(p)=\frac{1}{(2\pi)^4}\int f(x)e^{-ipx}{\de}^4p$},
by linearity
\begin{displaymath}
\Omega_{R,S}(f)=\int \, \tilde{f}(p)
\, \Omega_{R,S}(e^{ipx}) \, {\de}^4p
~.
\end{displaymath}

Let us remind that the time-to-the-right map defined by
(\ref{Rmap}) is given by:
\begin{eqnarray}
\Omega_R(e^{ipx}) = e^{-i\vec{p}\vec{\x}}e^{ip_0\x_0} \label{right}
\end{eqnarray}
while  the ``time-symmetrized" map (\ref{Smap}) is
\begin{eqnarray}
\Omega_S(e^{ipx}) = e^{ip_0{\x}_0/2}e^{-i\vec{p}\vec{\x}}e^{ip_0{\x}_0/2}
\label{sym}
\end{eqnarray}
(We are adopting conventions such that $px=p_0x_0-\vec{p}\vec{x}$,
with $p_\mu$ four real commuting parameters, and similarly
$p{\x}=p_0{\x}_0-\vec{p}\vec{\x}$.) We also note that the relation
between the two Weyl maps here considered is the following \be
\Omega_R(e^{ipx})=\Omega_S(e^{i\vec{p}e^{\frac{\lambda}{2}
p_0}\vec{x}-ip_0x_0}) \ee {\it i.e.} it is possible to go from
time-to-the-right to time-symmetrized ordering through a
four-momenta transformation. This will also play a key role in our
analysis.

Of course, it is reasonable to contemplate even more options in
addition to these two; however, the two options of Weyl map that
we consider, the one inspired by time-to-the-right ordering and
the one inspired by time-symmetrized ordering, can be viewed as
the most natural ones. In fact, the time coordinate has a
privileged role in the \kM\ commutation relations and it is
natural to think first of all to three options: time-symmetrized,
time-to-the-right and time-to-the-left. We are actually
considering also the time-to-the-left option; in fact, one can
obtain the $\Omega_L$ map \be
\Omega_L(e^{ipx})=e^{ip_0\x_0}e^{-ip\x} \label{timeleft}\ee from
the $\Omega_R$ map by substituting the variable $p_\mu$ inside the
integral with $S(-p_\mu)$, where $S$ is the antipode map defined
in (\ref{antipMR}). We will show that this relation between the
two maps $\Omega_R$ and $\Omega_L$ makes the implications of
time-to-the-left ordering almost indistinguishable from the ones
of the time-to-the-right ordering (and can be simply obtained from
these by substituting $\lambda$ with $-\lambda$). Then, for our
purposes it is will be sufficient to focus on these two examples
in order to illustrate the differences and the common features of
two different choices of Weyl maps.

Having specified our objective as the one of describing free scalar
fields in \kM\ we have not yet established the form of the action.
We will look for an action which realizes our objective of
a theory with symmetries described
by a 10-generator Hopf algebra.

As announced, our construction of the theory will for convenience
make use of Weyl maps, which allow us to keep track a various
properties in the familiar context of functions of auxiliary commutative
coordinates.
We are therefore ready for the first step in constructing
the action for a free scalar particle. Of course we need
a rule of integration.
We can assume a rule of integration that is  naturally expressed using
the $\Omega_R$ Weyl map
\begin{equation}
\int_R {\de}^4{\x}\;\Omega_R(f)=\int f(x)\,{\de}^4x \;\;.
\label{intR}
\end{equation}
which states that the integral of a right-ordered function of \kM\
corresponds exactly to the integral of the corresponding
commutative function. In this way  the right integral of a
right-ordered exponential corresponds to the standard delta
function: \be \frac{1}{(2\pi)^4}\int dx
\Omega(e^{ikx})=\delta(k)\label{delta} \ee This rule has been
proposed in \cite{MO} and largely investigated in literature (see
for example \cite{AmelinoMajid}). Our alternative choice of Weyl
map would naturally invite us to consider the integration rule
\begin{equation}
\int_S {\de}^4{\x}\; \Omega_S(f)=\int f(x)\,{\de}^4x\;\;.
\label{intS}
\end{equation}
Actually these integrals are
equivalent, {\it i.e.} $\int_R{\de}^4{\x}\;\Phi=\int_S{\de}^4{\x}\;\Phi$
for each element $\Phi$ of \kM. This is easily verified by expressing
the most general element of \kM\ both in its $\Omega_R$-inspired
form and its $\Omega_S$-inspired form
\begin{eqnarray}
\Phi = \int{\de}^4p\;\tilde{f}(p)\Omega_R(e^{ipx}) =
\int{\de}^4p\;\tilde{f}(p_0,pe^{-\lambda p_0/2})e^{-3\lambda
p_0/2}\Omega_S(e^{ipx})
\label{rightandsymm}
\end{eqnarray}
and observing that
\begin{displaymath}
\int_R{\de}^4{\x}\;\Phi=\int_S{\de}^4{\x}\;\Phi = (2\pi)^4\tilde{f}(0)
~.
\end{displaymath}
In our search of a maximally-symmetric theory with construction
based on $\Omega_R$ or $\Omega_S$ we therefore have a natural candidate
for the integration rules to be used: (\ref{intR}),
which can be equivalently reformulated as (\ref{intS}).
[Because of the equivalence we will omit indices $R$ or $S$ on
the integration symbol.]

We can now also start formulating an educated guess for the
general structure of the action we are seeking
\begin{displaymath}
S(\Phi)=\int{\de}^4{\x}\;\Phi(\Box_\lambda+M^2)\Phi
\end{displaymath}
where $\Phi$ is a generic real\footnote{ we have defined the
concept of ``reality" for a function of noncommuting coordinates
in Subsection~1.1.1 by introducing the conjugation $*$ } element
of \kM, $M^2$ is (real, dimensionful and) positive and
$\Box_\lambda$ is a (differential) operator which is still to be
specified (we need more guidance concerning our requirement of
obtaining a maximally-symmetric theory) but we know it should
reproduce the familiar D'Alembert operator in the $\lambda
\rightarrow 0$ commutative-spacetime limit. For each real element
$\Phi$, the action $S(\Phi)$ is a real number.

\section{Hopf-Algebra description of symmetries}

In the familiar context of theories in commutative spacetimes we
describe an external symmetry as a transformation of the
coordinates that leaves invariant the action of the theory, and we
shall insist on this property in the case of \kM\ spacetime. We
want to emphasize that the generalization of the  symmetry concept
that we give in the special case of \kM\ spacetime has a general
character and can be applied to other noncommutative spacetimes as
well.

In preparation for our analysis, let us consider the simple action
$S[\phi]$ of a free scalar field $\phi$ in commutative Minkowski
spacetime: \be S(\phi)=\int{\de}^4x\;\phi(\Box + M^2)\phi \ee
where the operator $\Box=\partial_0^2-\nabla^2$ is the familiar
D'Alembert operator.

We will review the Lie-algebra description of the
symmetries of this action and then show that, upon considering
appropriate coalgebraic properties, this can be
cast in Hopf-algebra language.

So we start with our free scalar field in commutative Minkowski spacetime,
and we observe that the most general
infinitesimal transformation that can
be considered is $x'_\mu=x_\mu+\epsilon A_\mu(x)$,
with $A_\mu$ four real functions of the coordinates.

Taking into account that the condition of relativity invariance for a scalar field is
\be
\phi'(x')=\phi(x),
\ee
we find that
\bea
\phi'(x')-\phi'(x)&=&\phi(x)-\phi'(x)\nn\\
(x'-x)^\mu\partial_\mu\phi'(x)&=&\phi(x)-\phi'(x)\nn\\
\epsilon A^\mu(x)\partial_\mu\phi'(x)&=&\phi(x)-\phi'(x)\label{covarcond}
\eea
from which it follows that
\be
\phi'(x)=\phi(x)-\epsilon A^\mu(x)\partial_\mu\phi'(x)
\ee
and then
\be
\partial_\mu\phi'(x)=\partial_\mu\phi(x)+O(\epsilon).
\ee So, the equality (\ref{covarcond}) at the leading order in
$\epsilon$ is: \be \phi'(x)-\phi(x)=-\epsilon
A^\mu(x)\partial_\mu\phi(x) \ee
In terms of the generator $T$ of the
transformation, $T=iA_\mu(x)\partial^\mu$,
one obtains $x'=(1-i\epsilon T)x$ and $\phi'=(1+i\epsilon T)\phi$.
[The action of $T$ over $\x$ is indicated by $T\x$.]

The variation of the action, at the leading order in $\epsilon$ is
\bea
S(\phi')-S(\phi)&=&i\epsilon\!\int\!{\de}^4x\left(T\{\phi
(\Box+M^2)\phi\}+\phi[\Box,T]\phi\right)\nn\\
&=&i\epsilon\!\int\!{\de}^4x\left(T{\mL}(x)+\phi[\Box,T]\phi\right)\nn
\eea
and therefore the action is invariant under $T$-generated transformations
if and only if
\bea
\int\!{\de}^4x\left(T{\mL}(x)+\phi[\Box,T]\phi\right) = 0
~.\label{twoparts}
\eea

In the case of the illustrative example we are here considering of
a free scalar field in commutative Minkowski spacetime there is a
well-established form of the (maximally-symmetric) action and one
can just verify that condition (\ref{twoparts}) is satisfied. In
cases in which the (possibly noncommutative) spacetime is given
and one is looking for a maximally-symmetric form of the action
that analysis can naturally progress in two steps. In the first
step one can determine the algebra whose elements $T$ satisfy
\begin{equation}\label{eq:sym}
\int{\de}^4x\;T\,{\mL}(x)=0
\end{equation}
for every scalar function ${\mL}(x)$, independently of the form of
the differential operator
contained in ${\mL}(x)$. In fact, consider an action $S[\phi]=\int dx\mL(x)$,
with a Lagrangian $\mL(x)=\phi${\cal {O}}$\phi$ containing a generic differential operator ${\cal {O}}$.
The condition that
$\int{\de}^4x\;T\,\{\phi {\cal {O}} \phi)\}=0$
for a generic $\phi$ is equivalent to the condition
$\int{\de}^4x\;T\,{\mL}(x) = 0$ for a generic ${\mL}$.

Then, in the second step, one can construct the
Lagrangian ${\mL}(x)$ of the theory in terms of $\phi(x)$ and
of a differential operator ${\cal {O}}$ imposing $[{\cal {O}},T]=0$
(in our illustrative example ${\cal {O}} = \Box-M^2$).

This observation will prove useful as we later look for a
maximally-symmetric action in $\kappa$-Minkowski.
For the case of a free scalar field in commutative Minkowski spacetime
the choice  ${\cal {O}} = \Box-M^2$ is well established and
leads to a maximally-symmetric action.
The symmetries of this action are described in terms of
the classical Poincar\'{e} algebra $\mP$, generated
by the elements
\begin{displaymath}
P_\mu=-i\partial_\mu\quad M_j=-\epsilon_{jkl}x_kP_l\quad N_j=-x_jP_0+x_0P_j
\end{displaymath}
which satisfy the commutation relations
\bea
& [P_\mu,P_\nu]=0\qquad [M_j,P_0]=0\qquad \cM{P} & \nn\\
& \cM{M}\qquad\cM{N} & \nn\\
& [N_j,P_0]=iP_j \quad [N_j,P_k]=i\delta_{jk}P_0 \quad
[N_j,N_k]=-i\epsilon_{jkl}M_l & \label{poincare} \eea The operator
$\Box=P_\mu P^\mu$ is the first Casimir of the algebra, and of
course satisfies $[\Box,T]=0$.

For this case of a maximally-symmetric theory in commutative
Minkowski spacetime it is conventional to describe the symmetries
fully in terms of the Poincar\'{e} Lie algebra. For \kM\
noncommutative spacetime we shall argue that a description in
terms of a Hopf algebra is necessary. In order to do this step we
want to first show that even in the commutative-Minkowski case
there is an underlying Hopf-algebra structure, but the
commutativity of functions in Minkowski spacetimes implies that
the additional structures present in the Hopf algebra  are all
"trivial" (in a sense that we will show below). We will then
observe that the noncommutativity of functions in noncommutative
\kM\ spacetime leads to a nontrivial Hopf-algebra structure, which
cannot be faithfully captured in the simpler language of Lie
algebras.

Now we show how a symmetry Lie algebra for the commutative Minkowski spacetime can be promoted to
Hopf algebra.
In order to do this, let us introduce
a map $\epsilon:\mP\to \complex$, such that
\begin{equation}\label{eq:Usym}
\int{\de}^4x\;U\,{\mL}(x)=\,\epsilon(U)\int{\de}^4x\;{\mL}(x)
\end{equation}
for each element $U\in\mP$ and for each function ${\mL}(x)$. It is
straightforward to verify that $\epsilon(\mathtt{1})=1$ ( we
indicate with $\mathtt{1}$ the identity transformation) and
of course from (\ref{eq:sym}) it follows that $\epsilon(T)=0$ for each
generator $T$ of the algebra.
Moreover, from (\ref{eq:Usym}) it follows that
\begin{displaymath}
\int\!{\de}^4x\,U(U'{\mL})=\epsilon(U)\!\!\int\!{\de}^4x\,U'{\mL}=
\epsilon(U)\epsilon(U')\!\!\int\!{\de}^4x\,{\mL}
\end{displaymath}
But the first member is also equal to $\epsilon(UU')\!\int\!{\de}^4x\,{\mL}$,
and therefore $\epsilon$ is an algebra morphism
\begin{displaymath}
\epsilon(UU')=\epsilon(U)\epsilon(U')
\end{displaymath}
This recursive formula allows us to
calculate $\epsilon$ for a generic element of the algebra.
[Also note that the application of $U$ to a constant
function $f=1$ gives $U{\cdot}1=\epsilon(U)$.]

For each $U\in\mP$ we can also introduce
a map $\Delta\!:\mP\to\mP\otimes\mP$ such
that for every $f(x)$ and $g(x)$
\begin{equation}\label{eq:cop}
U(f{\cdot} g)= (U_{(1)}f)(U_{(2)}g)
\end{equation}
where $\Delta(U)=\sum_i U_{(1)}^i\otimes U_{(2)}^i$
is written simply as $U_{(1)}\otimes U_{(2)}$ (Sweedler notation).

 From the associativity of the product of functions it follows straightforwardly
that the map $\Delta$ is  coassociative, {\it i.e.} $ U_{(1)}\otimes\Delta
(U_{(2)})=\Delta (U_{(1)})\otimes U_{(2)}$.
From the property $(UU'){\cdot}(fg)=U(U'{\cdot} fg)$ it follows that $\Delta$
is an algebra morphism, {\it i.e.} $\Delta(UU')=\Delta(U)\Delta(U')$,
which allows to calculate $\Delta$ recursively.
And finally by considering (\ref{eq:cop}) for $g=1$ one
obtains $Uf=U(f{\cdot} 1)= (U_{(1)}f)(U_{(2)}1)= (U_{(1)}f)\epsilon(U_{(2)})$
from which it follows that $U_{(1)}\epsilon(U_{(2)})=U$ (and similarly
$\epsilon(U_{(1)})U_{(2)}=U$).

In this way, these two maps $\epsilon$ and $\Delta$,
defined by (\ref{eq:Usym}) and (\ref{eq:cop}),
verify all the requests (\ref{homo}\ref{coaxi1}\ref{coaxi2}), and we can identify
them with the counit and the coproduct maps.
So, they make a generic symmetry algebra into
a \emph{bialgebra}, as we have showed in subsection 1.2.1.

One can easily verify that $\Delta(1)=1\otimes 1$, and $\Da{T}$ for each
generator of Poincar\'{e} algebra. This last property, which plays a key role
in allowing a description of the symmetries at the simple Lie-algebra
level (without any true need to resort to a full Hopf-algebra description)
is actually connected with the commutativity of function in Minkowski spacetime.
In fact, from $f {\cdot} g = g {\cdot} f$, one easily finds that $\Delta$ is symmetric
\begin{displaymath}
 U_{(1)}\otimes U_{(2)}= U_{(2)}\otimes U_{(1)}
\end{displaymath}
for all $U$ or, adopting math gergon, $\Delta$ is
\emph{cocommutative}. We say that a cocommutative coproduct is, in
some sense, ``trivial'' in order to emphasize it simple structure
with respect to the coproduct of the symmetry generators of a
noncommutative spacetime, which is in general not cocommutative.

Defining $S(1)=1$, $S(T)=-T$ for each generator and
$S(UU')=S(U')S(U)$ we obtain a map
satisfying $U_{(1)}S(U_{(2)})= S(U_{(1)})U_{(2)}=\epsilon(U)$.
The map $S$ is just the antipode introduced in (\ref{antipode})
This makes $\mP$ a Hopf algebra (see subsection 1.2.1).

The universal enveloping algebra  of a Lie algebra is then
equivalent to a Hopf algebra generated by primitive elements $T$:
\be \Delta(T)=T\otimes 1+1\otimes
T,\;\;\;\epsilon(T)=0,\;\;\;S(T)=-T \label{trivial} \ee We
classify this algebra as a ``trivial Hopf algebra".

For theories in commutative spacetime
the symmetries can always be described in terms of a trivial
Hopf algebra. In contemplating theories in noncommutative spacetime
it is natural to insist on the requirement that the symmetries
be described by a Hopf algebra. The Lie-algebra description cannot
be maintained, since it would not provide a sufficient set of rules
to handle consistently the laws of symmetry transformation
of products of (noncommutative) functions.
The requirement that symmetries be described in terms of a Hopf
algebra actually is a simple statement: the action of symmetry
transformations on products
of functions should be consistent with the fact that such
products are themselves functions, and, accordingly,
the laws of transformation of products of functions should
still only require the appropriate action of the generators
of the (Hopf) algebra.

Once the algebra properties are specified (action of symmetry
transformations on functions of the noncommutative coordinates)
the properties of the counit, coproduct and antipode can always be
formally derived, but these will not in general satisfy the Hopf
algebra criteria since they may require the introduction of new
operators, not included in the algebra sector. If this does not
occur (if the counit, coproduct and antipode that one obtains on
the basis of the algebra sector can be expressed fully in terms of
operators in the algebra) the Hopf-algebra criteria are
automatically satisfied.

\section{Symmetries of a theory in \kM\ spacetime}
\subsection{General strategy}

By straightforward generalization of the result (\ref{twoparts})
reviewed in the previous section, we pose
that a transformation $T$ will be a symmetry if (and only if)
\be
\int{\de}^4{\x}\;\left(T{\cdot}\left\{\Phi\left(\Box_\lambda-M^2\right)\Phi\right\}+
\Phi[\Box_\lambda,T]\Phi\right)=0 \label{twopartsnc}~.
\ee
When several such symmetry generators are available they may or may
not combine together to form a Hopf algebra.
When they do form a Hopf algebra, denote it generically by $\mA$,
and we attribute to $\mA$ the role of symmetry (Hopf) algebra.

As mentioned in the preceding section, the search of a maximally-symmetric
action can be structured in two steps.
In the first step one looks for a Hopf algebra (in our case a Hopf algebra
which has the Poincar\'{e} algebra as
classical limit) whose generators $T$ satisfy
\begin{equation}\label{eq:kMsym}
\int{\de}^4{\x}\;T{\mL}({\x})=0
\end{equation}
for each element ${\mL}({\x})$ of \kM.
In the second step one looks for an operator $\Box_\lambda$
that is invariant ($[\Box_\lambda,T]=0$) under the action of this algebra.

\subsection{Translations}
In introducing the concept of translations we of course want
to follow as closely as possible the analogy with the well-established
concepts that apply in the commutative limit $\lambda \rightarrow 0$.
Since we have defined functions in \kM\ in terms of the Weyl maps,
and since the Weyl maps are fully specified once given on
Fourier exponentials (then the map on generic functions simply
requires the introduction of standard Fourier transforms),
we can, when convenient, confine the discussion to the Fourier
exponentials.
And an ambiguity, which is deeply connected with noncommutativity,
confronts us immediately:
in commutative Minkowski
the translation generator acts according to
\bea
P_{\mu}(e^{ikx}) &=& k_{\mu} e^{ikx}
\label{eq1}
\eea
but this action of the translation generators cannot
be implemented on general functions of the \kM\ coordinates.
The root of the ambiguity can be exposed
by just considering the same function of  \kM\ coordinates
written in two ways, the time-to-the-right form
and the time-symmetrized form.
Let us first write a specific function $\Phi$ of the \kM\ coordinates
in the way suggested by the time-to-the-right Weyl map:
\begin{eqnarray}
\Phi = \int\de^4p\;\tilde{\phi}(p)\Omega_R(e^{ipx})
~.
\label{eq2}
\end{eqnarray}
where $\tilde{\phi}(k)$ is the inverse Fourier transform of $\phi(x)$.
On the basis of Eqs.~(\ref{eq1}) and (\ref{eq2})
it would seem natural to define translations in \kM\ as
generated by the operators $P^R_{\mu}$ such that
\bea
P^R_{\mu} \Omega_R(e^{ikx}) = k_{\mu} \Omega_R(e^{ikx})
~.
\label{eq3}
\eea
But, as already stated through Eq.~(\ref{rightandsymm}),
the same function $\Phi$ written in Eq.~(\ref{eq2})
using time-to-the-right ordering can also be equivalently
expressed in time-symmetrized form as
\begin{eqnarray}
\Phi = \int\de^4p\;\tilde{\phi}(p_0,pe^{-\lambda p_0/2})
e^{-3\lambda p_0/2}\Omega_S(e^{ipx})
\label{eq4}
\end{eqnarray}
and on the basis of Eqs.~(\ref{eq1}) and (\ref{eq4})
it would seem natural to define translations in \kM\ as
generated by the operators $P^S_{\mu}$ such that
\bea
P^S_{\mu} \Omega_S(e^{ikx}) = k_{\mu} \Omega_S(e^{ikx})
~.
\label{eq5}
\eea

Let us notice that we had already encountered an ``ordering ambiguity"
in introducing a law of integration in \kM, but there we eventually
realized that there was no ambiguity after all (the two approaches
to the law of integration led to identical results).
The ordering ambiguity we are facing now in defining translations
is certainly more serious. In fact, the two candidates as
translation generators $P^S_{\mu}$ and $P^R_{\mu}$
are truly inequivalent, as one can easily
verify by applying $P^S_{\mu}$ and $P^R_{\mu}$ to
a few examples of functions in \kM; in particular, if we apply the operator $P_R$ on the time-to-the-right
ordered exponential, we find :
\begin{eqnarray}
P^R_{\mu} (e^{-i\vec{k}\vec{\x}}e^{ik_0{\x}_0})
 & \! = \! &
P^R_{\mu} \Omega_R(e^{ikx}) =
k_{\mu} \Omega_R(e^{ikx}) =
k_{\mu} (e^{-i\vec{k}\vec{\x}}e^{ik_0{\x}_0})=
k_{\mu} ( e^{ik_0{\x}_0/2}e^{
-ie^{\frac{\lambda}{2} k_0}\vec{k}\vec{\x}}e^{ik_0{\x}_0/2}), \nn
\eea
whereas, if we apply the operator $P_S$ on the same \kM\ element, we have
\bea
P^S_{\mu} (e^{-i\vec{k}\vec{\x}}e^{ik_0{\x}_0})&=& e^{\frac{\lambda}{2} k_0} k_{\mu}
( e^{ik_0{\x_0}/2}e^{-ie^{\frac{\lambda}{2} k_0}\vec{k}\vec{\x}}e^{ik_0{\x}_0/2})\neq P^R_{\mu} (e^{-i\vec{k}\vec{\x}}e^{ik_0{\x}_0})~.
\label{eq6}
\end{eqnarray}

It is also easy to verify that both $P^S_{\mu}$ and $P^R_{\mu}$
satisfy condition (\ref{eq:kMsym}):
\begin{equation}
\int{\de}^4{\x}\; P^{R,S}_{\mu} {\mL}({\x})=0
\end{equation}
Moreover the quadruplet of operators $P^S_{\mu}$ and
the quadruplet of operators $P^R_{\mu}$ do separately
give rise to genuine Hopf algebras of translation-like
symmetry transformations.
Since, as mentioned in section 2.5, the exponentials $e^{-i\vec{k}\vec{\x}}e^{ik_0\x_0}$
form a basis of \kM,
the coproduct of the $P^R_{j}$ operators, $\Delta P_j^R$,
is obtained consistently from observing that
\bea
&&P^R_j\Omega_R(e^{ikx})\Omega_R(e^{ipx})
=-i\Omega_R(\partial_je^{i(k\oplus_R p)x})\nn\\
&&=-i\Omega_R((k\oplus_R p)_je^{i(k\oplus_R p)x})\nn\\
&&=[P^R_j\Omega_R(e^{ikx})][\Omega_R(e^{ipx})]
+[e^{-\lambda P^R_0}\Omega_R(e^{ikx})][P^R_j\Omega_R(e^{ipx})]
~,
\label{calccoprod}
\eea
where $p \oplus_R q \equiv (p_0+q_0, p_1 + q_1 e^{-\lambda p_0},
 p_2 + q_2 e^{-\lambda p_0}, p_3 + q_3 e^{-\lambda p_0})$ has been defined in (\ref{oplusR}).
Introducing the map $\Delta$ as in (\ref{eq:cop}), this relation implies that:
\be
\Delta P^R_j=P_j^R\otimes 1
+ e^{-\lambda P_0^R}\otimes P^R_j\label{coprodpr}
\ee
Following an analogous procedure one can derive
\begin{eqnarray*}
& \Da{P_0^R} &
\end{eqnarray*}
and the full structure of a four-generator Hopf algebra emerges.

The same goes through, with equal success, for the $P^S_{\mu}$
alternative.
The coproducts naturally take a different form,
\bea
\Delta P^S_0&=&P^S_0\otimes 1+1\otimes P^S_0\nn\\
\Delta P^S_j&=&P_j^S\otimes e^{\frac{\lambda}{2}P_0^S}
+ e^{-\frac{\lambda}{2}P_0^S}\otimes P^S_j
\label{coprodps}
\eea
but the full structure of a four-generator Hopf algebra
is again straightforwardly obtained.

We must therefore live with this ambiguity. As we look
for an example of maximally-symmetric theory in \kM\
both the option $P^R_{\mu}$ and the option $P^S_{\mu}$ must be
considered, {\it i.e.}
we can look for 10-generator extensions of either.

\subsection{Rotations}
Following the same idea that allows us to introduce translations
in \kM , we attempt now to obtain a 7-generator Hopf algebra,
describing four translation-like operators and three rotation-like
generators. Since, as just mentioned, our analysis of translations
alone led us to two alternatives, we are in principle prepared for
at least two alternative versions of the 7-generator Hopf-algebra
for translations and rotations: ($P^R,M^R$) and ($P^S,M^S$), with
$M^R$ and $M^S$ to be determined.

For what concerns the translations we have found that an
acceptable Hopf-algebra description was obtained by
straightforward ``quantization"
of the classical translations:
the $P^R_{\mu}$ translations where just obtained from the
commutative-spacetime translations through the $\Omega_R$
Weyl map and the $P^S_{\mu}$
translations where just obtained from the
commutative-spacetime translations through the $\Omega_S$
Weyl map.

It is natural to explore the possibility of describing also the
rotations by straightforward ``quantization"
\begin{eqnarray}
 && M^R_j{\cdot}\Omega_R(f) = \Omega_R (M_j f)
 =\Omega_R(i\epsilon_{jkl}x_k\partial_l f) \label{rotar}\\
 && M^S_j{\cdot}\Omega_S(f) = \Omega_S (M_j f)
 =\Omega_S(i\epsilon_{jkl}x_k\partial_l f)
 \label{rotas}
\end{eqnarray}
where $\partial_\mu=\partial/\partial x^\mu$,  $\partial_0=\partial^0$ and $\partial_j=-\partial^j$.
Again in defining these rotation generators we used the fact that any element of \kM\ (any ``function
in \kM\ spacetime") can be obtained through the
action of a Weyl map from some commutative
function $f(x)$.

This is actually another instance in which an ordering ambiguity
is only apparent. In fact, $M_j^S$ and $M_j^R$ act exactly in
the same way, as one can verify through the observation that
\bea
M_j^S\Omega_R(e^{ikx})&=&M_j^S\Omega_S(e^{ik'x})_{k'=(k_0,k_ie^{\lambda k_0/2})}\nn\\
=i\epsilon_{jnl}\Omega_S(x_n\partial_le^{ik'x})&=&-\epsilon_{jnl}\Omega_S(x_ne^{\lambda k_0/2}k_le^{ik'x})\nn\\
=i\epsilon_{jnl}k_l\Omega_S(\partial_{k^n}e^{ik'x})
&=&i\epsilon_{jnl}k_l\partial_{k^n}\Omega_S(e^{ik'x})\nn\\
=i\epsilon_{jnl}k_l\partial_{k^n}\Omega_R(e^{ikx})
&=&-\epsilon_{jnl}k_l\Omega_R(-i\partial_{k^n}e^{ikx})\nn\\
=i\epsilon_{jnl}\Omega_R(x_n\partial_le^{ikx})&=&M_j^R\Omega_R(e^{ikx})\nn
\eea We will therefore remove the indices $R$/$S$ on the rotation
generator, and denote it simply by $M_j$.

One easily verifies that the operators $M_j$, as defined by (\ref{rotar})
(and equivalently defined by (\ref{rotas})),
are good candidates for
the construction of (Hopf) algebras of
symmetries of a theory in \kM\ spacetime; in fact:
\begin{equation}
\int{\de}^4{\x}\; M_j {\mL}({\x})=0 ~.
\end{equation}

It is also straightforward to verify that
\bea
[M_j,M_k]=i\varepsilon_{jkl} M_l
\label{rotalgebra}
\eea
and, following a line of analysis analogous to the one
of Eq.~(\ref{calccoprod}),
that a trivial coproduct must be adopted for
these candidate rotation operators:
\bea
\Delta M_j=M_j\otimes 1+1\otimes M_j
\label{rotcoprod}
\eea
Therefore the triplet $M_j$
forms a 3-generator Hopf algebra
that is completely undeformed (classical) both in the algebra and
in the coalgebra sectors. (Using the intuitive description introduced
earlier this is a trivial rotation Hopf algebras, whose structure could
be equally well  captured by the standard Lie algebra of rotations.)

There is therefore a difference between the translations sector
and the rotations sector. Both translations and rotations can
be realized as straightforward (up to ordering) quantization
of their classical actions, but while for rotations even the
coalgebraic properties are classical (trivial coalgebra)
for the translations we found
a nontrivial coalgebra sector (\ref{coprodpr})
(or alternatively (\ref{coprodps})).

Up to this point we have two candidates, $P^R_\mu$ and $P^S_\mu$,
for a  4-generator Hopf algebra of translation symmetries and a
candidate, $M_j$, for a 3-generator Hopf algebra of rotation
symmetries. These can actually be put together straightforwardly
to obtain two candidates for 7-generator translations-rotations
symmetries ($P^R_\mu,M_j$) and ($P^S_\mu,M_j$). It is sufficient
to observe that \bea [M_j,P^{R,S}_{\mu}]{\cdot}
\Omega(e^{ikx})&=&\varepsilon_{jkl}\Omega([x_k\partial_{\mu}
-\partial_{\mu}x_k]\partial_l e^{ikx})\nn\\
&=&-\delta_{\mu k}\varepsilon_{jkl}\Omega(\partial_le^{ikx})
\eea
from which it follows that
\be
[M_i,P^{R,S}_j]=i\varepsilon_{ijk}P^{R,S}_k,\;\;
[M_i,P^{R,S}_0]=0
\ee
{\it i.e.} the action of rotations on energy-momentum
is undeformed. Accordingly, the generators $M_j$
can be represented as differential operators over
energy-momentum space in the familiar
way, $M_j=i\varepsilon_{jkl}P_k\partial_{P_l}$.
[This also provides another opportunity
to verify (\ref{rotalgebra}).]

\subsection{Boosts}
In the analysis of translations and boosts
we have already encountered two different situations: rotations
in \kM\ spacetime are essentially classical in all respects,
while translations
have a ``classical" action
(straightforward $\Omega$-map ``quantization" of the
corresponding classical action)
but have nontrivial coalgebraic properties.
As we now intend to include also boosts, and obtain 10-generator
symmetry algebras, we encounter another possibility:
for boosts not only the coalgebra sector is nontrivial but
even the action on functions in \kM\ cannot be obtained
by ``quantization" of the classical action.

Because of its profound implications for our analysis,
we devote this subsection to the analysis of classical boosts,
showing that they do not lead to acceptable symmetries.
For simplicity we only look for
boosts $N_j^R$
that could combine with the translations $P^R_\mu$ and rotations $M_j$
to give us a 10-generator symmetry algebra.
The assumption that the action of these boosts could be ``classical"
means
\bea
N_j^R\Omega_R(f)=\Omega_R(N_jf)
=\Omega_R(i[-x_0\partial_j+x_j\partial_0]f])
~.
\eea
It is easy to see that these boosts combine with the rotations $M_j$ to close  a healthy
undeformed Lorentz algebra, and that adding also the
translations $P_\mu^R$ one obtains
the undeformed Poincar\'{e} algebra (\ref{poincare}).
In fact, for example:
\bea
[N_j,P_k]\Omega_R(e^{ikx})&=&\Omega_R([N_jk_k+i\partial_kN_j]e^{ikx})=\Omega_R((\delta_{jk}\partial_0])]e^{ikx})=\delta_{jk}\partial_0\Omega_R(e^{ikx})\nn\\
&=&iP_0\Omega_R(e^{ikx})
\eea
For this algebra we have introduced the coproduct of $M_j$, that turns out to be not deformed, and the coproduct of $P_\mu$, that instead turns out to be deformed.
Now we attempt to find a coproduct for $N_j$ that close the Hopf algebra structure of $(P_\mu,M_j,N_j)$ in a compatible way.
However, these algebras cannot be extended (by introducing a suitable coalgebra sector) to obtain a full Hopf algebra of symmetries
of theories in our noncommutative \kM\ spacetime.
In fact, we find an inconsistency in the coproduct of
these boosts $N_j^R$, which signals an obstruction originating
from an inadequacy in the description of the action
of boosts on (noncommutative) products of \kM\ functions.

In order to work out the coproduct of the boosts $N_j^R$, we will use the following relations.
\bea
\Omega_R(e^{ipx})\x_0&=&\x_0\Omega_R(e^{ipx})-\lambda p \x\Omega_R(e^{ikx}),\label{eq:uno}\\
x_j\Omega_R(e^{ikx})&=&e^{\lambda k_0}\Omega_R(e^{ikx})x_j\label{eq:due}
\eea
They will prove useful also in the next sections.
The first relation can be easily computed in $1+1$--dimensions as follows
\bea
[\Omega_R(e^{ipx}),\x_0]&=&[e^{-ip\x}e^{ip_0\x_0},\x_0]=[e^{-ip\x},\x_0]e^{ip_0\x_0}=\sum_{n}\frac{(-i)^n}{n!}p^n[\x^n,\x_0]e^{-ip_0\x_0}\nn\\
&=&-i\lambda\sum_{n}\frac{(-i)^n}{(n-1)!}p^n\x^{n}e^{ik_0\x_0}=-\lambda p\x\sum_n\frac{(-i)^n}{n!}p^nx^ne^{ik_0\x_0}\nn\\
&=&-\lambda p\x e^{-ip\x}e^{ip_0\x_0}=-\lambda p \x\Omega_R(e^{ikx})
\eea
While, in order to prove the second relation we write:
\be
\x_j\Omega_R(e^{ikx})=\x_je^{-ik\x}e^{ik_0\x_0}=e^{-ik\x}\x_je^{ik_0\x_0}
\ee
Making a series expansion of the exponential in $\x_0$, we have:
\be
\x_je^{ik_0\x_0}=\sum_n\frac{i^n}{n!}k_0^n \x_j\x_0^n\label{gio'}
\ee
We observe that
\be
\x_j\x_0^n=[\x_j,\x_0]\x_0^{n-1}+\x_0\x_j\x_0^{n-1}=(-i\lambda+\x_0)\x_j\x_0^{n-1}
\ee
Defining $S_n\equiv \x_j\x_0^{n}$, we have the recursive formula:
\be
S_n=(-i\lambda+\x_0)S_{n-1}
\ee
from which follows that $S_n=(-i\lambda +\x_0)^n\x_j$. Thus the (\ref{gio'}) is:
\bea
\x_je^{ik_0\x_0}&=&\sum_n\frac{i^n}{n!}k_0^n(-i\lambda +\x_0)^n\x_j=\sum_n\frac{i^n}{n!}k_0^n\sum_l\frac{n!}{(n-l)!l!}(-i\lambda)^{n-l}\x_0^l\x_j\nn\\
&=&\sum_l\frac{i^l}{l!}(k_0\x_0)^l\sum_n\frac{1}{(n-l)!}k_0^{n-l}(\lambda)^{n-l}\x_j=e^{\lambda k_0}e^{ik_0\x_0}\x_j
\eea
and the (\ref{eq:due}) it easy proved.

The coproduct of $N_j$ can be inferred from observing that:
\bea
N_j^R \Omega_R(e^{ikx})\Omega_R(e^{ipx})&=&\Omega_R(N_je^{i(k\oplus_R p)x})=\Omega_R[(x_0(k\oplus_R p)_j-x_j(k+p)_0)e^{i(k\oplus_R p)x}]\nn\\
&=&k_j\Omega_R(e^{ikx})\Omega_R(e^{ipx}){\x}_0-k_0{\x}_j\Omega_R(e^{i(k\oplus_R p)x})\nn\\
&&+e^{-\lambda k_0}p_j\Omega_R(e^{i(k\oplus_R
p)x}){\x}_0-p_0{\x}_je^{\lambda
k_0}\Omega_R(e^{ikx})\Omega_R(e^{ipx}) \eea This expression can be
rewritten using the  relations (\ref{eq:uno}) and (\ref{eq:due})
that we have found above: \bea
N_j^R \Omega_R(e^{ikx})\Omega_R(e^{ipx})&&=k_j\Omega_R(e^{ikx}){\x}_0\Omega_R(e^{ipx})-k_0{\x}_j\Omega_R(e^{i(k\oplus_R p)x})\nn\\
&&-\lambda k_jp_l\Omega_R(e^{ikx}){\x}_l\Omega_R(e^{ipx}) \nn\\
&&+e^{-\lambda k_0}p_j\Omega_R(e^{i(k\oplus_R p)x}){\x}_0-p_0e^{\lambda k_0}\Omega_R(e^{ikx}){\x}_j\Omega_R(e^{ipx})\nn\\
&&=\Omega_R([k_jx_0-k_0x_j]e^{ikx})\Omega_R(e^{ipx})\nn\\
&&+\Omega_R(e^{-\lambda k_0}e^{ikx})\Omega_R([p_jx_0-p_0x_j]e^{ipx})\nn\\
&&-\lambda\Omega_R(k_je^{ikx})\Omega_R(p_lx_le^{ipx})+2\Omega_R(\sinh(\lambda k_0)e^{ikx})\Omega_R(p_0x_je^{ipx}).\nn
\eea
From this one concludes that the coproduct is
\begin{eqnarray}
\Delta(N_j^R) &=& N_j^R\otimes 1+e^{-\lambda P_0^R}\otimes N_j^R+\nn\\ &&
+2\sinh\lambda P_0^R \otimes{\x}_jP_0^R-\lambda P_j^R\otimes\vec{\x}\vec{P}_R
\label{oldboostcoprod}
\end{eqnarray}
The problem is that $\Delta(N_j^R)$ is not an element of the algebraic tensor
product, {\it i.e.} it is not a function only of the elements $M,N,P$.
It would
be necessary to introduce new elements to close the Hopf algebra; for example,
the well known \emph{dilatation operator} ${\cal D}=\vec{\x}\vec{P}_R$, which
is independent from the other generators.
Therefore, as anticipated, the ``classical" choice $N_j^R$ cannot be combined
with $M^R$ and $P^R_\mu$ to obtain a 10-generator symmetry algebra.
From (\ref{oldboostcoprod}) one might imagine that perhaps an 11-generator
symmetry algebra, including the dilatation operator ${\cal D}$,
could be contemplated but this is also not possible since one can
easily verify that ${\cal D}$ does not satisfy condition (\ref{eq:kMsym}),
\begin{equation}
\int{\de}^4{\x}\;  {\cal D} {\mL}({\x}) \neq 0 ~.
\end{equation}
${\cal D}$ cannot be a symmetry of a scalar action in \kM.

Thus the ``classical" choice $N_j^R$ is inadequate.
One can easily verify that all these points that we have
 made about the ``classical" $N_j^R$ and the ($P^R_\mu$,$M_j$,$N^R_j$)
algebra also apply to the ``classical" $N_j^S$
and the ($P^S_\mu$,$M_j$,$N^S_j$) algebra.

Now we show that we are able to construct a 10-generator
symmetry-algebra extending the 7-generator symmetry algebra
($P^R_\mu$,$M_j$) but this requires nonclassical boosts. We
indicate such a deformed boost generators with the symbol
${\mN}_j^R$.

On the basis of the problem encountered above
it is clear that our key task is to find
an operator ${\mN}_j^R$ such that $\Delta({\mN}_j^R)$ is
an element of $\mA\otimes\mA$, where $\mA$ is the algebra generated
by $(P^R_\mu,M_j,{\mN}_j^R)$.

We look for a symmetry algebra that preserves the Lorentz algebra
as a sub-algebra. This means that the commutators among
$M_j,\mN_j$ must be expressed only in terms of $M_j$ and $\mN_j$,
without the presence of the $P_\mu$ operators. On the basis of
dimensional considerations one can see that this algebra remain
undeformed, \emph{i.e.} $\lambda$-independent. In fact $M_j$ and
$\mN_j$ are dimensionless, and the presence of $\lambda$ in the
commutation relations of the Lorentz sector would involve the
presence of $P_\mu$ as well (since $\lambda$ has the dimension of
a length and the product $\lambda P_\mu$ represents the only
possibility to introduce a deformation in a dimensionless way).
So, in order to deform the Lorentz-sector commutation relations
one should renounce to the property that the Lorentz-sector
generators form a subalgebra. On the other hand we require that in
the commutative limit $\lambda=0$ the symmetry algebra is given by
the classical Poincar\'e algebra (\ref{poincare}), that represent
the symmetry algebra of the Minkowski commutative spacetime.
Therefore, the subalgebra request and the commutative limit fix
the Lorentz sector to remain classical, exactly as in the
commutative Minkowski spacetime case.



Let us start by observing that by imposing that the deformed boost
generator ${\mathcal{N}}_j$ transforms as a vector under rotation
(see Appendix~B), one finds that the most general form for
${\mN}_j$ is: \bea {\mathcal{N}}_j \Omega(\phi(x))
=\Omega\{[ix_0A(-i\partial_x)\partial_j
+\lambda^{-1}x_jB(-i\partial_x) -\lambda x_lC(-i\partial_x)
\partial_{l} \partial_{j} -i\epsilon_{jkl} x_k
D(-i\partial_x)\partial_{l}]\phi(x)\} \label{boost} \eea where
$A,B,C,D$ are unknown functions of $P^R_\mu$ (in the classical
limit $A=i,D=0$; moreover, as $\lambda \rightarrow 0$ one obtains
the classical limit if $\lambda C \rightarrow 0$ and $B
\rightarrow \lambda P_0$).

Imposing \be {\mN}^R_j[\Omega(e^{ikx})\Omega(e^{ipx})]
=[{\mN}^R_{(1),j}\Omega(e^{ikx})][{\mN}^R_{(2),j}\Omega(e^{ikp})]
\ee one clearly obtains some constraints on the functions
$A,B,C,D$ that must be satisfied in order to obtain the desired
result that the operator ${\mN}_j^R$ is such that
$\Delta({\mN}_j^R)$ is an element of $\mA\otimes\mA$. These
constraints are discussed and analyzed in the Appendix B. The
final result is \be {\mN}_j^R\Omega_R(f)=\Omega_R(-[ix_0\partial_j
+x_j(\frac{1-e^{2i\lambda
\partial_0}}{2\lambda}-\frac{\lambda}{2}\nabla^2) +\lambda
x_l\partial_l\partial_j]f)\label{boostRfin} \ee

It is easy to verify that the Hopf algebra
$(P^R_\mu,M_j,{\mN}_j^R)$ satisfies all the requirements for a
candidate symmetry-algebra for theories in \kM\ spacetime. In
particular from the differential form (\ref{boostRfin}) it is easy
to verify  the following deformed cross relations among the boost
and the translation generators: \bea
{}[\mN_j,P_0]=iP_j^R,&&[\mN_j^R,P_k^R]=i\delta_{jk}\ts\frac{1-e^{-2\lambda
P_0}}{2\lambda}+\frac{\lambda}{2}(P^{R})^2\td-i\lambda P_j^RP_k^R
\eea The coproduct of $\mN_j$ has been obtained in Appendix B and
corresponds to: \be \Delta(\mN_j)^R=\mathcal{N}_j^R\otimes
1+e^{-\lambda P_0}\otimes
\mathcal{N}_j^R+\lambda\varepsilon_{jkl}P_k^R\otimes M_l. \ee Thus
the "right" basis $(P^R_\mu,M_j,{\mN}_j^R)$ is just the
Majid-Ruegg bicrossproduct basis of \kkP , that we have discussed
in section 1.3.

Analogously on finds that the deformed boost generators $\mN_S$
are: \be {\mN}_j^S\Omega_S(f)=\Omega_S(-[ix_0\partial_j
+x_j(\frac{\sinh(i\lambda\partial_0)}{\lambda}+\frac{\lambda}{2}\nabla^2)
+\frac{\lambda}{2}x_l\partial_l\partial_j]e^{i\frac{\lambda}{2}\partial_0}f)
\label{boostSfin2} \ee combine with the translations $P^S_\mu$ and
the rotations $M_j$ to give a genuine symmetry Hopf algebra
$(P^S_\mu,M_j,{\mN}_j^S)$.

Although introduced following different formulations (respectively
the action on right-ordered functions and the action on
symmetrically-ordered functions) ${\mN}_j^R$ and ${\mN}_j^S$ are
actually identical. This is easily seen by observing that \be
{\mN}_j^S\Omega_R(f)=\Omega_R(-[ix_0\partial_j
+x_j(\frac{1-e^{2i\lambda \partial_0}}{2\lambda}
-\frac{\lambda}{2}\nabla^2)+\lambda x_l\partial_l\partial_j]f)
\label{boostSfin1} \ee and comparing with (\ref{boostRfin}). We
therefore remove the indices $R$/$S$ and use the unified notation
${\mN}_j$.

In summary we have two bases candidate for the Hopf algebra of
 10-generator Poincar\'{e}-like symmetries for \kM: $(P^R_\mu,M_j,{\mN}_j)$
 and $(P^S_\mu,M_j,{\mN}_j)$.
The $(P^S_\mu,M_j,{\mN}_j)$ is a bicrossproduct basis of \kkP\
that has not been introduced in literature before; we will
illustrate it better in the following in order to discuss its
properties.

\subsection{An action for the free scalar field in \kM}
We have completed what we had announced as ``step 1" of our analysis:
we have constructed explicitly
candidates (actually not one but two candidates)
of 10-generator symmetry Hopf algebras for theories in \kM\ spacetime,
whose generators, generically denoted by $T$, satisfy
\begin{equation}\label{eq:kMsymbis}
\int{\de}^4{\x}\;T{\mL}({\x})=0
\end{equation}
for every element ${\mL}({\x})$ of \kM\ ({\it i.e.} for every
function ${\mL}({\x})$ of the \kM\ coordinates).

We are finally ready for the second step: introducing a
differential operator $\Box_\lambda$, whose classical $\lambda
\rightarrow 0$ limit is the familiar D'Alembert operator, and such
that the action \be
S(\Phi)=\int{\de}^4{\x}\;\Phi(\Box_\lambda+M^2)\Phi
\label{actionfinal} \ee is invariant under one of the realizations
of  Hopf-algebra symmetry we have constructed
$(P^R_\mu,M_j,{\mN}_j)$ and $(P^S_\mu,M_j,{\mN}_j)$. We therefore
must verify that, for some choice of $\Box_\lambda$,
$[\Box_\lambda,T]=0$ for every $T$ in the Hopf algebra.

For the $(P^R_\mu,M_j,{\mN}_j)$ case, guided by the intuition that
$\Box_\lambda$ should be a scalar with respect to
$(P^R_\mu,M_j,{\mN}_j)$ transformations, one is led to the
proposal \be \Box_\lambda=-\frac{2}{\lambda^2}\left(\cosh(\lambda
P_0^R)-1\right)+ e^{\lambda P_0^2}P_R^2\label{boxfin} \ee

In fact, it is easy to verify that with this choice of
$\Box_\lambda$ the action (\ref{actionfinal}) is invariant under
the $(P^R_\mu,M_j,{\mN}_j)$ transformations. Therefore, we have
finally managed to construct an action describing free scalar
fields in \kM\ that enjoys 10-generator (Hopf-algebra) symmetries
$(P^R_\mu,M_j,{\mN}_j)$.

This same choice of differential operator $\Box_\lambda$ is also
acceptable from the perspective of the alternative
$(P^S_\mu,M_j,{\mN}_j)$ symmetries. It turns out that the
$\Box_\lambda$ given in (\ref{boxfin}) is also a scalar with
respect to the $(P^S_\mu,M_j,{\mN}_j)$ transformations
($[\Box_\lambda,T]=0$ for every $T$ in $(P^S_\mu,M_j,{\mN}_j)$)
and the action (\ref{actionfinal}) with $\Box_\lambda$ given in
(\ref{boxfin}) is invariant also under the $(P^S_\mu,M_j,{\mN}_j)$
transformations.

\section{Aside on the new \kkP\ basis that emerged from
the analysis}

In the analysis reported in the previous sections, in order to
contributing to the understanding the symmetries of physical
theories in \kM , we have stumbled upon an interesting new basis
of the so-called \kkP\ Hopf algebras.

We have recognized that the Hopf algebra $(P^R_\mu,M_j,N_j)$, to
which we were led by our time-to-the-right ordering convention,
turns out to be a well known example of \kkP\ Hopf algebra, the
Majid-Ruegg algebra first discussed in
Refs.~\cite{MajidRuegg,lukieAnnPhys}. Instead the
$(P^S_\mu,M_j,N_j)$ algebra, which we ended up considering
starting from the time-symmetrized ordering convention, is
actually a new example of \kkP\ Hopf algebra, which had not
previously emerged in the literature.

We want to note here some of the key characteristics of our
$(P^S_\mu,M_j,N_j)$ Hopf algebra. The algebra sector of
$(P^S_\mu,M_j,N_j)$ is characterized by commutation relations \bea
\qs P_{\mu}^S,P_{\nu}^S\qd&=&0\nn\\
\qs M_j,M_k\qd&=&i\varepsilon_{jkl}M_l\;\;\qs {\mN}_j,M_k\qd=
i\varepsilon_{jkl}{\mN}_l \;\;\qs {\mN}_j,{\mN}_k\qd=-i\varepsilon_{jkl}M_l\nn\\
\qs M_j,P^S_0\qd&=&0\;\;\;[M_j,P_k^S]=i\epsilon_{jkl}P_l^S\nn\\
\qs {\mN}_j,P^S_0\qd&=&ie^{-\frac{\lambda}{2}P^S_0}P_i^S\nn\\
\qs {\mN}_j,P_k^S\qd&=&i e^{-\frac{\lambda}{2}P^S_0}\qs\ts
\frac{\sin(\lambda P^S_0)}{\lambda}
+\frac{\lambda}{2}\vec{P^S}^2\td\delta_{jk}-\frac{\lambda}{2}P_j^SP_k^S\qd\nn
~, \eea with ``mass" Casimir \be C(P^S)=\cosh(\lambda
P^S_0)-\frac{\lambda^2}{2}\vec{P^S}^2 ~, \ee while the co-algebra
sector is characterized by \bea \Delta(P^S_0)&=&P^S_0\otimes
1+1\otimes P^S_0\;\;\;\Delta(P_i^S) =P_i^S\otimes
e^{\frac{\lambda}{2}P^S_0}
+e^{-\frac{\lambda}{2}P^S_0}\otimes P^S_j\nn\\
\Delta(M_i)&=&M_i\otimes 1+1\otimes M_i\nn\\
\Delta({\mN}_j)&=&{\mN}_j\otimes 1+e^{-\lambda P^S_0}\otimes
{\mN}_j
+\lambda\epsilon_{jkl}e^{-\frac{\lambda}{2}P^S_0}P_j\otimes M_k
\eea

Just like the Majid-Ruegg $(P^R_\mu,M_j,N_j)$ algebra, also
$(P^S_\mu,M_j,N_j)$ is a bicrossproduct Hopf
algebra~\cite{MajidRuegg,lukieAnnPhys}. Since the  Majid-Ruegg
$(P^R_\mu,M_j,N_j)$ algebra is often referred to as the
``bicrossproduct basis", it may be natural to call ``type-2
bicrossproduct basis" our $(P^S_\mu,M_j,N_j)$ Hopf algebra.

One recognizes that $\Delta(P^S_0)$, $\Delta(P_i^S)$ and $C(P^S)$
for the Hopf algebra $(P^S_\mu,M_j,N_j)$ are identical to the ones
of the so-called \kkP\ ``standard basis" Hopf algebra~\cite{LNRT},
introduced in section 1.3. However, other characteristics of
$(P^S_\mu,M_j,N_j)$ are different from the ones of the ``standard
basis", which in particular does not leave the Lorentz sector
algebra undeformed and then a sub-algebra.

\section{Reducing the ambiguity to the choice of
translation generators} The action (\ref{actionfinal}) with
$\Box_\lambda$ given in (\ref{boxfin}) is invariant both under
$(P^R_\mu,M_j,{\mN}_j)$ transformations and under
$(P^S_\mu,M_j,{\mN}_j)$ transformations. It is of course not
proper to state that it is invariant under 14-generator
$(P^R_\mu,P^S_\mu,M_j,{\mN}_j)$ transformations (this 14-generator
algebra is not a Hopf algebra and its generators are not truly
independent), but rather we have an ambiguity in the description
of the symmetries of the action we have constructed for a free
scalar field in \kM. As discussed we are here formulating the
ambiguity in terms of two realizations of the Hopf-algebra
symmetry, $(P^R_\mu,M_j,N_j)$ and $(P^S_\mu,M_j,N_j)$, which are
profoundly connected with two different choices of ordering in
\kM\ (in the sense codified in the Weyl maps $\Omega_R$ and
$\Omega$). We anticipate that the ambiguity should actually be
between more than two options: if one considered other types of
ordering conventions (ignoring the fact that these other
alternatives might not be ``natural") for functions in \kM\ one
would find other candidates $P^*_\mu$ as translation generators
and, correspondingly, other candidate bases for the 10-generator
Hopf algebra of Poincar\'{e}-like symmetries for \kM\ of the type
$(P^*_\mu,M_j,{\mN}_j)$.

 One could consider assuming that the two bases
  $(P^R_\mu,M_j,N_j)$ and $(P^S_\mu,M_j,N_j)$ are actually
equivalent descriptions of the symmetries of the theory, but this
would only postpone the question at the level of describing
energy-momentum in our theory. We are used to associate
energy-momentum with the translation generators and it is not
conceivable that a given operative definition of energy-momentum
could be equivalently described in the context of the
$(P^R_\mu,M_j,N_j)$ and $(P^S_\mu,M_j,N_j)$ Hopf algebra bases. If
the properties of $P^R_\mu$ describe the properties of a certain
operative definition of energy-momentum then for that same
operative definition of energy-momentum a description in terms of
$P^S_\mu$ will turn out to be unacceptable. The difference would
be easily established by using, for example, the different
dispersion relations that $P^R_\mu$ and $P^S_\mu$ satisfy. In fact
the dispersion relation for a free scalar particle in \kM\ is
suggested by (\ref{boxfin}) to be\footnote{This suggestion will be
proved in Chapter 4 in which we will obtain the equation of motion
for a free scalar particle through an action principle}: \be
\cosh(\lambda P_0^R)-\frac{\lambda^2}{2} e^{\lambda
P_0^2}P_R^2=\cosh(\lambda m) \label{righttran}\ee in terms of the
$P_R$ generators, but in terms of the $P_S$ generators it takes
the form:
$$\cosh(\lambda
P_0^S)-\frac{\lambda^2}{2}P_S^2=\cosh(\lambda m),
$$
where $m$ is the mass at rest of the particle. The two relations
are clearly different in terms of the different translation
generators, so the ambiguity concerning translations leads to an
ambiguity concerning the dispersion relations for particles in \kM
. We have illustrated in Chapter˜1 that the dispersion relations
have a meaningful physical property, and could, in particular,
have observable consequences in
astrophysics~\cite{gacEMNS97,gactp,gacgianfran} and in
cosmology~\cite{jurekCOSMO,maguCosmoDispRel}). Thus we expect that
only one dispersion relation is observable and only one choice of
translation generators is the physically correct choice. However
our characterization of the symmetries does not allow to select a
preferred choice of translations.

We want to notice that the choice of a time-to-the-left ordering
formalism (\ref{timeleft}), leading to the introduction of the
$P_L$ translation generators: \be
P^L_{\mu}\Omega_L(e^{ikx})=k^\mu\Omega_L(e^{ikx})\ee would suggest
the following dispersion relation:
$$ \cosh(\lambda P_0^L)-\frac{\lambda^2}{2} e^{-\lambda
P_0^2}P_L^2=\cosh(\lambda m)
$$
that can be obtained from the (\ref{righttran}) simply by
substituting $\lambda$ with $-\lambda$, as mentioned above.

While clearly our analysis leaves this question unanswered, we
have reduced and clarified the ambiguity that one finds in
previous \kM\ literature concerning the description of symmetries.
As mentioned in the introduction of this chapter, in the
literature there has been discussion (see, {\it e.g.},
Refs.~\cite{lukieAnnPhys,Kow02-NST}) of a large class of Hopf
algebra bases as candidates for the description of the symmetries
of \kM.~These candidates were identified using a
Hopf-algebra-duality criterion. The commutation relations that
characterize these different Hopf algebra bases fulfilling the
duality condition take a very different form in the different
algebras, just like very different commutation relations
characterize our algebras $(P^R_\mu,M_j,N_j)$ and
$(P^S_\mu,M_j,N_j)$. In the previous literature the differences
between the algebras were perceived in very general terms, so much
so that, in terms of our notation, our two algebras would have
been distinguished as $(P^R_\mu,M^R_j,N^R_j)$ and
$(P^S_\mu,M^S_j,N^S_j)$. Instead our two algebra bases involve the
same generators for rotations and boosts. These generators have
different commutation rules with translations, but only because
the translations $P^R_\mu$ are genuinely different from the
translations $P^S_\mu$. When acting on the same entity (in
particular when acting on functions of the \kM\ coordinates) the
generators for rotations and boosts of the two bases act in
exactly the same way, they are identical operators.

We have therefore clarified that the difference between
alternative bases for the 10-generator algebra of symmetries
merely reflect a different concept of translations.


\section{More on the description of translations}
The results reported in the previous Sections
were based on the concept of symmetry codified
in (\ref{twopartsnc}). This is a natural symmetry requirement, which however
deserves a few more comments, which will also lead us to a noteworthy
characterization of translations.
Our discussion of this point will be more easily followed as we consider
a specific action. Let us focus on the simple example
\be
S(\phi)=\int{\de}^4{\x}\;\phi^2 ~.
\ee
It is natural to describe infinitesimal translations in terms
of
\bea
{\x}&\rightarrow&{\x}'={\x} - \alpha \e\nn\\
\phi({\x})&\rightarrow& \phi'({\x})
=\phi({\x})+i\alpha T\phi({\x})+O(\alpha^2)\nn
\eea
where $T=-i{\e}^{\mu}\partial_{\mu}$ is the
generator of translations and $\alpha\in R$ is an expansion parameter.

On the basis of an analogy with corresponding analyses in commutative
spacetimes there are actually two possible starting points
for a description of $T$ as a symmetry of the action:
\be
\delta_{I} S(\phi)= i \int{\de}^4{\x}\;T{\cdot}\phi^2 = 0 \label{sostanza}
\ee
and
\be
\delta_{II} S(\phi)=S(\phi')-S(\phi) = 0 \label{forma}
\ee
In the context of theories in commutative spacetimes the conditions
(\ref{sostanza}) and (\ref{forma})
are easily shown to be equivalent.
But in a noncommutative space-time this is not necessarily the case.
Let us start by considering the simplest possibility, the case
of commutative transformation parameters $\epsilon$. In this
case by expanding (\ref{sostanza}) and (\ref{forma}) to
first order one finds
\begin{eqnarray*}
\delta_{II} S(\phi) =
S(\phi+i\alpha\e P\phi)-S(\phi) &=& i\alpha\int{\de}^4{\x}\Big\{({\e}
P\phi)\phi+\phi({\e} P\phi)+i\alpha({\e} P\phi)({\e} P\phi)\Big\}=
\\ &=& i\alpha\int{\de}^4{\x}\Big\{({\e} P\phi)\phi+\phi({\e}
P\phi)\Big\}+o(\alpha) \\
\delta_{I} S(\phi)= \int{\de}^4{\x}i\alpha{\e} P\phi^2 &=&
i\alpha\int{\de}^4{\x}
\Big\{{\e}_\mu(P^\mu\phi)\phi+{\e}_0\phi(P^0\phi)+{\e}_j(e^{-\lambda
P_0}\phi)(P^j\phi)\Big\} ~,
\end{eqnarray*}
which clearly indicates that the condition $\delta_{I} S(\phi)=0$
does not imply $\delta_{II} S(\phi)=0$ and {\it vice versa}.
If we want to preserve the double description (\ref{sostanza})
and (\ref{forma}) of symmetry under translation transformations
we must therefore introduce noncommuative transformation parameters.
In fact, it is easy to verify that assuming
\begin{displaymath}
[\e_0,x_\mu]=0
 ~,~~~ \Phi\e_j=\e_j (e^{-\lambda P_0}\Phi) ~,
\end{displaymath}
which follow from
\begin{displaymath}
[\e_j,\x_0]=i\lambda\e_j ~,~~~ [\e_j,\x_k]=0 ~,
\end{displaymath}
one finds that the conditions (\ref{sostanza}) and (\ref{forma}) are
equivalent. Interestingly this choice of noncommutativity of the
transformation parameters allows to describe them as differential
forms\footnote{Note that this is one of the two differential calculi
introduced in Ref.~\cite{oeckdiff}.}, $\e_\mu=\de\x_\mu$,
and therefore the condition for invariance of the action
under translation transformations can be cast in the
form
\begin{displaymath}
S(\Phi+i\de\x P\Phi)-S(\Phi)=i\de\x_\mu\int\de^4\x\;P^\mu\Phi^2
\end{displaymath}

It appears plausible that other
choices of noncommutative transformation parameters would preserve
the double description (\ref{sostanza})
and (\ref{forma}) of symmetry under translation transformations.
But it appears likely that this connection with differential
forms has some deep meaning: in order to preserve the
possibility to describe translation symmetries according to (\ref{forma})
(having already verified that these symmetries satisfied the first
criterion (\ref{sostanza}))
we ended up adopting the following description of translations
\bea
{\x}_\mu&\rightarrow&{\x}_\mu'={\x}_\mu + \de\x_\mu \nn\\
\phi({\x})&\rightarrow& \phi'({\x})
=\phi({\x})+ i\de\x_\mu P^\mu\Phi \nn
\eea
where the $\de\x_\mu$ describe the proper concept, as previously
established~\cite{oeckdiff}, of differential forms for our noncommutative
spacetime and the $P^\mu$ act as previously
described ($P_\mu\Omega_R(f)=\Omega_R(-i\partial_\mu f)$).
This is rather satisfactory from a conceptual perspective, since
even in commutative spacetime an infinitesimal translation
is most properly described as ``addition" of a differential form.

The differentials satisfy nontrivial commutation relations~\cite{oeckdiff}
\begin{displaymath}
[\de\x_0,\x_\mu]=0\quad[\de\x_j,\x_k]=0\quad [\de\x_j,\x_0]=i\lambda\de\x_j
\end{displaymath}
as required for our translations to preserve the $\kappa$-Minkowski
commutation relations (again with ${\x}_\mu'={\x}_\mu + \de\x_\mu$)
\bea
[{\x}_j,{\x}_k]=0  \rightarrow  [{\x}_j',{\x}_k']=0~,~~~~
[{\x}_j,{\x}_0]=i\lambda {\x}_j \rightarrow [{\x}_j',{\x}_0']=i\lambda {\x}_j'
\nn
\eea

An infinitesimal translation
associates to each element $\Phi$ of \kM\ ( which we here denote by $M_\kappa$) the element
$\Phi' \equiv \Phi+\de\Phi$ of an algebra that we denote by $M_\kappa\oplus \Gamma$,
with product rule
\bea
(\Phi+\de\Phi)(\Psi+\de\Psi)
=\Phi\Psi+\Phi{\cdot}\de\Psi+\de\Phi{\cdot}\Psi=\Phi\Psi+\de(\Phi\Psi).
\label{prodmod}
\eea
This algebra is isomorphic to $\kappa$-Minkowski through the map $1+\de$, which
is an algebra-isomorphism. Then an infinitesimal translation transforms an
element of $\kappa$-Minkowski in an element of a ``second copy" of
$\kappa$-Minkowski. It is a transformation internal to the
same {\underline{abstract}} algebra.
This abstract algebra {\underline{is}} our ``space of
functions of the spacetime coordinates".

One can easily verify that the equivalence of (\ref{sostanza})
and (\ref{forma}) emerged as a result of the fact that, while for our
translation generators $P_\mu$ the Leibniz rule clearly does not hold,
for the infinitesimal translation ``$\de$"
the Leibniz rule holds:
\begin{displaymath}
\de(\Phi\Psi) = (\Phi\Psi)'- \Phi\Psi = \de\Phi \, \Psi + \Phi \, \de\Psi
\end{displaymath}
(where again  $\Phi' \equiv \Phi+\de\Phi$ and we used
the product rule (\ref{prodmod})).

This network of results and interpretations provides the conceptual
ground for a description of translational symmetry
in $\kappa$-Minkowski spacetime.
Simply by inspection of the $\kappa$-Minkowski commutation
relations one already concludes that classical translations
cannot be a symmetry of this spacetime.
The question then is whether translations are a lost/broken symmetry
of this spacetime or instead they are simply deformed by
the $\kappa$-Minkowski noncommutativity.
We have shown that one can construct theories in $\kappa$-Minkowski
spacetime that enjoy a deformed/quantum (Hopf-algebra) translational
symmetry.

Again by inspection of the $\kappa$-Minkowski commutation
relations one can see that instead classical rotations can
be implemented as a symmetry. But for boosts something analogous
to what happens for translations occurs: classical boosts are not
a symmetry of $\kappa$-Minkowski, but, as we showed, there is
a quantum/deformed version of boosts that are symmetries.


\chapter{Deformed Dirac equation from a 5D bicovariat differential calculus}

In the commutative case (see for example~\cite{WeinbergBook}) one
can obtain the wave equation of a particle requiring that the free
particle states be solutions of a system of partial differential
equations, which is required to contain the Klein-Gordon equation.
The Dirac construction of the equation for spin-$1/2$ particles
represents a well-known example of this procedure.

In this chapter we will show that a $\kappa$-deformed Dirac
equation can be obtained in \kM\ noncommutative spacetime
following closely the line of analysis adopted by Dirac in the
conventional case of commutative Minkowski spacetime. We will see
that the choice of a differential calculus on \kM\ plays a key
role in this construction.

We start this chapter showing that the ``maximally''-symmetric
action obtained in the previous chapter allows us to write a wave
equation for a scalar particle in \kM . This equation, which can
be viewed as a $\kappa$-deformed Klein-Gordon equation, furnishes
the physical condition that will be used in the analysis of the
spinorial equation.

\section{Deformed equation for a scalar particle in \kM }

In the analysis reported in this Chapter, we introduce a field in
\kM\ using the \emph{right-ordered} map (\ref{Rmap}) discussed in
Chapter 2. As shown earlier, choosing the $\Omega_R$ map is
equivalent to choosing the prescription on the \kM\ functions, in
which the time variable is always to the right with respect to the
spatial variables: \be \Omega_R(f)=\int d^4k \tilde{f}(k)\,
\Omega_R(e^{ikx})=\int d^4k \tilde{f}(k)\,
e^{-i\vk\vec{\x}}e^{ik_0{\x}_0} \label{kMelem} \ee where
$\tilde{f}(k)=\frac{1}{(2\pi)^4}\int d^4k\,f(x)e^{-ikx}$ is the
classical inverse Fourier transform of the commutative function
$f(x)$. Since in this Chapter we consider only the
\emph{time-to-the-right} ordering, from here after we omit the
label $R$.

Let us start by analyzing the possibility to obtain a Klein-Gordon
equation for scalar particles in \kM . This can be done applying a
generalized "action principle" to the  action $S[\Phi]$ for a free
scalar field $\Phi$ (of \kM )  found in the previous Chapter.

In the previous Chapter, we have found that the symmetry algebra
of the theory described by $S[\Phi]$ is the \kkP\ algebra, which
can be represented by a large number of Hopf-algebra realizations
with 10 Poincar\'e-like generators. Each realization (or basis) of
\kkP\ is naturally connected with an ordering prescription in \kM
. In particular, the Majid-Ruegg basis~(\ref{kappapoincare}) has
generators ($\mP_{\mu},M_i,{\mN}_i$), that act in the following
way on the right-ordered elements of \kM : \bea
&&\mP_{\mu}\Omega(\phi(x))=\Omega[-i\partial_{\mu}\phi(x)]\nn\\
&&M_j\Omega(\phi(x))=\Omega[i\epsilon_{jkl}x_k\partial_l\phi(x)]\nn\\
&&{\mN}_j\Omega(\phi(x))=\Omega(-[ix_0\partial_j+x_j(\frac{1-e^{2i\lambda
\partial_0}}{2\lambda}-\frac{\lambda}{2}\nabla^2)+\lambda
x_l\partial_l\partial_j]\phi(x))\label{bicrossbasis} \eea It is
also convenient to remind that the commutation relations among
them are: \bea
\qs \mP_{\mu},\mP_{\nu}\qd&=&0,\nn\\
\qs M_j,M_k\qd&=&i\varepsilon_{jkl}M_l,\;\;\;\qs {\mN}_j,M_k\qd=i\varepsilon_{jkl}{\mN}_l,
\;\;\;\qs {\mN}_j,{\mN}_k\qd=-i\varepsilon_{jkl}M_l,\nn\\
\qs M_j,\mP_0\qd&=&0,\;\;\;[M_j,\mP_k]=i\epsilon_{jkl}\mP_l\nn\\
\qs {\mN}_j,\mP_0\qd&=&i\mP_j,\nn\\
\qs {\mN}_j,\mP_k\qd&=&i\qs\ts \frac{1-e^{-2\lambda
\mP_0}}{2\lambda}+\frac{\lambda}{2}\vec{\mP}^2\td\delta_{jk}-\lambda
\mP_j\mP_k\qd\nn \eea with the co-algebra sector given by: \bea
\Delta(\mP_0)&=&\mP_0\otimes 1+1\otimes \mP_0,\;\;\;\Delta(\mP_i)=\mP_i\otimes 1+e^{-\lambda \mP_0}\otimes \mP_j\nn\\
\Delta(M_j)&=&M_j\otimes 1+1\otimes M_j\nn\\
\Delta({\mN}_j)&=&{\mN}_j\otimes 1+e^{-\lambda \mP_0}\otimes
{\mN}_j-\lambda\epsilon_{jkl}\mP_k\otimes M_l \eea and the mass
Casimir operator: \be C_{\lambda}(\mP)=\cosh(\lambda
\mP_0)-\frac{\lambda^2}{2}e^{\lambda
\mP_0}\vec{P}^2.\label{Rcasimir} \ee

The "maximally-symmetric" action (\ref{actionfinal}) for a free
scalar theory in \kM\ is the following:
 \be S[\Phi]=\int
d{\x}\;\Phi({\x})[\Box_{\lambda}+M_{KG}^2]\Phi({\x}) \ee where the
integral is the right integral (\ref{intR}) and $\Box_{\lambda}$
is the generalization of the D'Alembert operator
$\Box=(\partial_0^2-\nabla^2)$: \be
\Box_{\lambda}=-\frac{2}{\lambda^2}\qs\cosh(\lambda
\mP_0)-1\qd+e^{\lambda \mP_0}\vec{\mP}^2\ee and $M_{KG}$ is a mass
parameter. Notice that the relation between the $\Box_\lambda$
operator and $C_\lambda(\mP)$ is the following: \be
\Box_\lambda=\frac{2}{\lambda^2}[1-C_\lambda(\mP)]\label{boxdirac}
\ee

The action $S[\Phi]$ leads to a  $\kappa$-deformed Klein-Gordon
equation for a free-spinless particle. It can be obtained through
requiring that the variation of the action, induced by the
variation of the field $\delta\phi$, be zero for all $\delta\phi$,
exactly as in the commutative case. In fact, consider the
following variation of the action: \be S[\Phi+\delta
\Phi]-S[\Phi]=\int  d{\x}\;\{
(\Phi+\delta\Phi)[\Box_{\lambda}+M_{KG}^2](\Phi+\delta\Phi)-\Phi[\Box_\lambda
+M_{KG}^2]\Phi\} \ee At the first order in $\delta\phi$ it is: \be
S[\Phi+\delta \Phi]-S[\Phi]\approx\int
d{\x}\;\{\delta\Phi[\Box_{\lambda}+M_{KG}^2]\Phi+\Phi[\Box_\lambda
+M_{KG}^2]\delta\Phi\}. \ee Introducing the explicit expressions
of the \kM\ elements \bea
\Phi(\x)&=&\frac{1}{(2\pi)^2}\int d^4k\, \tilde{\phi}(k)\,\Omega(e^{ikx})\nn\\
\delta\Phi(\x)&=&\frac{1}{(2\pi)^2}\int d^4q
\,\tilde{\delta\phi}(q)\,\Omega(e^{iqx}),\nn \eea we obtain \be
S[\Phi+\delta \Phi]-S[\Phi]\approx\int
d{\x}\;\{\Omega(e^{iqx})[\Box_{\lambda}+M_{KG}^2]\Omega(e^{ikx})+\Omega(e^{ikx})[\Box_\lambda
+M_{KG}^2]\Omega(e^{iqx})\}\tilde{\phi}(k)\tilde{\delta\phi}(q).
\ee and taking into account (\ref{boxdirac}): \bea S[\Phi+\delta
\Phi]-S[\Phi]&\approx&\frac{2}{\lambda^2(2\pi)^4} \int d^4kd^4q
d{\x}\;\{\Omega(e^{iqx})
[1-C_\lambda(\mP)+\frac{\lambda^2}{2}M_{KG}^2]\Omega(e^{ikx})+\nn\\
&&+\Omega(e^{ikx})[1-C_\lambda(\mP)+
\frac{\lambda^2}{2}M_{KG}^2]\Omega(e^{iqx})\}\tilde{\phi}(k)\tilde{\delta\phi}(q)\nn\\
&=&\frac{2}{\lambda^2(2\pi)^4}\int
d{\x}d^4kd^4q\;\{[2-C_\lambda(k)-C_\lambda(q)+\lambda^2M_{KG}^2]\}\Omega(e^{i(k\oplus
q)})\tilde{\phi}(k)\tilde{\delta\phi}(q)\nn \eea where the sum
$k\oplus q=(k_0+q_0,k_j+e^{-\lambda k_0}q_j)$ has been introduced
in (\ref{Roplus}). Then, using the (\ref{delta}), we perform the
integral in $dx$: \bea
S[\Phi+\delta \Phi]-S[\Phi]&\approx&\frac{2}{\lambda^2}\int d^4kd^4p\;
\{[2-C_\lambda(k)-C_\lambda(p)+\lambda^2M_{KG}^2]\}
\delta(k\oplus p)\tilde{\phi}(k)\tilde{\delta\phi}(p)\nn\\
&=&\frac{2}{\lambda^2}\int
d^4k\;\{[2-C_\lambda(\bar{k})-C_\lambda(k)+\lambda^2M_{KG}^2]\}
\tilde{\phi}(k)\tilde{\delta\phi}(\bar{k}) \eea where
$\bar{k}=-k_0,-e^{\lambda k_0}k_j$. Noticing that
$C_\lambda(\bar{k})=C_\lambda(k)$, we have: \bea S[\Phi+\delta
\Phi]-S[\Phi]&\approx&\frac{4}{\lambda^2}\int
d^4k\;\{[1-C_\lambda(k)+\frac{\lambda^2}{2}M_{KG}^2]\}
\tilde{\phi}(k)\tilde{\delta\phi}(\bar{k})\nn
\eea This variation is zero for all $\delta \phi$ if and only if:
\be
\ts 2\lambda^{-2}\qs \cosh(\lambda k_0)-1\qd-e^{\lambda
k_0}\vec{k^2}-M_{KG}^2\td\tilde{\phi}(k)=0.\label{eq:KG} \ee This
is the wave equation for a scalar particle in \kM\ written in the
"momentum" space (the energy is $E\equiv k_0$ and the spatial
momenta are $\vec{p}\equiv\vec{k}$).
From (\ref{eq:KG}) we obtain the following dispersion relation for
a free scalar particle in \kM\ \bea \cosh(\lambda
E)-\frac{\lambda^2}{2}e^{\lambda
E}\vec{p}^2=1+\frac{\lambda^2}{2}M_{KG}^2.\label{kmdisprel} \eea
We notice that the mass parameter $M_{KG}$ is not the mass at rest
because for $\vec{p}=0$ the energy $E\neq M_{KG}$. But we can see
that $M_{KG}$ and the mass at rest $m$ are related by the relation
$M_{KG}=\sqrt{2(\cosh(\lambda m)-1)}/\lambda$.

\section{Spin-$1/2$ particles in \kM} Now we want to write a
Dirac equation in \kM\ following the method adopted by Dirac in
the commutative case. The Dirac procedure consists in writing a
linear partial differential equation with arbitrary coefficients
$\gamma^\mu$ ($\mu=0,...,3$): \be (i\gamma^{\mu}\partial_{\mu}+m
I)\psi(x)=0\label{diraccomm} \ee where $m$ is the particle mass,
$\psi$ is a n-plet of fields, and $\gamma^{\mu}$ are ($n\times n$)
hermitian matrices to be determined imposing the dispersion
relation (that works as physical condition). This equation turns
out to be consistent with the physical condition if and only the
matrices $\gamma^\mu$ are the standard generators of the Clifford
algebra: \be \{\gamma^\mu,\gamma^\nu\}=2\eta^{\mu\nu}.
\label{Clifford}\ee In this way the lowest dimension is $n=4$. We
remind (see \cite{ItzyZub}) that the Dirac equation
(\ref{diraccomm}) with the choice of the matrices (\ref{Clifford})
satisfies \emph{automatically}\footnote{since $\gamma^\mu$
transform as a four-vector under the standard Lorentz rotations}
the property of invariance under the action of the standard
Lorentz-symmetry. This invariance property is in fact a
fundamental request for any relativistic wave equation.

Also in \kM , as in the commutative case, we introduce a
Lorentz-spinor wave function $\Psi(\x)$, whose components are of
the form \be \Psi_r({\x})=\int d^4k \,\tilde{\psi}_r(k)\; e^{-ik
{\x}}e^{ik_0{\x_0}}, \ee  and involve the product of an element of
the algebra of functions on spacetime
($e^{-ik{\x}}e^{ik_0{\x_0}}$) and a finite vector space containing
the spin degrees of freedom $\tilde{\psi}_r(k)$.

The Lorentz spinor $\Psi(\x)$ represents the space of physical
states in \kM\ and it will satisfy a wave equation with a
(deformed) $\Dirac_{\lambda}$ operator: \be
[\Dirac_{\lambda}+M_DI]\Psi(\x)=0.\label{eq:kmdirac} \ee where $I$
is the identity matrix and $M_{D}$ is a mass parameter which we
expected to be some simple function of the mass at rest $m$.

 As
already announced, we want to find a suitable form for the Dirac
operator $\Dirac_\lambda$ in \kM\ following the procedure adopted
in the commutative case. We have stressed that one of the key
points of the standard Dirac construction
 is the presence of the differential calculus (\emph{i.e.} the presence of the derivatives $\partial_\mu$)
 in the equation (\ref{diraccomm})
. Therefore, in preparation for our analysis of a Dirac equation
in \kM , it is useful to review briefly some aspects of the
differential calculus in such a space.

\subsection{Choice of a differential calculus on \kM }
In the commutative case there is only one ``natural" differential
calculus, which involves the ordinary derivatives, which enters
the Dirac equation (\ref{diraccomm}). In this case, the exterior
derivative operator $d$ of a commutative function $f(x)$  is the
usual one:
 \bea d f(x)&=&dx^\mu\partial_\mu f(x)\nn\\
 &=&idx^\mu P_\mu f(x) \eea
where, in the second line, we have expressed the vector fields
$\partial_\mu$ in terms of the standard translation generators
$P_\mu=-i\partial_\mu$ (see Chapter 3). In this way it is clear
that the $\partial_\mu$ transform \emph{covariantly} under the
standard Lorentz algebra (generated by $M_j,N_j$): \bea {}
[M_j,P_0]=0,&&[M_j,P_k]=i\epsilon_{jkl}P_l\nn\\
{}[N_j,P_0]=iP_j,&&[N_j,P_k]=i\delta_{jk}P_0 \eea

In the case of \kM , instead, the introduction of a differential
calculus is a more complex problem~\cite{Sitarz}\cite{Masl}. In
fact in a noncommutative spacetime the differential calculus is in
general non unique and there are several differential calculi that
can be constructed. In the analysis that follows we focus on one
possible choice of differential calculus in \kM , the
"five-dimensional differential calculus", introduced by Sitarz~
\cite{Sitarz}. We have reported the Sitarz construction in the
Appendix~C. In this five-dimensional (5D) differential calculus
the exterior derivative operator $d$ of a generic \kM\ element
$F({\x})=\Omega(f(x))$ can be written in the form (\ref{result1}):
\bea &&dF({\x})={\bf dx}^a \,{\D}_a(\mP)F({\x}),\;\;\;\;a=0,\dots,4\nn\\
&&{\D}_a(\mP)=\ts\frac{i}{\lambda}[\sinh(\lambda\mP^0)+
\frac{\lambda^2}{2}e^{\lambda \mP^0}\mP^2], i\mP_je^{\lambda
\mP_0}, -\frac{1}{\lambda}(1-\cosh(\lambda
\mP_0)+\frac{\lambda^2}{2}\mP^2 e^{\lambda \mP_0})\td,
\;\;\;j=1,2,3\nn\\
\label{5Ddiffcalc1} \eea where $\mP_\mu$ are the translation
generators of the Majid-Ruegg \kkP\ basis~(\ref{kappapoincare}),
whose action on a right-ordered function of \kM\ is
$\mP_\mu(e^{-ik\x}e^{ik_0\x_0})=k_\mu(e^{-ik\x}e^{ik_0\x_0})$. The
commutation relations between the one-form generators ${\bf dx}^a$
and the \kM\ generators $\x^{\mu}$ are: \bea {} [{\bf
dx}^0,\x^0]=-i\lambda
{\bf dx}^4,&&[{\bf dx}^0,\x^j]=i\lambda {\bf dx}^j\nn\\
{}[{\bf dx}^j,\x^0]=0,&& [{\bf dx}^j,\x^k]=i\delta_{jk}({\bf dx}^0+{\bf dx}^4)\nn\\
{}[{\bf dx}^4,\x^{\mu}]=-i\lambda {\bf dx}^\mu&&
\label{comreldiffcalc} \eea The introduction of such a 5D calculus
in our 4D spacetime may at first appear to be surprising, but it
can be naturally introduced on the basis of the fact that the that
\kkP /\kM\ framework can be obtained (and was originally obtained
~\cite{LNRT}) by contraction of a 5D (quantum-deformed) anti-de
Sitter algebra. The fifth one-form generator is here denoted by
"$dx^4$", but this is of course only a formal notation, since
there is no fifth \kM\ coordinate ${\x}_4$. And the peculiar role
of $dx_4$ in this differential calculus is also codified in the
fact that, the last component ${\D}_4(\mP)$ is essentially the
Casimir (\ref{Rcasimir}) of \kkP : \be
\D_4(\mP)=\lambda^{-1}(C_\lambda(\mP)-1). \ee

This differential calculus is characterized by interesting
transformation properties under the action of the Lorentz sector
of \kkP . In fact taking into account of (\ref{kappapoincare}) one
finds that: \bea {}
[M_j,\D_0(\mP)]=0,&[M_j,\D_k(\mP)]=i\epsilon_{jkl}\D_l(P),&[M_j,\D_4(\mP)]=0\nn\\
{}[\mN_j,\D_0(\mP)]=i\D_j(\mP),&[\mN_j,\D_k(\mP)]=i\delta_{jk}\D_0(\mP),&[\mN_j,\D_4(\mP)]=0
\eea Thus the operators $\D(\mP)_\mu$ transform under \kkP\ action
in the same way as the $P_\mu$ operators transform under the
standard Poincar\'e action, while $\D_4(\mP))$ is invariant.

This differential calculus originates in ~\cite{Sitarz} by the
request that it remain invariant under the action of the \kkP\
action, \emph{i.e} the commutation relations
(\ref{comreldiffcalc}) that characterize it remain invariant under
the action of the \kkP\ generators: \bea
{}[\mathbf{dx}^a,\x^\mu]=\upsilon^{a\mu}_\rho \mathbf{dx}^\rho
&\to&\mathrm{T}\, [\mathbf{dx}^a,\x^\mu]=\upsilon^{a\mu}_\rho
\;\mathrm{T}\,\mathbf{dx}^\rho \eea where $\mathrm{T}$ denotes
globally the \kkP\ generators ($\mP_\mu,M_j,\mN_j$). A
differential calculus in which the commutation relations between
the $1-$form generators and the \kM\ generators remain $invariant$
under the action of symmetry algebra (\kkP\ in our case), is
called "covariant" differential calculus.
 In~\cite{Masl} it has been
demonstrated that the 5D differential calculus of Sitarz is  the
unique covariant one with respect to the left action of \kkP\
group.

There are other (non-covariant) differential calculi candidate for
\kM , and we will consider one of them in Section~3. But, in light
of the good properties  said above, the covariant 5D differential
calculus appears to be the natural candidate in order to construct
the Dirac equation in \kM .

\subsection{Constructing the deformed Dirac equation}

Let us now consider the Dirac equation (\ref{eq:kmdirac}) in \kM\
spacetime. If the differential calculus is given by
(\ref{5Ddiffcalc1}) the most general parametrization of the Dirac
equation in \kM\ is: \be \ts
i\tilde{\gamma}^\mu\D_\mu(\mP)+\tilde{\gamma}^4\D_4(\mP)+M_{D}I\td\Psi({\x})=0,
\label{pde:dirac} \ee where $\tilde{\gamma}^a,\;a=0,...,4$ are
five hermitian matrices to be determined. Let us observe that,
while in commutative Minkowski one can safely assume that the
matrices are just numbers (independent of any of the variables
that characterize the system), the presence of the scale $\lambda$
in \kM\ forces us to allow for a possible dependence of the
$\tilde{\gamma}^a$ on the mass $m$ ($\tilde{\gamma}^a(\lambda
m)$).
 Moreover we require that, in the
commutative limit $\lambda\to0$, the operator $\Dirac_\lambda$
reduces to the standard Dirac operator: \be
\Dirac=i\gamma^\mu\partial_\mu=-\gamma^\mu P_\mu,
\label{limitD}\ee
 with $\gamma^\mu$ the $4\times 4$ standard generators
of the Clifford algebra. Making  the commutative limit of
$\Dirac_\lambda$ we obtain: \bea
\lim_{\lambda=0}\,\Dirac_\lambda&=&-\tilde{\gamma}^\mu P_\mu-
\tilde{\gamma}^4(0)\lambda [(P_0^2-P^2)+O(\lambda)]\nn \eea Thus,
comparing this expression with the (\ref{limitD}), we find that
for $\lambda=0$ the matrices $\tilde{\gamma}^\mu$ and the mass
parameter $M_D$ must be : \bea \tilde{\gamma}^\mu(\lambda m)&\to& \gamma^\mu\nn\\
M_D(\lambda m)&\to& m, \label{classlim}\eea while the commutative
limit of $\tilde{\gamma}^4$ must be finite, in fact the term that
comes from the fifth "spurious" element of the differential
calculus must disappear: \be \tilde{\gamma}^4(0)\lambda
[(P_0^2-P^2)+O(\lambda)]=0 \ee thus, in the limit $\lambda\to 0$,
$\tilde{\gamma}^4$ cannot be singular.

We remind that in the commutative case the Dirac operator $\Dirac$
must satisfy two fundamental conditions in order to give rise to
the relativistically correct Dirac equation~\cite{ItzyZub}. Here
we generalize these conditions to the \kM\ case and use them in
order to determine $\Dirac_\lambda$:
\begin{itemize}
\item i) \emph{Physical condition}: $\Dirac_{\lambda}$ must be
such that the components of $\Psi$ must satisfy the
$\kappa$-deformed KG equation (\ref{eq:KG}), so that a ``plane
wave'' on shell, \emph{i.e.} with momenta ($E,\vec{p}$) satisfying
the dispersion relation (\ref{kmdisprel}) \be \cosh(\lambda
E)-\frac{\lambda}{2}e^{\lambda E}p^2=\cosh(\lambda m),
\label{Ep}\ee be a solution of the (\ref{eq:kmdirac}). The most
general form of the ``plane wave'' on-shell is: \be
u(\vec{p})e^{-ip_j\x_j}e^{iE\x_0}+v(\vec{p})e^{-iS(p_j)}e^{iS(E)\x_0}\label{decomp}
\ee where $S(E,\vec{p})=(E,-e^{\lambda E}\vec{p})$ is the antipode
map, which generalizes the inversion operation in \kM . In fact
both $e^{-ip_j\x_j}e^{iE\x_0}$ and $e^{-iS(p_j)}e^{iS(E)\x_0}$ are
solution of the $\kappa$-deformed KG equation (if $E=E(p)$
satisfies the dispersion relation (\ref{Ep})). Thus, the following
equations must be satisfied: \bea
&&(\Dirac_{\lambda}-M_DI)u_r(\vec{p})e^{-ip_j\x_j}e^{iE(p)\x_0}=0\nn\\
&&(\Dirac_{\lambda}-M_DI)v_r(\vec{p})e^{-iS(p_j)\x_j}e^{-iE(p)\x_0}=0.\nn\
\eea where the first equation generalizes the Dirac equation for
particles, while the second generalizes the Dirac equation for
antiparticles.

\item ii) \emph{Covariance property}: in order to assure that the
Dirac equation (\ref{eq:kmdirac}) transforms covariantly under the
action of the symmetry algebra, the operator $\Dirac_\lambda$ must
commute with all the  generators of the spinorial representation
of the symmetry algebra (\emph{i.e} of \kkP ):
\be
[\mathrm{T},\Dirac_{\lambda}]=0 \;\;\;
\ee where $\mathrm{T}$ denotes globally the generators of the
spinorial representation of \kkP .
\end{itemize}

As underlined above, in the commutative case the physical
condition is sufficient to determine uniquely the Dirac equation,
and the covariance property ii) is automatically satisfied. Thus,
we first impose the physical condition i), looking for all
possible choices of $\tilde{\gamma}^a$, such that the "plane-wave"
components (\ref{decomp}) is a solution of the Dirac equation in
\kM . To be simple, we focus only on the equation for particles,
with wave equation given by:
 \be
u_r(\vec{p})e^{-ip\x}e^{iE\x_0}, \label{diracpw} \ee with ($E,p$)
obeying the dispersion relation (\ref{Ep}).

 It is convenient for our analysis to rewrite the equation
(\ref{pde:dirac}) in the following form: \be \ts
i\D_0(\mP)+i\al^i\D_i(\mP)+\al^4\D_4(\mP)+\bt
M_D\td\Psi(\x)=0\label{pde:dirac1} \ee where we have introduced
the notation\footnote{Notice that $\tilde{\gamma}^0$ cannot be
zero because in the limit $\lambda=0$ it does not vanish, thus its
inverse is defined} $\bt\equiv(\tilde{\gamma}^0)^{-1},
\al^i\equiv(\tilde{\gamma}^0)^{-1}\tilde{\gamma}^i,
\al^4\equiv(\tilde{\gamma}^0)^{-1}\tilde{\gamma}^4$.

Next we observe that, by multiplying Eq.~(\ref{pde:dirac1}) by the
operator $i\D_0-(i\al^i\D_i+\al^4\D_4+\bt M_D)$ \be \ts
i\D_0-(i\al^i\D_i+\al^4\D_4+\bt M_D)\td\ts i\D_0
+i\al^i\D_i+\al^4\D_4+\bt M_D\td \Psi(\x)=0 \ee one obtains \be
[\D^2_0+\ts i\al^i\D_i+\al^4\D_4+\bt M_D\td^2]\Psi(\x)=0 ~.
\label{square} \ee

The requirement of consistency with the deformed KG equation
translates into the condition that a choice of $\Psi$ given by
plane waves (\ref{diracpw}) with on-shell energy-momentum
$(E(p),p)$, such that $\cosh(\lambda
E)-\frac{\lambda^2}{2}e^{\lambda E}\vp^2 =\cosh(\lambda m)$,
should be a solution of our sought deformed Dirac equation. This
allows to obtain from (\ref{square}) the following $n-plet$ of
equations: \bea
 \left[\D^2_0(E,p)\right.&-&(\al^i)^2(\D_i(E,p))^2+(\al^4)^2\D_4^2+\bt^2M_D^2+ \nonumber\\
&-&\sum_{i<j}\{\al^i,\al^j\}\D_i(E,p)\D_j(E,p)+i[\{\al^i,\al^4\}\D_4
+M_D\{\al^i,\bt\}]\D_i(E,p)+\nonumber\\
&+&\left. M_D\{\al^4,\bt\}\D_4\right] u(\vec{p})=0
\label{condizione} \eea

Using the fact that ordinary space-rotation symmetry should still
be preserved the equations (\ref{condizione}) straightforwardly
lead to the consistency requirements:
 \bea
&&\sum_{i<j}\{\al^i,\al^j\}\D_i(E,p)\D_j(E,p)=0\\
&&[\{\al^i,\al^4\}\D_4+M_D\{\al^i,\bt\}]\D_i(E,p)=0\\
&&\sum_{i} (\al^i)^2 (D_i(E,p))^2 \propto \sum_{i} (D_i(E,p))^2
\eea From this we conclude that \bea
\{\al^i,\al^j\}&=&0\;\; i\neq j\label{constr1}\\
\{\al^i,\al^4\}\D_4(k)&=&-M_D\{\al^i,\bt\}\label{constr2}\\
(\al^1)^2&=&(\al^2)^2 = (\al^3)^2\label{constr2bis} \eea We can
now use these results to write equation (\ref{condizione}) as
follows: \be
 [\D^2_0(E,p)-|\vec{\D}(E,p)|^2 {\cal G}+\D_4^2(\al^4)^2
 +M^2\bt^2+M_D\{\al^4,\bt\}\D_4]u(\vec{p})=0 \label{eq:KG1}
\ee where we introduced the notation ${\cal G}$ for the common
value (see (\ref{constr2bis})) of the $(\al^i)^2$ matrices, ${\cal
G} \equiv (\al^1)^2 =(\al^2)^2 = (\al^3)^2$

Next we can use the fact that, on the basis of
(\ref{5Ddiffcalc1}), we know that the $D_a(k)$ have the following
on-shell expressions: \bea
\D_0(E,p)&=&\frac{i}{\lambda}[e^{\lambda E}-\cosh(\lambda m)]\nonumber\\
\D_i(E,p)&=&ie^{\lambda E}p_i\nonumber\\
\D_4(E,p)&=&\frac{1}{\lambda}[\cosh(\lambda m)-1] ~. \eea This
allows us to rewrite (\ref{eq:KG1}) as \bea
&-&\lambda^{-2}[e^{2\lambda E}+\cosh^2(\lambda m)
-2\cosh(\lambda m)e^{\lambda E}]I+\nonumber\\
&+&{\cal G} e^{2\lambda E}|\vp|^2+\nonumber\\
&+&(\al^4)^2\lambda^{-2}(1-2\cosh(\lambda m)+\cosh^2(\lambda m))+M_D^2\bt^2+\\
&+&\lambda^{-1}M_D\{\al^4,\bt\}(\cosh(\lambda m)-1)=0 \nonumber ~,
\eea which can also be cast in the form
\bea &&-\sin^2(\lambda m)I+\lambda^2[{\cal G}-I]e^{2\lambda
E}|\vp|^2
+\nonumber\\
&+&(\al^4)^2(1-2\cosh(\lambda m)+\cosh^2(\lambda
m))+\lambda^2M_D^2\bt^2
+\nonumber\\
&+&\lambda M_D\{\al^4,\bt\}(\cosh(\lambda m)-1)=0 ~, \label{joc61}
\eea using again the dispersion relation.

Since Eq.(\ref{joc61}) must hold for every arbitrary value of
$\vp$ we can deduce that \be {\cal G} = I \ee and \be
\sinh^2(\lambda m)I-(\al^4)^2(\cosh(\lambda m)-1)^2
-\lambda^2M_D^2\bt^2-M_D\lambda\{\al^4,\bt\}(\cosh(\lambda m)-1)=0
\ee which can be conveniently rewritten as \be \sinh^2(\lambda
m)I=(\al^4(1-\cosh(\lambda m))+\lambda M_D \bt)^2 ~.
\label{constr3} \ee

At this point we have reduced our search of consistent deformed
Dirac equations to the search of matrices $\{ \al^j,\al^4, \beta
\}$ such that the following requirements
(\ref{constr1}-\ref{constr2}-\ref{constr3}) are satisfied: \bea
\{\al^j,\al^k\} &=& 2\delta^{jk} I\nn\\
\{\al^j,\al^4\}[\cosh(\lambda m)-1]&=&-\lambda M_D \{\al^j,\bt\}
\label{const2}\\
\sinh^2(\lambda m)I&=&(\al^4(\cosh(\lambda m)-1)+\lambda M_D
\bt)^2 \label{const3} \eea

In deriving from these requirements an explicit result for the
matrices $\{ \al^j,\al^4, \beta \}$ it is convenient to first
consider the case in which $M_D{\cdot} m\neq0$, the case of
massive particles. This allows us to introduce the matrix $A$ \be
A \equiv \frac{\left(\al^4(\cosh(\lambda m)-1)+\lambda M_D \bt
\right)}{ \sinh(\lambda m)} \label{A} \ee which allows to cast the
deformed Dirac equation in the following form:
 \be
[i\D_0(k)+i\al^i\D_i(k) + \D_4(k)\frac{\sinh(\lambda
m)}{\cosh(\lambda m)-1}A-M_D[\frac{\lambda\D_4(k)}{\cosh(\lambda
m)-1} -1]\bt]\tilde{\psi}(k)=0 \label{jodirac1} \ee From
(\ref{const2}) and (\ref{const3}) it follows that $\{A,\al^i\}=0$
and $A^2=I$.

We are at this point ready to obtain the most general deformed
Dirac equation in \kM. In fact, imposing that the $E$ and $p$ are
connected by the dispersion relation (on shell)
Eq.~(\ref{jodirac1}) simplifies to \be [i\D_0(E,p)+i\D_i(E,p)\al^i
+\frac{\sinh(\lambda m)}{\lambda}A]u(\vec{p})=0 \label{7} \ee
where the matrices $\al^j,A$ satisfy the conditions\footnote{From
these conditions one can also infer that the $n {\times} n$
matrices we are seeking must have $n$ even and $n \ge 4$ (not
smaller than $4 {\times} 4$ matrices). In fact, from the
anticommutation relations it follows that $TrA=0$, $A^2=1$, and
$detA={\pm}1$ which requires $n$ to be even. The case $n=2$ is
also excluded since there are only $3$ independent anticommuting
$2 {\times} 2$ matrices (Pauli matrices). We take $n=4$ just as in
the $\lambda \rightarrow 0$ (commutative-Minkowski) limit.} \be
\{\al^j,\al^k\}=2\delta^{jk} I,\;\;\;,\{\al^j,A\}=0,\;\;\;A^2=I
\label{alfajo} \ee
From (\ref{alfajo}) one finds that the matrices
$A,\,\tilde{\gamma}^j\equiv(\tilde{\gamma}^0)^{-1}\al^j$ satisfy
the Clifford relations: \be
A^2=-(\tilde{\gamma}^j)^2=1,\;\;\{A,\tilde{\gamma}^j\}=0,\;\;\;j=1,2,3.
\ee {\it i.e.} they must be the usual (undeformed!) Dirac
matrices. And we find (by multiplying (\ref{7}) by $A$ and making
use of the standard notation for the Dirac matrices $\gamma^0
\equiv A$ and $\gamma^j \equiv \tilde{\gamma}^j$) that in terms of
the usual Dirac matrices there is a unique solution to our problem
of finding the most general deformed Dirac equation in \kM : \be
\ts \frac{e^{\lambda E}-\cosh(\lambda m)}{\lambda}\gamma^0+p_j
e^{\lambda E}\gamma^j-\frac{\sinh(\lambda m)}{\lambda} I\td
u(\vec{p})=0\label{on-shell} \ee The equation for the
antiparticles wave will be: \be \ts \frac{e^{-\lambda
E}-\cosh(\lambda m)}{\lambda}\gamma^0-p_j
\gamma^j-\frac{\sinh(\lambda m)}{\lambda} I\td
v(\vec{p})=0\label{on-shellap}\ee In the $\lambda=0$ limit these
equations reduce to the standard ones: \bea \ts
E\gamma^0+p_j\gamma^j-m I\td u(\vec{p})=0 \\
\ts E\gamma^0+p_j\gamma^j+m I\td v(\vec{p})=0. \eea

 We want to notice
that this result is consistent with the one~\cite{aaa} of the
recently-proposed schemes for a DSR~\cite{bruno} (see
Section~1.4). It is interesting however underlying that the result
(\ref{on-shell},\ref{on-shellap}) has been obtained with a
specific choice of differential calculus in \kM . Thus this
particular choice allows the agreement with the result obtained in
the DSR framework.

\subsection{Invariance under \kkP\ action and off-shell Dirac equation}
 We have established that the on-shell Dirac equation in \kM\ must
take the form (\ref{on-shell},ref{on-shellap}), but, since the
matrix $\beta$ is still undetermined, we do not yet have a
definite description off shell (and it is the Dirac equation off
shell in the energy-momentum sector that encodes all the
properties of the ``spacetime formulation" of the Dirac equation).
In order to determine the matrix $\bt$ it is necessary to impose
the covariance of our deformed Dirac equation in the \kkP\ sense
(condition ii)).

This will require us to introduce a representation of the
$\kappa$-Poincar\'{e} (Hopf) algebra for spin-$1/2$ particles.

In preparation for this analysis we first briefly review the
analogous analysis for the classical Poincar\'{e} (Lie) algebra. A
key point is that for the Lorentz sector the representation can be
described as the sum of two parts: \be
M^T_j=M_j+m_j\;\;\;N^T_j=N_j+n_j \ee where $(M_j,N_j)$ is a
spinless unitary representation of $O(3,1)$ and acts in the
"outer" space of the particle (the one that codifies the momenta
and orbital momenta of the particle), whereas $(m_j,n_j)$ is a
finite-dimensional representation of $O(3,1)$ and acts in the
"inner" space (spin indices). A representation of the whole
Poincar\'{e} group is then obtained by introducing four
translation generators $P_{\mu}=-i\partial_{\mu}$ that act only in
the "outer" space of the particle. The spinorial representation of
the classical Poincar\'{e} algebra is therefore given by \be
P_{\mu}^T=P_{\mu}\;\;\;M^T_j=M_j+m_j\;\;\;N^T_j=N_j+n_j \ee which
of course satisfy the following familiar commutation relations:
 \bea
&&[P^T_\mu,P^T_\nu]=0 \nn\\
&& \qs M_j^T,M_k^T\qd=\!\! i\epsilon_{jkl}M_l^T ~,~~~[N_j^T,N_k^T]
=-i\epsilon_{jkl}M_l^T ~,~~~[M_j^T,N_k^T]=i\epsilon_{jkl}N_l^T\nn\\
&&\qs M_j^T,P^T_0\qd=0\;\;\;\;\; [M_j^T,P^T_k]=i\epsilon_{jkl}P^T_l\nn \\
&&\qs N_j^T,P^T_0\qd=iP^T_j
\;\;\;\;\;[N_j^T,P^T_k]=i\delta_{jk}P^T_0.\nn \eea The
differential form of the spinless realization is given by: \be
P_\mu=-i\partial_\mu,\;\;\;M_j=\epsilon_{jkl}x_kP_l,\;\;\;N_j=x_jP_0-x_0P_j
\ee and the finite dimensional realization can be expressed in
terms of the familiar $\gamma$ matrices in the following way: \be
m_j=\frac{i}{4}\epsilon_{jkl}\gamma_k\gamma_l ~,~~~ n_j
=\frac{i}{2}\gamma_j\gamma_0 \label{n}\ee The action of the global
generators of Poincar\'{e}  over a Dirac spinor is:
 \bea
M^T_j&&\psi_r(x)=\int dk\; [(M_j e^{ik x} )\psi_r(k)
+ e^{ik x} (m_j\psi_r(k))]\nn\\
N^T_j&&\psi_r( x)=\int dk\; [(N_j e^{ik x} )\psi_r(k)
+ e^{ik x} (n_j\psi_r(k))]\nn\\
P_{\mu}^T&&\psi_r( x)=\int dk\; (P_{\mu} e^{ik x}) \psi_r(k)\nn
\eea and the Dirac operator is of course an invariant: \be
[P_{\mu}^T,\Dirac]=[M_j^T,\Dirac]=[N_j^T,\Dirac]=0 \ee

We intend to obtain analogous results for spinors and the Dirac
operator in \kM. Our deformed Dirac operator must be invariant,
\be [{\mathcal{P}}_{\mu}^T,\Dirac_{\lambda}]
=[{\mathcal{M}}_j^T,\Dirac_{\lambda}]
=[{\mathcal{N}}_j^T,\Dirac_{\lambda}]=0 \label{covariance}\ee
under the action of \kkP\ generators
${\mathcal{P}}^T,{\mathcal{M}}^T,{\mathcal{N}}^T$, which satisfy
the following commutation relations:  \bea
\qs{\mathcal{P}}_{\mu}^T,{\mathcal{P}}_{\nu}^T\qd&=&0\nn\\
\qs{\mN}_j^T,{\mN}_k^T\qd&=&-
i\epsilon_{jkl}{\mM}_l^T,\;\;\;\qs{\mN}_j^T,{\mM}_k^T \qd=
i\epsilon_{jkl}{\mN}_l^T\nn\\
\qs{\mM}_j^T,{\mM}_k^T \qd&=&
i\epsilon_{jkl}{\mM}_l^T\nn\\
\qs{\mM}_j^T,{\mathcal{P}}_0^T\qd&=&0\;\;\;\qs{\mM}_j^T,{\mathcal{P}}_k^T\qd
=i\epsilon_{jkl}{\mathcal{P}}_l^T\nn\\
\qs{\mN}_j^T,{\mathcal{P}}_0^T \qd&=&
iP_j^T\;\;\;\qs{\mN}_j^T,{\mathcal{P}}_k^T\qd=i\delta_{jk}[\frac{1-e^{-
2\lambda{\mathcal{P}}_0^T}}{2\lambda}+\frac{\lambda}{2}({\mathcal{P}}^T)^2]
-i\lambda{\mathcal{P}}_j^T{\mathcal{P}}_k^T\nn \eea

Consistently with the results obtained so far we expect that it
will not be necessary to deform the rotations: \be {\mM}^T_j =
M^T_j = M_j+m_j \ee and in fact this works perfectly, as one can
easily verify.

Boosts in general require a deformation, and we already know from
the earlier points of our analysis that the differential form of
the spinless realization must be given by the operators ${\mN}_j$
introduced in (\ref{bicrossbasis}).  We are therefore seeking a
suitable spinorial realization of boosts ${\mN}_j^T$, in terms of
$\mN_j$ and $n_j$, such that the covariance condition of the Dirac
operator (\ref{covariance}) be satisfied. This problem of the
search of $\mN^T_j$ is analyzed in Appendix~\ref{appspinrepr}. A
key point is represented by the transformation laws of the $\D$'s
\bea [{\mN}_j,\D_0]&=&i\D_j,\;\;\qs {\mN}_j,\D_k\qd=
i\D_0\delta_{jk},\;\;\qs {\mN}_j,\D_4\qd=0.\nn \eea The final
result is that the spinorial realization of boosts is given by:
\bea {\mathcal{N}}_j^T={\mathcal{N}}_j+n_j \eea in which the
finite-dimensional realization of the boosts $n_j$ remains
classical. Moreover, in order to assure (\ref{covariance}), there
are only three consistent possibilities for the matrix $\bt$,
which are $\bt=0,\gamma^0,\gamma^0\gamma^5$.

Actually only $\bt=\gamma^0$ is acceptable; in fact, both for
$\bt=0$ and for $\bt=\gamma^0\gamma^5$ it is easy to check that
our deformed (off-shell) Dirac equation would not reproduce the
correct $\lambda \rightarrow 0$ (classical-spacetime) limit.

We are left with a one-parameter family ($M_D$ is the parameter)
of deformed Dirac equations \be \qs
i\gamma^0\D_0(\mP)+i\gamma^i\D_i(\mP)
+(\D_4(\mP)\frac{\sinh(\lambda m)-\lambda M_D}{\cosh(\lambda
m)-1}+M_D)I\qd \Psi({\x})=0 \label{final5D} \ee As far as we can
see the free parameter $M_D$ does not have physical consequences
(it clearly does not affect the on-shell equation), and it appears
legitimate to view it as a peculiarity associated with the nature
of the (five-dimensional) differential calculus in \kM. Our
formalism allows $M_D =m f(\lambda m)$ with $f$ such that $M_D
\rightarrow m$ whenever $\lambda \rightarrow 0$ and $M_D
\rightarrow 0$ whenever $m \rightarrow 0$. In particular, we can
see from (\ref{A}) that the choice $M_D =\lambda^{-1}\sinh(\lambda
m)$ corresponds to the choice of $\al^4=0$ (because we have
established that $A=\bt=\gamma^0$). We want to notice that in this
case the fifth component $\D_4(\mP)$, coming from the differential
calculus, would not be involved.


\subsection{Massless particles}
Since the on-shell Dirac equation in \kM\ is just the one already
obtained in Subsection~4.2.2, clearly the case of on-shell
massless particles ($m \rightarrow 0$) in \kM\ is also consistent
with the corresponding result already discussed in
Subsection~4.2.2.

Concerning a space-time formulation of the deformed Dirac equation
for massless particles we simply observe that (\ref{final5D}) has
a well-defined $m \rightarrow 0$ limit: \be \qs
i\gamma^0\D_0(\mP)+i\gamma^i\D_i(\mP) \qd \Psi({\x})=0
\label{final5Dnomass} \ee which is therefore well suited for the
description of massless spin-$1/2$ particles.

\subsection{Aside on a possible ambiguity in the derivation of the
Dirac equation in \kM}
 When we introduced
 \bea
d F({\x})=\mathbf{dx}^a{\D}_a(\mP)F({\x})\;\;\;\;a=0,\dots,4 \eea
with \bea {\D}_a(\mP)=\ts\frac{i}{\lambda}[\sinh(\lambda \mP^0)
+\frac{\lambda^2}{2}e^{\lambda \mP^0}\mP^2],i\mP_i e^{\lambda
\mP_0},-\frac{1}{\lambda}(1-\cosh(\lambda \mP_0)
+\frac{\lambda^2}{2}\mP^2 e^{\lambda \mP_0})\td\nn\\
\eea we overlooked an equally valid way of introducing the
exterior derivative operator $d$ of a generic \kM\ element
$F({\x})=\Omega(f(x))$ in terms of the 5D differential calculus:
 \bea
dF({\x})=\E(\mP)_a F({\x}) \mathbf{dx}^a \;\;\;\;a=0,\dots,4 \eea
The vector fields $\E_a(\mP)$ have been determined in Appendix~C
and turns out to be given by the (\ref{result}): \bea
\E_a(\mP)=\ts \frac{i}{\lambda}[\sinh(\lambda \mP^0)
-\frac{\lambda^2}{2}e^{\lambda \mP^0}\mP^2],i\mP_j,
\frac{1}{\lambda}(1-\cosh(\lambda \mP_0)
+\frac{\lambda^2}{2}\mP^2 e^{\lambda \mP_0})\td\nn\\
\eea

There is however a simple relation between $\D(k)$ and $\E(k)$
deformed derivatives: \be \E(k)=-\D(S(k))
\;\;\;S(k)=(-k_0,-e^{\lambda k_0}k_i)\label{relaz} \ee where $S$
is the antipode map, which generalizes the inversion operation in
the way that is appropriate for \kM\ studies, and one can easily
verify that there is no real ambiguity due to the choice of
formulation of the exterior derivative operator $d$. The same
physical Dirac theory is obtained in both cases.

We can rewrite the (\ref{pde:dirac}) in the following way: \be \ts
i\E_0(\mP)+i\al'^i\E_i(\mP)+\al'^4\E_4(\mP)+\bt' M_D'\td
\Psi'(\x)=0 \ee and we can repeat the same procedure of  Section
4.2.2, obtaining a result similar to the (\ref{jodirac1}): \be \qs
i\E_0(\mP)\gamma^0+i\E_i(\mP)\gamma^i-M_D'\gamma^0\bt'[\frac{\lambda
\E_4(\mP)}{1 -\cosh(\lambda m)}-1]-\frac{\E_4(\mP)\sinh(\lambda
m)}{1-\cosh(\lambda m)}I\qd \Psi'({\x})=0 \label{kdiracx2} \ee
Also in this case $\bt'$ can be determined imposing the invariance
of the Dirac operator under the spinorial representation of \kkP .

The spinorial representation of the algebra symmetry of the Dirac
equation (\ref{kdiracx2}) is (see Appendix~D): \be
{\mN}_j^T={\mN}_j+e^{-\lambda P_0}n_j
-\lambda\epsilon_{jkl}P_jm_l\label{kkps2} \ee and $\bt'$ can be
chosen (as in the previous case) like $\bt'=0$ or $\bt'=\gamma^0$
or $\bt'=\gamma^0\gamma^5$. The commutative limit $\lambda=0$
agrees only with $\bt'=\gamma^0$, thus the equation in terms of
$\E(\mP)$ is the following: \be \qs
i\E_0(\mP)\gamma^0+i\E_i(\mP)\gamma^i+[M'_D(1-\frac{\lambda\E_4(\mP)}{1-\cosh(\lambda
m)})-\frac{\E_4(\mP)\sinh(\lambda m)}{1-\cosh(\lambda m)}] I\qd
\Psi'({\x})=0 \ee

It is interesting comparing the two Dirac equation obtained. \\
The first Dirac equation (\ref{final5D}) in the energy-momentum
space is: \be \qs i\D_0(k)\gamma^0+i\D_i(k)\gamma^i +M_D
[\frac{\lambda\D_4(k)}{1-\cosh(\lambda
m)}+1]I-\D_4(k)\frac{\sinh(\lambda m)}{1 -\cosh(\lambda m)}I\qd
\tilde{\psi}(k)=0, \label{dirone}\ee while the second Dirac
equation (\ref{kdiracx2}), taking into account (\ref{relaz}),
becomes \be \qs i\D_0(S(k))\gamma^0
+i\D_i(S(k))\gamma^i-M_D'[\frac{\lambda\D_4(S(k))}{1
-\cosh(\lambda m)}+1]I+\frac{\D_4(S(k))\sinh(\lambda m)}{1
-\cosh(\lambda m)}I\qd \tilde{\psi}'(k)=0. \label{dirtwo}\ee Thus
the second equation can be obtained from the first via the
following substitution: \be \D(P)\to \D(S(k)),\;\;\;
M_D\to-M_D',\;\;\; m\to -m. \ee

On shell the two equation are given by: \bea \qs \frac{e^{\lambda
E}-\cosh(\lambda m)}{\lambda}\gamma^0 +e^{\lambda
E}p_j\gamma^j-\frac{\sinh(\lambda m)}{\lambda} I\qd u(\vec{p})&=&0 \nn\\
\qs \frac{e^{-\lambda E}-\cosh(\lambda m)}{\lambda}\gamma^0
-p_j\gamma^j-\frac{\sinh(\lambda m)}{\lambda} I\qd v(\vec{p})&=&0 \nn\\
&& \nn\\
\qs \frac{e^{-\lambda E}-\cosh(\lambda m)}{\lambda}\gamma^0
-p_j\gamma^j+\frac{\sinh(\lambda m)}{\lambda} I\qd u'(\vec{p})&=&0\nn\\
\qs \frac{e^{\lambda E}-\cosh(\lambda m)}{\lambda}\gamma^0
+e^{\lambda E}p_j\gamma^j+\frac{\sinh(\lambda m)}{\lambda} I\qd
v'(\vec{p})&=&0 \eea The careful reader can see that the strict
relation between these two couple of equations leads to the
description of the same spin $1/2$ particle  theory in \kM .

In summary we can write the differential operator $dF(\x)$ in two
different but equivalent ways putting the one-forms $dx_a$ at the
left or at the right of the deformed derivatives: \be
dF({\x})=dx^{a}\D_a(P)F=\E_a(P)Fdx^{a}, \ee the two forms give
rise to the two forms the Dirac equation (\ref{dirone}) and
(\ref{dirtwo}).
These equations are invariant under the two spinorial
representations of \kkP\ that are different in the boost
generators ${\mN}_j^T$: \bea
{\mN}_j^T&=&{\mN}_j+n_j\nn\\
\bar{{\mN}}_j^T&=&{\mN}_j+e^{-\lambda P_0}n_j
-\lambda\epsilon_{jkl}P_jm_l.\nn \eea The second representation of
the boost generators can be obtained from the first by exchanging
the tensor factors $SO(3,2)\otimes SO(3,2)$ of the spinless
representation times the finite representation of the 5D de-Sitter
algebra before the contraction to \kkP\ (see for example
\cite{NST} where a similar mechanism is shown using the standard
basis of \kkP ).

This fact could reflect some connection between the one-form
generators $dx_a$ and the generators of the Clifford algebra
$\gamma_{\mu}$ through which we express the finite representation
of $SO(3,2)|_{s=1/2}$ in the tensor product \bea
dF=dx^{a}\D_a(P)F&\leftrightarrow& SO(3,2)|_{spinless}\otimes
S0(3,2)|_{s=1/2}\nn\\
dF=\E_a(P)Fdx^{a}&\leftrightarrow& SO(3,2)|_{s=1/2}\otimes
S0(3,2)|_{spinless}, \eea where the notation $SO(3,2)|_{s=0}$
denotes the spinless representation of the 5D de-Sitter algebra
and the notation $SO(3,2)|_{s=1/2}$ denotes the finite dimensional
representation of 5D de-Sitter algebra in terms of the standard
$\gamma_{\mu}$ matrices.

\section{An obstruction for a Dirac equation in \kM\ based
on a four-dimensional differential calculus} In alternative to the
five-dimensional calculus which we have so far considered some
studies (see, {\it e.g.}, Ref.~\cite{AmelinoMajid}) of  \kM\
spacetime have used a four-dimensional differential
calculus\footnote{This four-dimensional differential calculus was
originally obtained as a generalization of a two-dimensional
differential calculus over two-dimensional \kM\ \cite{MO}.}: \be
[{\x}_{\mu},dx_j]=0, \;\;[\x_{\mu},dx_0]=-i\lambda dx_{\mu} \ee
One can then express the derivative operator of the element
$\psi({\x})$ of \kM\ in the following way: \be
d\Psi=\tilde{\partial}_{\mu}\Psi({\x}) dx^{\mu} \ee where
$\tilde{\partial}_{\mu}$ are deformed derivatives that act on the
time-to-the-right-ordered exponential as follows: \bea
&&\tilde{\partial}_je^{-ik{\x}}e^{ik_0{\x}_0}
=\partial_je^{-ik{\x}}e^{ik_0{\x}_0}=
i k_je^{-ik{\x}}e^{ik_0{\x}_0}\nn\equiv d_j(k)e^{-ik{\x}}e^{ik_0{\x}_0}\\
&&\tilde{\partial}_0e^{-ik{\x}}e^{ik_0{\x}_0}=\frac{i}{\lambda}(1-
e^{-\lambda k_0})e^{-ik{\x}}e^{ik_0{\x}_0}\equiv d_0(k)
e^{-ik{\x}}e^{ik_0{\x}_0}\nn \eea

We would like to proceed with this four-dimensional calculus just
as done for the five-dimensional calculus: we write a general
parametrization of a deformed Dirac equation, \be \ts
id_0(k)+id_i(k) \rho^i+M_D' \sigma \td \tilde{\psi}_{\kappa}(k) =0
\label{dirack4} \ee where $\rho^i,\sigma$ are four matrices
(constant or at most dependent on $\lambda m$) to be determined by
imposing that an on-shell ``plane wave" (with $\cosh(\lambda
E)-\lambda^2e^{-\lambda E}k^2=\cosh(\lambda m)$) is solution of
the deformed Dirac equation and by imposing covariance in the
\kkP\ sense.

The requirement that an on-shell plane wave is a solution leads to
\bea
 &&\qs d^2_0(E,p)-(\rho^i)^2(d_i(E,p))^2+{M_D'}^2\sigma^2+\right.\nonumber\\
&&-\left.\sum_{i<j}\{\rho^i,\rho^j\}d_i(E,p)d_j(E,p)+M_D'\{\rho^i,\sigma\}
d_i(E,p)\qd u(\vec{p})=0 \label{condizione4} \eea from which one
derives as necessary conditions: \be
\{\rho^i,\rho^j\}=0,\;\;\{\rho^i,\sigma\}=0,\;\;
(\rho^1)^2=(\rho^2)^2=(\rho^3)^2 \label{joc64} \ee These
conditions are necessary but not sufficient, and actually there is
no choice of the matrices $\rho^i,\sigma$ of the type that we are
seeking that allows to satisfy (\ref{condizione4}) for all values
of the momentum $p$. To see this let us use (\ref{joc64}) to
rewrite (\ref{condizione4}) as \be (d_0^2(E,p)I-d_i(E,p)^2 {\cal
Q}+{M_D'}^2\sigma^2)u(\vec{p})=0 \label{con} \ee where ${\cal Q}
\equiv (\rho^1)^2=(\rho^2)^2=(\rho^3)^2$. In this Eq.~ (\ref{con})
we are left with two unknown matrices, ${\cal Q},\sigma$, to be
determined, and it is easy to see that there is no choice of
${\cal Q},\sigma$ that allows to satisfy (\ref{con}) for all
values of the momentum $p$. For example, by looking at the form of
the equation for $p=0$ (and $E=m$) one is forced to conclude that
\be \sigma^2= - d_0^2(m)I/M_D^2(m) \ee but then, with this choice
of $\sigma^2$, Eq.~(\ref{con}) turns into an equation for ${\cal
Q}$ which does not admit any solution of the type we are seeking:
\be (d_0^2(E)-d_0^2(m)-d_i^2 {\cal Q})=-(1-e^{-\lambda E})^2
+(1-e^{-\lambda m})^2+\lambda^2p^2 {\cal Q}=0 \ee {\it i.e.}
(using again the dispersion relation) \be {\cal Q} =
[(1-e^{-\lambda E})^2-(1-e^{-\lambda m})^2][( e^{2\lambda
E}+1)-2e^{\lambda E}\cosh(\lambda m)]^{-1} \ee

What we have found is that there is no choice of
energy-momentum-independent matrices $\rho^i,\sigma$ that can be
used in to obtain a consistent Dirac equation for \kM . The
analogous problem for the 5D calculus did have a perfectly
acceptable solution. Here, with the four-dimensional differential
calculus, we would be led to consider energy-dependent matrices
$\rho^i,\sigma$ but this is unappealing on physical grounds and in
any case the fact that this awkward assumption can be avoided in
the five-dimensional calculus appears to be a good basis for
preferring the five-dimensional calculus over the four-dimensional
calculus.


\chapter{Conclusions}
In this thesis we have investigated some issues that are relevant
for the possibility to construct physical theories on the \kM\
noncommutative spacetime. As discussed in Chapter 1, this
spacetime exhibits some interesting properties that make it a good
candidate for a quantum-spacetime description of Quantum Gravity
in the zero-curvature limit.

We have analyzed three main problems that are at the basis of the
construction of a well-defined field theory in \kM : the
definition of a formulation of the notion of "field", which should
describe a particle as in the commutative case; the description of
the symmetries of theories on such a spacetime; and the
construction of the equations of motion for particles.\\

 The first point, analyzed in Chapter 2, concerns the
generalization of the notion of classical field from the
commutative Minkowski to the noncommutative \kM\ spacetime.
Following the procedure adopted in the case of Quantum Mechanics,
which is based on the introduction of Weyl systems/maps, we have
generalized the Weyl formalism to the more general case of
Lie-algebra type noncommutative space, in which \kM\ is included.
In this way we have introduced fields in \kM\ through a
generalized Weyl map. The generalization of the Weyl formalism has
allowed us to develop a general method to construct deformed
products in Lie-algebra type of noncommutative spacetimes,
possibly including central extension. These products are
generalizations of the well-known Moyal-Weyl star product of
Quantum Mechanics. As in the case of Quantum Mechanics (or
canonical noncommutative spacetimes), they may be useful in the
analysis of field theory also in \kM\ .  We have clarified that
the different star products used in literature in order to
describe \kM\ correspond to choosing certain Weyl maps (equivalent
to choosing a description of the \kM\ functions with an ordering
prescription of the time variable with respect to the spatial
ones). In this way, our analysis has clarified the definition of
different star products for \kM\  that had been employed in
literature. This enabled us to perform a comparative study of the
different star products, and to discover several relations among
them. We have also discovered a relation between the structure of
the star products and the structure of  the coproducts and other
features which characterize the $\kappa$-Poincar\'{e} quantum
algebra.

 The analysis of the second point, \emph{i.e.} the analysis of the
symmetries of our space, has provided insight on some questions,
and raised new questions, for the study of physical theories in
certain types of noncommutative spacetimes. In Chapter 3 we have
introduced a concept of noncommutative-spacetime symmetry, which
follows very closely the one adopted in commutative spacetimes,
and is analyzed most naturally in terms of a Weyl map introduced
in Chapter 2. We found that for a specific simple theory, a theory
describing a free scalar field in \kM, it was impossible to find a
formulation that would admit invariance under ordinary (classical)
Poincar\'{e} transformations.  More importantly for the key
objectives of our analysis, we did find (Section~3.2)
10-generators symmetries of our description of a free scalar field
in \kM, and these symmetries admitted formulation in terms of
Hopf-algebra (quantum) versions of the Poincar\'{e} symmetries.
Although our analysis allowed us to reduce the amount of ambiguity
in the description of the symmetries of theories in these
noncommutative spacetimes, we are left with a choice between
different realizations of the concept of translations in the
noncommutative spacetime. We have clarified that such an ambiguity
might have to be expected on the basis of the type of coordinate
noncommutativity here considered, but it remains to be seen
whether by appropriate choice of the action of the theory one can
remove the ambiguity, {\it i.e.} construct a theory which is
invariant under one specific type of translations and not under
any other type. The fact that in this first exploratory
application of our description of symmetries we only considered a
free scalar theory in \kM\ spacetime might be a significant
limitation. In fact, our concept of symmetries applies directly to
theories and not to the underlying spacetime on which the theories
are introduced, and therefore one may expect different results for
different theories (even restricting our attention to theories
10-generator Poincar\'{e}-like symmetries). This might represent
an opportunity for attempts to solve the issue of the ambiguity
concerning the description of translations. One hypothesis that
deserves investigation in future studies is the one that perhaps
the theory we considered does not have enough structure to give
proper physical significance to energy-momentum. One natural
context in which to explore this issue might be provided by
attempting to construct gauge theories in \kM\ spacetime following
the approach here advocated. In the study of gauge theories in
canonical noncommutative spacetime it has emerged that gauge
transformations and spacetime transformations are deeply
connected. It therefore seems important to aim for the
construction of gauge theories in \kM\ spacetime, which might have
a deep role in clarifying the status of the energy-momentum
observables.

The third point concerns the construction of the equations of
motion for some theories that one can introduce in \kM . In
Chapter 4, we have focused on the theories for scalar and
spin-$1/2$ free particles. The equation of motion for scalar free
particles is obtained (Section~4.1) starting from the
maximally-symmetric action constructed in Chapter 3 for a scalar
theory, and using a natural extension of the variational principle
to \kM . In the case of spin-$1/2$ particles a wave equation has
been constructed following the Dirac original procedure for
spinorial particles. In particular, our analysis has revealed that
the correspondence between the usual Dirac operator (that enters
the standard Dirac equation in the commutative case) and the
ordinary differential calculus on the commutative Minkowski space
has an analogous in the correspondence between a deformed-Dirac
operator (that is showed to enter in the deformed-Dirac equation
in \kM ) and a five-dimensional differential calculus in \kM .
Thus, this result has emphasized the role of such a differential
calculus (the unique one having certain covariance properties, as
discussed in Section~4.2) with respect to the others that one can
construct on \kM . A noteworthy observation about the
deformed-Dirac equation obtained with our technique in \kM\ is the
agreement with the Dirac wave equation obtained in the framework
of DSR theories (discussed in Subsection~1.4). This result
encourages the idea on the possibility that the \kM\ spacetime
might provide an example of quantum spacetime in which DSR are
present.

Our work opens interesting questions in the study of the
construction of theories on \kM\ noncommutative spacetime. As
already mentioned, the necessity of facing the study of gauge
theories might represent a key point in reducing the ambiguity
concerning the translation characterization. On the side of the
construction of the equations of motion for particles in \kM\ our
analysis has been restricted only to free theories, on which we
have not encountered any conceptual or formal obstruction. The
next step would be the description of interacting theories using
the tools (Weyl maps, integrals, star products) that we have
introduced in order to deal with the free theories. It would be
interesting to investigate in which way (if any) the predictions
on the energy-momentum space of a theory for two interacting
particles would be connected with the coproduct structures that
describe the composition of representations from a Quantum Group
perspective.

\addcontentsline{toc}{chapter}{Acknowledgements}
\chapter*{Acknowledgements}

First and foremost, I wish to thank my advisors Giovanni
Amelino-Camelia and Fedele Lizzi, for their guidance and
encouragement, and for their contribution to my thesis. I must
thank Michele Arzano, Francesco D'Andrea and Alessandro Zampini,
who have been co-adventurers with me in discovering the results
reported in this work. I also thank Gianluca Mandanici and
Patrizia Vitale: I have much benefitted from the conversations
with them. A special thank is deserved to Gaetano Fiore for the
extraordinary patience and care in reading my thesis. I thank the
University of Naples ``Federico II" and INFN (Sez. Naples) for
giving me the opportunity of participating in Schools and
Conferences. In particular I thank the PhD coordinator Antonio
Sciarrino. I want also thank Guido Celentano that has been
infinitely helpful to me in his effort to make the bureaucracies
tolerable. I wish to thank my graduation-thesis advisor Francesco
Guerra for the help and advises that has given me in these years.
Finally, I thanks the PhD staff of the University of Roma ``La
Sapienza" for their great hospitality.


\newpage

\setcounter{section}{0}
\appendix{Calculation of the $*_3$ and $*_4$ products of two exponentials\label{appstar34}}
In this appendix we give some details of the calculations of the
product of two (ordinary) exponential functions.
We start from the form~\eqn{eq:Moyal} of the Moyal product of two functions $f,g$ on $\real^{6}$:
\be
f(u)\star g(u)=\frac{1}{(2\pi)^6}\int d^6s d^6t e^{isu}e^{-i\frac{\lambda}{2}sJ_6t}\tilde{f}(s-t)\tilde{g}(t)
\ee
where $u=(q_1,...p_{3})$ and $J$ denotes the antisymmetric matrix:
$$
{J_{6}}=\left(
\begin{array}{ccc}
0& &I_3\\
-I_3& &0
\end{array}
\right)
$$
Let us express $u=(\vec{q},\vec{p})$ and introduce the following
notation:
\bea
u&=&(\vq,\vp)\nn\\
s&=&(\vs_1,\vs_2)\nn\\
t&=&(\vt_1,\vt_2)
\eea
in which all vectors belong to $\real_3$.
We find an equivalent form of the Moyal product in terms of the functions $f(u)$ and $g(u)$:
\bea
f(u)\star g(u)&=&\frac{1}{(2\pi)^{12}}\int d^6s d^6t^6\sigma d^6\tau e^{isu}e^{-i\frac{\lambda}{2}sJ_6t}e^{-i\sigma (s-t)}e^{-i\tau t}f(\sigma)g(\tau)\nn\\
&=&\frac{1}{(2\pi)^{12}}\int d^6s d^6td^6\sigma d^6\tau e^{is(u-\frac{\lambda}{2}J_6t-\sigma)+it(\sigma-\tau)}f(\sigma)g(\tau)\nn\\
&=&\frac{1}{(2\pi)^{6}}(\frac{2}{\lambda})^6\int d^6td^6\sigma d^6\tau \delta^{(6)}(t+\frac{2}{\lambda}J_6(u-\sigma))e^{it(\sigma-\tau)}f(\sigma)g(\tau)\nn\\
&=&\frac{1}{(2\pi)^{6}}(\frac{2}{\lambda})^6\int d^6\sigma d^6\tau e^{-i\frac{2}{\lambda}(\sigma-\tau)J_6(u-\sigma)}f(\sigma)g(\tau)\nn
\eea
which, with the substitutions $\tau=u+t, \sigma=u-\frac{\lambda}{2}J_6s$, can be put in the following form:
\be
f(u)\star g(u)=(2\pi)^{-6}\int d^6sd^6t
f(u-\frac{\lambda}{2}J_{6}s)g(u+t)e^{-ist} \label{moyal}
\ee
Using the integral form for a Moyal product~(\ref{moyal}), the deformed (six
dimensional) product of two exponential is:
\be
e^{ik^{\mu}x_{\mu}}\star e^{il^{\mu}x_{\mu}}=(2\pi)^{-6} \int
d^6sd^6t e^{ik^{\mu}x_{\mu}(u-\frac{\lambda}{2}
J_6s)}e^{il^{\mu}x_{\mu}(u+t)} e^{-ist}
\ee
with (using~\eqn{jsmap} for the last step):
\be
ik^{\mu}x_{\mu}(u)\equiv ikx=ik^0x_0-i\vk{\cdot}\vx=-i(k^0\vq
{\cdot}\vp+\vk {\cdot}\vq)
\ee
the arguments of the $x$'s become
\be
u-\frac{\lambda}{2}J_6s=(\vq-\frac{\lambda}{2}\vs_2,\vp-\frac{\lambda}{2}\vs_1)
\ee
\bea
ik^{\mu}x_{\mu}(u-\frac{\lambda}{2}J_6s)
&=&-i[k^0(\vec{q}-\frac{\lambda}{2}\vec{s_2})
(\vec{p}+\frac{\lambda}{2}\vec{s_1})+\vec{k}{\cdot}(\vec{q}-\frac{\lambda}{2}\vec{s_2})]\nonumber \\
&=&-i(k^0\vq{\cdot}\vp+\vk{\cdot}\vq)+i\frac{\lambda}{2}(k^0(\vs_2{\cdot}\vp-\vq{\cdot}\vs_1)+\vk{\cdot}\vs_2)-
i\frac{\lambda^2}{4}k^0\vs_1{\cdot}\vs_2 \nonumber \\
&=&ikx+i\frac{\lambda}{2}(k^0(\vs_2{\cdot}\vp-\vq{\cdot}\vs_1)+\vk{\cdot}\vs_2)+i\frac{\lambda^2}{4}k^0\vs_1{\cdot}\vs_2
\eea
\bea
il^{\mu}x^{\mu}(u+t)&=&-i[l^0(\vec{q}+\vt_1)(\vp+\vec{t_2})+\vec{l}{\cdot}(\vec{q}+\vec{t_1})]\nn\\
&=&ilx-i(l^0(\vq{\cdot}\vt_2+\vp{\cdot}\vt_1+\vt_1{\cdot}\vt_2)+\vl{\cdot}\vt_1)
\eea
and
\bea
e^{ik^{\mu}x_{\mu}}&\star
&e^{il^{\mu}x_{\mu}}=(2\pi)^{-6}e^{i(k+l)x} \int
d\vec{s}_1d\vec{s}_2d\vec{t}_1d\vec{t}_2
e^{i\frac{\lambda}{2}(k^0(\vs_2{\cdot}\vp-\vq{\cdot}\vs_1)+\vk{\cdot}\vs_2)+i\frac{\lambda^2}{4}k^0\vs_1{\cdot}\vs_2
}\nn\\
&&e^{-i(l^0(\vq{\cdot}\vt_2+\vp{\cdot}\vt_1+\vt_1{\cdot}\vt_2)+\vl{\cdot}\vt_1)}
e^{-i(\vec{s}_1{\cdot}\vec{t}_1+\vec{s}_2{\cdot}\vec{t}_2)}\eea
Reordering the exponentials:
\bea
e^{ik^{\mu}x_{\mu}}\star
e^{il^{\mu}x_{\mu}}&=&(2\pi)^{-6}e^{i(k+l)x}
\int d\vec{s}_1d\vec{s}_2d\vec{t}_1d\vec{t}_2e^{-i\vs_1(\frac{\lambda}{2}k^0\vq+\frac{\lambda^2}{4}k^0\vs_2-\vt_1)}\nn\\
&&e^{-i\vt_2(l^0\vt_1+l^0\vq+\vs_2)}
e^{i\frac{\lambda}{2}\vs_2(k^0\vp+\vk)-i\vt_1{\cdot}(\vl+l^0\vp)}
\eea
at this point we can make the integration in the $\vs_1$ e $\vt_2$
variables:
\bea
\frac{1}{(2\pi)^3}\int d\vs_1
e^{i\vs_1(-\frac{\lambda}{2}k^0\vq+\frac{\lambda^2}{4}k^0\vs_2-\vt_1)}&=&
\delta^{(3)}(-\frac{\lambda}{2}k^0\vq+\frac{\lambda^2}{4}k^0\vs_2-\vt_1)\nn\\
\frac{1}{(2\pi)^3}\int d\vt_2
e^{-i\vt_2(l^0\vq+l^0\vt_1+\vs_2)}&=&\delta^{(3)}(l^0\vq+l^0\vt_1+\vs_2)
\eea
to obtain:
\bea
e^{ik^{\mu}x_{\mu}}\star e^{il^{\mu}x_{\mu}} &=&e^{i(k+l)x}\int
d\vs_2d\vt_1\delta^{(3)}
(-\frac{\lambda}{2}k^0\vq+\frac{\lambda^2}{4}k^0\vs_2-\vt_1)\delta^{(3)}(l^0\vq+l^0\vt_1+\vs_2)\nonumber \\
&&e^{i\frac{\lambda}{2}\vs_2(k^0\vp+\vk)-i\vt_1{\cdot}(\vl+l^0\vp)}
\eea
and making integral in  $d\vt_1$ we have:
\bea
e^{ik^{\mu}x_{\mu}}\star e^{il^{\mu}x_{\mu}} &=&e^{i(k+l)x}\int
d\vs_2
e^{i\frac{\lambda}{2}\vs_2{\cdot}(k^0\vp+\vk)-i\frac{\lambda^2}{4}k^0\vs_2(\vl+l^0\vp)
+i\frac{\lambda}{2}k^0\vq(\vl+l^0\vp)}\nn\\
&&\delta^{(3)}(\vs_2(1+\frac{\lambda^2}{4}k^0l^0)+l^0(1-\frac{\lambda}{2}k^0)\vq)
\eea
Using $a$ and $b$ defined in~\eqn{defab}:
\bea
a(k^0,p^0)&=&1+\frac{\lambda^2}{4}k^0p^0\nn\\
b(k^0)&=&1-\frac{\lambda}{2}k^0,
\eea
we make the last integration to obtain:
\be
\delta(\vs_2a(k^0,l^0)+l^0b(k^0)\vq)=
\frac{1}{|a(k^0,l^0)|^3}\delta(\vs_2+l^0\frac{b(k^0)}{a(k^0,l^0)}\vq)
\ee
which fixes $\vs_2=-l^0\frac{b}{a}\vq$, so that the integral becomes:
\bea
&&\frac{1}{|a(k^0,l^0)|^3}e^{-i\frac{\lambda}{2}l^0\frac{b}{a}
\vq{\cdot}(k^0\vp+\vk)+i\frac{\lambda^2}{4}k^0l^0\frac{b}{a}\vq(\vl+l^0\vp)+
i\frac{\lambda}{2}k^0\vq(\vl+l^0\vp)}\nn\\
&=&\frac{1}{|a(k^0,l^0)|^3}e^{-i\frac{\lambda}{2}l^0\frac{b}{a}
(-k^0x_0+\vk{\cdot}\vx)+i\frac{\lambda^2}{4}k^0l^0\frac{b}{a}(\vl{\cdot}\vx-l^0x_0)+
i\frac{\lambda}{2}k^0(\vl{\cdot}\vx-l^0x_0)}\nn\\
&=&\frac{1}{|a(k^0,l^0)|^3}e^{i\frac{\lambda}{2}l^0\frac{b}{a}
kx-i\frac{\lambda^2}{4}k^0l^0\frac{b}{a}lx-i\frac{\lambda}{2}k^0lx}\nn\\
&=&\frac{1}{|a(k^0,l^0)|^3}e^{i\frac{\lambda}{2a}
(l^0bk-\frac{\lambda}{2}k^0l^0bl+\frac{\lambda}{2}k^0la)x}\nn\\
&=&\frac{1}{|a(k^0,l^0)|^3}e^{i\frac{\lambda}{2a}(l^0b(k^0)k-k^0b(-l^0)l)x}
\eea The final result is: \be e^{ikx}\star
e^{ilx}=\frac{1}{|a(k^0,l^0)|^3}e^{i(k+l)x}e^{i\frac{\lambda}{2a(k^0,l^0)}
(l^0b(k^0)k-k^0b(-l^0)l)x} =\frac{1}{|a(k^0,l^0)|^3}e^{i(k\oplus_3
l)x}\label{new} \ee It results also $\bar k=-k$. From
relation~\eqn{new} can be seen that the function $e^{ikx}$ is not
unitary for the product~$*_3$, and to make it unitary one should
renormalize it dividing by
$|a(k^0,k^0)|^{3/2}$, thus finding~\eqn{W3def}.\\

Following the same procedure of the previous calculation
we work out of the $*_4$ product of two exponentials.
In this case we consider the map given by
relation~\eqn{seleneprod'}, which gives rise to the product~$*_4$.
Again with the use of~\eqn{moyal} we calculate the product among
exponential functions:
\be
f(u)=e^{ikx},\ \ \ g(u)=e^{ilx}
\ee
where $x=x(u)$ and $u,s,t$ are defined as in the previous
appendix. We have then:
\bea
ikx(u-\frac{1}{2}\lambda
Js)&=&-i\left(k^0\sum_{i=1}^3(p_i+\frac{\lambda}{2} s_{1i})+
\sum_{i=1}^3k_ie^{q_i-\frac{\lambda}{2} s_{2_i}}\right)\nn\\
&=&k^0x_0+i\sum_{i=1}^3\left(-\frac{\lambda}{2}k^0s_{1i}-k_ix_ie^{-\frac{\lambda}{2}s_ {2i}}\right)\\
ilx(u+t)&=&-i\sum_{i=1}^3\left(l^0(p_i+t_{2i})+l_ie^{q_i+t_{1i}}\right)\nn\\
&=&l^0x_0-i\sum_{i=1}^3[l^0t_{2i}+l_ix_ie^{t_{1i}}]
\eea
Performing the integral:
\bea
e^{ikx}*_4e^{ilx}=&e^{i(k^0+l^0)x_0}&\int ds dt
e^{i\sum_i\left(-\frac{\lambda}{2}k^0s_{1i}-k_ix_i
e^{-\frac{\lambda}{2}s_{2i}}\right)}
e^{i\sum_i^3\left(-l^0t_{2i}-l_ix_ie^{t_{1i}}\right)}e^{-i\sum_i^3(s_{1i}t_{1i}+s_{2i}t_{2i})}\nonumber\\
=&e^{i(k^0+l^0)x_0}&\prod_i^3\int ds_{2_i}dt_{1_i}\delta(-\frac{\lambda}{2}k^0-t_{1i})
\delta(l^0+s_{2i})e^{-ik_ix_ie^{-\frac{\lambda}{2}s_{2i}}}e^{-il_ix_ie^{t_{1i}}}\nn\\
=&e^{i(k^0+l^0)x_0}&\prod_i^3[e^{-ik_ix_ie^{\frac{\lambda}{2}l^0}}e^{-il_ix_ie^{-\frac{\lambda}{2}k^0}}]\nn\\
=&e^{i(k^0+l^0)x_0}&[e^{-i\vk\vx e^{\frac{\lambda}{2}l^0}}e^{-i\vl\vx e^{-\frac{\lambda}{2}k^0}}]\nn\\
=&e^{i(k\oplus_4 l)x}&
\eea

\appendix{Deformed boost generators\label{appdefboosts}}
In this appendix we report some aspects of the analysis necessary
in order to construct the boost generators ${\mathcal{N}}^R$.
Analogous techniques can be used for ${\mathcal{N}}^S$, but here
we focus on ${\mathcal{N}}^R$, and, since we therefore always
refer to the ``right-ordered" $\Omega_R$ map, the label $R$ is
omitted.

The starting point for the analysis reported in Subsection~3.3.4
was the 7-generator Hopf algebra with $P_{\mu}$
and $M_j$ generators, which we note again here for convenience:
\bea
[P_\mu,P_\nu]=0 ~, ~~~~~[M_j,P_0]=0, &\;\;& [M_j,P_l]=i\varepsilon_{jlm}P_m
 \nn\\
\Delta P_{\mu}=P_{\mu}\otimes 1+e^{\lambda P_0(\delta_{\mu 0}-1)}\otimes P_{\mu}&&\Delta M_j
=M_j\otimes 1+1\otimes M_j\nn
\eea
and they act on the right-ordered map in the following way:
\bea
P_{\mu}\Omega(e^{ikx})=\Omega(-i\partial_{\mu}e^{ikx})&&M_j\Omega(e^{ikx})=\Omega(i\epsilon_{jkl}x_k\partial_le^{ikx})\label{eq:appendix}
\eea
As showed in Subsection~3.3.4, one cannot extend this 7-generator algebra
to a 10-generator algebra by adding ordinary (classical) boost generators.
But one can obtain a 10-generator Hopf algebra by introducing deformed
boost generators ${\mathcal{N}}$, and in particular it is possible
to do so while leaving the Lorentz-sector commutation relations
unmodified
\bea
&&[M_j,M_k]=i\varepsilon_{jkl}M_l\nn\\
&&\qs {\mathcal{N}}_j,M_k\qd=i\varepsilon_{jkl}{\mathcal{N}}_l\nn\\
&&\qs {\mathcal{N}}_j,{\mathcal{N}}_k\qd=-i\varepsilon_{jkl}M_l ~.
\label{com:NN}
\eea
We intend to show this explicitly here.

Let us start observing that
the most general form in which the commutation
relations among ${\mathcal{N}}_j$
and $P_l$ can be deformed (consistently with the underlying space-rotation
symmetries) is
\bea
&&[{\mathcal{N}}_j,P_0]=iA(P)P_j\nn\\
&&\qs {\mathcal{N}}_j,P_l\qd=i\lambda^{-1}B(P)\delta_{jl}+i\lambda
C(P)P_jP_l +iD(P)\varepsilon_{jlm}P_m\label{com:PN} \eea where
$A,B,C,D$ are unknown dimensionless functions of $\lambda P_0$ and
$\lambda^2\vec{P}^2$. In the classical limit
$A(P)=1,\;\;\lambda^{-1}B(P)=P_0,\;\;D=0$ and $C$ can take any
form, as long as it is finite, since $\lambda C(P)$ must vanish in
the classical $\lambda \rightarrow 0$ limit.

From these relations it is easy to find that ${\mathcal{N}}_j$ can be
represented as differential operators inside the $\Omega$ map:
\bea
{\mathcal{N}}_j \Omega(\phi(x))=\Omega\{[-ix_0A(-i\partial_x)\partial_j
-\lambda^{-1}x_jB(-i\partial_x)+\lambda x_lC(-i\partial_x)\partial_l\partial_j
+i\epsilon_{jkl} x_k D(-i\partial_x)\partial_{l}]\phi(x)\}~. \nn
\eea
Using (\ref{eq:appendix}), this expression can be put in the form
\bea
{\mathcal{N}}_j \Omega(\phi(x))=\Omega\{[x_0A(P)P_j
-\lambda^{-1}x_jB(P)-\lambda x_lC(P)P_lP_j+M_jD(P)]\phi(x)\}~ \nn
\eea
and, using the relation (\ref{eq:uno}) $\Omega_R(x_0f)=\Omega_R(f)\x_0=[{\x}_0-\lambda\vec{\x}\vec{P}]\Omega_R(f)$, on obtains
\bea
{\mathcal{N}}_j \Omega(\phi)&=&[\x_0P_jA(P)-\lambda \x_lP_lP_jA(P)
-\lambda^{-1}\x_jB(P)-\lambda x_lP_lP_jC(P)+M_jD(P)]\Omega(\phi)~ \nn\\
&=&[\x_0P_jA-\lambda^{-1}\x_jB-\lambda \x_lP_lP_j(A+C)
+M_jD]\Omega(\phi)\nn\\
&=&[\x_0P_jA-\lambda^{-1}\x_jB \pm\lambda\x_j P_lP_l(A+C)-\lambda\x_lP_jP_l(A+C)
+M_jD]\Omega(\phi)\nn\\
&=&[\x_0P_jA-\lambda^{-1}\x_j(B-\lambda^2P^2(A+C))-\lambda(\x_lP_j-\x_jP_l)P_l(A+C)
+M_jD]\Omega(\phi)\nn\\
&=&[\x_0P_jA-\lambda^{-1}\x_j(B-\lambda^2P^2(A+C))+\lambda\epsilon_{jkl}M_kP_l(A+C)
+M_jD]\Omega(\phi)\nn
\eea
Then, one can rewrite ${\mathcal{N}}_j$ as
\be
{\mathcal{N}}_j=\Big(-\lambda^{-1}{\x}_jZ(P)+{\x}_0P_j\Big)A(P)+\lambda
V(P)\e_{jkl}P_kM_l+D(P)M_j \label{ZAVD}
\ee
where $V\equiv A+C$ and $Z \equiv (B-\lambda^2\vec{P}^2V)A^{-1}$.

In preparation for the rest of the analysis, we note here that
\begin{eqnarray*}
&& {\x}_j\Omega_R(f)=\Omega_R(x_jf)=\Big(e^{\lambda P_0}\Omega_R(f)\Big){\x}_j
\end{eqnarray*}
and we introduce the useful notations $g_p=e^{-ip_1{\x}_1}e^{ip_0{\x}_0}$
and
$p \cplus q=(p_0+q_0,p_1+q_1e^{-\lambda p_0},0,0)$, so that $g_p
g_q=g_{(p \cplus q)}$. For a generic scalar function it will be
implictly assumed that it depends on the operators $P_0$ and $P^2$
(so for example, with $A$ we will denote $A(P_0,P^2)$). A notation
of type $A(q_0,q^2)$ will be reserved to functions which depend on
a real four-vector $q$.

Since
\begin{displaymath}
M_1g_p=P_2g_p=P_3g_p=0
\end{displaymath}
it is easy to verify that
\be
{\mathcal{N}}_1\,g_{(p\cplus q)}=A(p\cplus q)\Big(-\lambda^{-1}Z(p\cplus
q){\x}_1+{\x}_0(p_1+q_1e^{-\lambda p_0})\Big)\,g_{(p\cplus q)} \label{unidim}
\ee
and we can remove ${\x}_0$ from the previous expression using the identity (for $q=0$)
\begin{displaymath}
p_1{\x}_0g_p=(\lambda^{-1}{\x}_1Z+{\mathcal{N}}_1A^{-1})g_p
\end{displaymath}
In fact
\bea
\x_0(p_1\cplus q_1)g_{p\cplus q}&=&(p_1\x_0g_p)g_q+e^{-\lambda p_0}q_1\x_0g_pg_q\nn\\
&=&(p_1\x_0g_p)g_q+e^{-\lambda p_0}g_p(q_1\x_0g_q)-\lambda e^{-\lambda p_0}q_1p_1\x_1g_pg_q\nn\\
&=&[(\lambda^{-1}{\x}_1Z+{\mathcal{N}}_1A^{-1})g_p]g_q+e^{-\lambda p_0}g_p[\lambda^{-1}{\x}_1Z+{\mathcal{N}}_1A^{-1}]g_q-\lambda e^{-\lambda p_0}q_1p_1\x_1g_pg_q\nn\\
&=&\lambda^{-1}{\x}_1[Z(p)+e^{-2\lambda p_0}Z(q)-\lambda^2e^{-\lambda p_0}q_1p_1] g_pg_q
+[{\mathcal{N}}_1A^{-1}g_p]g_q+e^{-\lambda p_0}g_p[{\mathcal{N}}_1A^{-1}g_q]\nn\\
\eea
Substituting this term in (\ref{unidim}) we have:
\bea
{\mathcal{N}}_1\,g_{(p\cplus q)}&=&\lambda^{-1}\x_1A(p\cplus q)[-Z(p\cplus q)+ Z(p)+e^{-2\lambda p_0}Z(q)-\lambda^2e^{-\lambda p_0}q_1p_1]g_{p\cplus q}\nn\\
&&+A(p\cplus q)\Big([{\mathcal{N}}_1A^{-1}g_p]g_q+e^{-\lambda
p_0}g_p[{\mathcal{N}}_1A^{-1}g_q]  \Big)\,g_{(p\cplus q)} \nn \eea
that can be put in a more compact form:
\begin{eqnarray*}
{\mathcal{N}}_1\,g_{(p\cplus q)} &=&+\lambda^{-1}{\x}_1
A(p\cplus q)\left\{-Z(p\cplus q)+Z(p)+e^{-2\lambda p_0}Z(q)
-\lambda^2e^{-\lambda p_0}p_1q_1\right\}g_{p+\cplus q}\nn\\
&&+({\mathcal{N}}_1
A^{-1}A_{(1)}g_p)(A_{(2)}g_q)+(e^{-\lambda
P_0}A_{(1)}g_p)({\mathcal{N}}_1 A^{-1}A_{(2)}g_q)
\end{eqnarray*}
We notice that it is impossible to eliminate ${\x}_1$
from this expression without reintroducing ${\x}_0$,
and therefore, if we
want the second member to be function
only of ${\mathcal{N}}_j$ (and of $P_0$, $P_1$ and
operators with null action on $g_p$),
the factor which multiplies ${\x}_1$ must
vanish identically. For this, it is necessary and sufficient that
\begin{equation}\label{eq:cond}
Z\Big(p_0+q_0,(p_1+q_1e^{-\lambda p_0})^2\Big)=Z(p_0,p_1^2)+e^{-2\lambda
p_0}Z(q_0,q_1^2)-\lambda^2e^{-\lambda p_0}p_1q_1
\end{equation}
for each $p$ and $q$.

We must now find a solution of (\ref{eq:cond}).
Applying $\partial_{p_1}\partial_{q_1}$ to (\ref{eq:cond}) we obtain
\begin{displaymath}
\partial_{p_1}\partial_{q_1} Z\Big(p_0+q_0,(p_1
+q_1e^{-\lambda p_0})^2\Big)
=-\lambda^2e^{-\lambda p_0}
\end{displaymath}
In the first member we can substitute $\partial_{q_1}$
with $e^{-\lambda p_0}\partial_{p_1}$, and
calculate the resulting expression in $q_0=q_1=0$,
leading us to
\begin{displaymath}
\partial_{p_1}^2Z(p_0,p_1^2)=-\lambda^2
\end{displaymath}
This can be integrated to
\begin{displaymath}
Z(p_0,p_1^2)=W(\lambda p_0)+\beta\lambda p_1-\lambda^2p_1^2/2
\end{displaymath}
with $\beta$ and $W$ arbitrary.

Since $Z$ is function only of $p_0$ and $p_1^2$, $\beta$
must be zero. Substituting in (\ref{eq:cond}) we get the condition
\begin{displaymath}
W(\lambda p_0+\lambda q_0)=W(p_0)+e^{-2\lambda p_0}W(\lambda q_0)
\end{displaymath}
For $p_0=q_0=0$ we have $W(0)=2W(0)$, that is $W(0)=0$.

Applying
$\partial_{\lambda q_0}$, using the identity $\partial_{\lambda q_0}W(\lambda
p_0+\lambda q_0)=\partial_{\lambda p_0}W(\lambda p_0+\lambda q_0)$ and
calculating in $q_0=0$ we obtain
\begin{displaymath}
\partial_{\lambda p_0}W(\lambda p_0)
=e^{-2\lambda p_0}\partial_{\lambda q_0}W(\lambda q_0)\big|_{q_0=0}
\end{displaymath}
which integrated, with the condition $W(0)=0$, gives
\begin{displaymath}
W(\lambda p_0)=\frac{1-e^{-2\lambda p_0}}{2}\,\partial_{\lambda p_0}W(\lambda
p_0)\big|_{p_0=0} ~.
\end{displaymath}

From the request that ${\mathcal{N}}_1$ has the correct classical limit,
 we derive $\partial_{\lambda p_0}W(0)=1$, and this allows us to
 write
 \begin{displaymath}
Z(p_0,p_1^2)=\frac{1-e^{-2\lambda p_0}}{2}+\frac{\lambda^2}{2}p_1^2 ~.
\end{displaymath}

Since $Z$ is a scalar, the knowledge of $Z(p_0,p_1^2)$ is equivalent to the
knowledge of $Z(p_0,p^2)$ for a generic $\vec{p}$
(just renaming $p_1^2\to p^2$).
We have therefore established that
\begin{displaymath}
{\mathcal{N}}_j=\tilde{{\mathcal{N}}}_jA+\lambda (A+C)\e_{jkl}P_kM_l+DM_j
\end{displaymath}
where $A,C,D$ are arbitrary dimensionless scalar function of
$P_\mu$ and
\begin{displaymath}
\tilde{{\mathcal{N}}}_j=-{\x}_j\left\{\frac{1-e^{-2\lambda
P_0}}{2\lambda}+\frac{\lambda}{2}P^2\right\}+{\x}_0 P_j
\end{displaymath}

It is straightforward
to verify that the 10 generators $({\mathcal{N}},M,P)$ close a Hopf algebra,
for any choice of
the triplet $(A,C,D)$. In fact, the coproduct of ${\mathcal{N}}_j$ is
\begin{displaymath}
\Delta {\mathcal{N}}_j=\left({\mathcal{N}}_j\otimes 1+e^{-\lambda P_0}\otimes
{\mathcal{N}}_j+\lambda\e_{jkl}P_k\otimes M_l\right)\Delta A
+\lambda \e_{jkl}(\Delta (A+C))(\Delta
P_k)(\Delta M_l)+\Delta D{\cdot}\Delta M_j
\end{displaymath}
where $(\Delta A,\Delta C,\Delta D)$ are known tensors,
for each choice of $(A,C,D)$.

The request of a classical Lorentz subalgebra ($M_j$,${\mathcal{N}}_j$)
leads to the conditions
\be
D=0\quad\qquad [\lambda A\partial_0
-2(C+\lambda^2 P^2D)\partial_{\vec{P}^2}+\lambda^2 D] C=-\lambda^2
\label{condjoc}
\ee
The deformed boost operators considered in
Subsection~3.3.4 correspond to the $A=1,C=D=0$ solution of (\ref{condjoc})
({\it i.e.} correspond to $\tilde{{\mathcal{N}}}_j$)
and take the form
\begin{displaymath}
{\mathcal{N}}_j\Omega_R(f)=\Omega_R\ts \qs -ix_0\partial_j
-x_j\ts\frac{1-e^{2i\lambda \partial_0}}{2\lambda}
-\frac{\lambda}{2}\nabla^2\td
+\lambda x_l\partial_l\partial_j\qd f\td ~.
\end{displaymath}

\appendix{Covariant differential calculus on \kM \label{app5Ddc}}

In this appendix we describe Sitarz's proposal of a ``covariant''
differential calculus on \kM . This calculus has been shown to be
the unique covariant differential calculus on \kM\ in~\cite{Masl}.

In order to give this construction we will use the time-to-the
right ordering prescription on the \kM\ elements (the other
prescriptions are indeed equivalent for this construction). Thus
we will extend the commutative function $f(x)$ to \kM\  through
the right Weyl map introduced in (\ref{Rmap}): \bea
F(\x)&=&\Omega_R(f(x))=\frac{d^4k}{(2\pi)^2}\int\ft(k)\Omega_R(e^{ikx})\nn\\
\Omega_R(e^{ikx})&=&e^{-ik\x}e^{ik_0\x_0}\nn
\eea
where the function $\ft(k)$ is the classical Fourier transform.
Since we therefore always refer to the $\Omega_R$ map,
the label $R$  will be omitted from here after.

In Chapter 3 we have given a characterization of the symmetries of
\kM\ and we have found that the Majid Ruegg basis of \kkP\ is a
possible physical basis candidate of generators for \kM . It is
characterized by some commutation relation which we note again
here for convenience: \bea
\qs \n^j,\mP^0\qd&=&i\mP^j\\
\qs \n^j,\mP^k\qd&=&i[\frac{1}{2\lambda}(1-e^{-2\lambda
\mP^0})+\frac{\lambda}{2}\mP^2]\delta^{jk}-i\lambda\mP^j\mP^k \eea
The differential representation of the boosts generators on the
momentum space is: \bea
\Ns^j\, F(\mP)&=&\{i\mP^j\partial_{\mP^0}+i[\frac{1}{2\lambda}(1-e^{-2\lambda \mP^0})+
\frac{\lambda }{2}\mP^2]\partial_{\mP^j}-i\lambda \mP^j\mP^l\partial_{\mP^l}\}F(\mP)\\
&=&\{i\mP^j\partial_{\mP^0}+i B(\mP)\partial_{\mP^j}-i\lambda
\mP^j\mP^l\partial_{\mP^l}\}F(\mP) \eea The mass Casimir is: \be
C_{\lambda}(\mP)=\cosh (\lambda
\mP^0)-\frac{\lambda^2}{2}e^{\lambda \mP^0}\mP^2 \ee

The action of the Majid-Ruegg-basis generators is:
\bea
&&\mP^{\mu}\, \Omega(f(x))=\Omega(-i\partial^{\mu}f(x)),\;\;\;
M^i\, \Omega(f(x))=i\epsilon^{ikl}\Omega(x^k\partial^lf(x))\nn\\
&&\Ns^i\, \Omega(f(x))
=\Omega\ts
[ix^0\partial_{x^i}-x^i(\frac{1}{2\lambda}(1-e^{2i\lambda
\partial_{x^0}})-\frac{\lambda }{2}\nabla^2)-\lambda
x^j\partial_{x^j}\partial_{x^i}]f(x)\td \eea $\Ns^j$ leave
invariant the commutation relations of \kM . \bea
\Ns^j\, \x^0=i\x^j &&\Ns^j\,\x^k=i\delta^{jk}\x^0\label{azionecoord}\\
\Ns^j\, (\x^0)^2=2i \x^j\x^0-2\lambda&&\Ns^j\, (\x^k)^2=2i \x^k\x^0\delta^{kj}-
\lambda\x^j-2\lambda\x^k\delta^{jk}\nn\\
\Ns^j\, \x^k\x^0=i(\delta^{jk}(\x^0)^2+\x^j\x^k)&&\Ns^j\,
\x^0\x^k=i(\delta^{jk}(\x^0)^2+\x^j\x^k)-\lambda\delta^{jk}\x^0\nn
\eea The form invariant over the action of the Magid-Ruegg basis
of \kkP\ is: \be s^2=\x_{\mu}\x^{\mu}-3i\lambda \x^0 \ee

Now we proceed t the construction of the differential calculus
following the Sitarz technique~\cite{Sitarz}. To be simple, let us
work in $2$D \kM\ generated by the coordinates $(t,x)$, with
$[t,x]=i\lambda x$. In order to reproduce the construction of
Sitarz we introduce a $3$D differential calculus with generators
$dt,dx,\phi$.

Assume that the commutation relations between the \kM\ generators and the generators of the one-forms can be written in the following way:
\bea
\qs dt,t\qd&=&Adt+Bdx+U\phi\nn\\
\qs dt,x\qd&=&Cdt+Ddx+P\phi\nn\\
\qs dx,t\qd&=&Edt+Fdx+Q\phi\nn\\
\qs dx,x\qd&=&Gdt+Hdx+R\phi\nn\\
\qs \phi,t\qd&=&Idt+Ldx+S\phi\nn\\
\qs \phi,x\qd&=&Vdt+Wdx+T\phi \eea where $A,...,T$ are unknown
coefficients which depend only by $\lambda$ and which have the
dimension of a length.

Assume also that the action of the \kkP generators on the fifth
one-form generator be: \be \mP\,\phi=M\,\phi=\mN\,\phi=0 \ee

Some conditions on the parameter $A,...,T$ can be obtained differentiating the commutation relations of \kM :
\be
d \ts [t,x]=i\lambda x\td  \to [dt,x]+[t,dx]=i\lambda dx
\ee
that fixes the following relations:
\be
E=C,\; Q=P,\; F=D-i\lambda \label{EQF}
\ee
The request of covariance condition under the boost action is represented by the following equations:
\bea
&&\Ns\,\qs dt,t\qd=\Ns\,(Adt+Bdx+U\phi)\label{prima}\\
&&\Ns\,\qs dt,x\qd=\Ns\,(Cdt+Ddx+P\phi)\label{secon}\\
&&\Ns\,\qs dx,t\qd=\Ns\,(Edt+Fdx+Q\phi)\\
&&\Ns\,\qs dx,x\qd=\Ns\,(Gdt+Hdx+R\phi)\\
&&\Ns\,\qs \phi,t\qd=\Ns\,(Idt+Ldx+S\phi)\\
&&\Ns\,\qs \phi,x\qd=\Ns\,(Vdt+Wdx+T\phi)\label{sesta} \eea These
equations represent a system of equations for the coefficients
$A,..,T$. In fact, consider for example the first equation
(\ref{prima}). The left side can be rewritten using
(\ref{azionecoord}): \bea
{}\mN\,[dt,t]&=&\mN[dt\,t-tdt]=(\mN_{(1)}dt)(\mN_{(2)}t)-(\mN_{(1)}t)(\mN_{(2)}dt)=d(\mN_{(1)}t)(\mN_{(2)}t)-(\mN_{(1)}t)d(\mN_{(2)}t)\nn\\
&=& d(\mN t)t +d(e^{-\lambda \mP_0}t)(\mN t)-[(\mN t)dt+ e^{\lambda \mP_0}t d(\mN t)]\nn\\
&=& idx\,t +i d(t+i\lambda )x-[ix dt+ i(t+i\lambda )] dx\nn\\
&=&i[dx,t]+i[dt,x]+\lambda dx\nn\\
 &=&iEdt+iF dx+iQd\phi+iCdt+iDdx+iP\phi+\lambda dx\nn
\eea which can be rewritten taking into account (\ref{EQF}) in the
following way: \bea
{}\mN\,[dt,t]&=&2iCdt+2i(D-i\lambda)dx+2iPd\phi\label{1.1} \eea
While the right side of (\ref{prima}) is: \be \Ns\,(Adt+Bdx+U
d\phi)=iAdx+iBdt\label{1.2} \ee Comparing (\ref{1.1}) and
(\ref{1.2}) we find \be B=2C,\;A=2D-i2\lambda,\;P=0 \label{EQFBAP}
\ee

Following the same procedure for the other equations
(\ref{secon}-\ref{sesta}), we find the following conditions:
\bea
&&A=B=C=E=F=H=P=Q=S=T=0\nn\\
&&D=G=i\lambda\nn\\
&&I=W,\;\;L=V\;\;R=-U\nn \eea Therefore the invariance conditions
(\ref{prima}-\ref{sesta}) do not determine all the unknown
coefficients ($V,W,,U$ remain unknown).
However, using the Jacoby Identities: \bea
[t,[dt,x]]+[dt,[x,t]]+[x,[t,dt]&=&0\nn\\
\qs t,[dx,x]\qd+[dx,[x,t]]+[x,[t,dx]]&=&0\nn\\
\qs t,[d\phi,x]\qd+[d\phi,[x,t]]+[x,[t,d\phi]&=&0 \eea we
determine $V=0$ and $WU=\lambda^2$. Thus we have: \bea
\qs dt,t\qd&=&U\phi\nn\\
\qs dt,x\qd&=&i\lambda dx\nn\\
\qs dx,t\qd&=&0\nn\\
\qs dx,x\qd&=&i\lambda dt-U\phi\nn\\
\qs \phi,t\qd&=&Wdt\nn\\
\qs \phi,x\qd&=&Wdx \eea Since $U\neq 0$, we scale $\phi\to
\frac{\lambda}{U}\phi$ and we find: \bea
\qs dt,t\qd&=&-\lambda\phi\nn\\
\qs dt,x\qd&=&i\lambda dx\nn\\
\qs dx,t\qd&=&0\nn\\
\qs dx,x\qd&=&i\lambda dt+\lambda\phi\nn\\
\qs \phi,t\qd&=&\lambda dt\nn\\
\qs \phi,x\qd&=&\lambda dx
\eea

Thus the exterior derivative operator $d$ of a \kM\ function can
be written in the following form: \be d \Omega(e^{ikx})=\E(\mP)_a
\Omega(e^{ikx})\tau^a=\Omega(\E(k)e^{ikx})\tau^a\label{EDO1} \ee
where $\tau^a=(dt,dx,\phi)$.

Through a little bit of algebra the vector fields $\E(\mP)$ can be
determined. In order to do this let us introduce the forms
$\psi,\psi'$ such that: \be \psi=dt-id\phi\;\;\psi'=dt+id\phi \ee
The (\ref{diffcalc}) can be written in terms of $\psi,\psi'$ in
the following way: \bea
\qs d\psi,t\qd=-i\lambda d\psi,&&[d\psi,x]=0\nonumber\\
\qs dx,t\qd=0,&&[dx,x]=i\lambda d\psi\nonumber\\
\qs d\psi',t\qd=i\lambda d\psi',&&[d\psi',x]=2i\lambda dx
\eea

In terms of these variables it is easy to compute the following commutation relations through seris expansions (in $a,b \in R$):
\bea
[dx,e^{ax}]=i\lambda \partial_{x}e^{ax}\psi\;&\qs \psi,e^{ax}\qd=0\;&\qs \psi',e^{ax}\qd=2\lambda(i\partial_{x}e^{ax}dx-\frac{\lambda}{2}\nabla^2 e^{ax}\psi)\nn\\
\qs dx,e^{bt}\qd=0\;&\qs \psi,e^{bt}\qd=(e^{-i\lambda\partial_t}-1)e^{bt}\psi\;&\qs \psi',e^{bt}\qd=(e^{i\lambda\partial_t}-1)e^{bt}\psi'\nn
\eea
From which it follows:
\bea [dx,e^{ax}e^{bt}]&=&[dx,e^{ax}]e^{bt}=i\lambda\partial_{x}e^{ax}\psi e^{bt}=i\lambda e^{-i\lambda\partial_t}\partial_{x}e^{ax}e^{bt}\psi\nn\\
\qs \psi,e^{ax}e^{bt}\qd&=&e^{ax}[\psi,e^{bt}]+[\psi,e^{ax}]e^{bt}=(e^{-i\lambda\partial_t}-1)e^{ax}e^{bt}\psi\nn\\
\qs \psi',e^{ax}e^{bt}\qd&=&e^{ax}[\psi',e^{bt}]+[\psi',e^{ax}]e^{bt}\nn\\
&=&(e^{i\lambda\partial_t}-1)e^{ax}e^{bt}\psi'+2\lambda[i\partial_{x}e^{ax}dx-\frac{\lambda}{2}\nabla^2 e^{ax}\psi]e^{bt}\nn\\
&=&(e^{i\lambda\partial_t}-1)e^{ax}e^{bt}\psi'+2i\lambda \partial_xe^{ax}e^{bt}dx+\lambda^2e^{-i\lambda\partial_t}\nabla^2e^{ax}e^{bt}\psi\nn
\eea
Choosing $a=-ik$ and $b=ik_0$, and reminding that $\Omega(e^{ikx})=e^{-ikx}e^{ik_0t}$, we have that:
\bea
[dx,\Omega(e^{ikx})]&=&i\lambda \Omega( e^{-i\lambda\partial_t}\partial_{x}e^{ikx})\psi=\lambda e^{\lambda E}k\Omega(e^{ikx})\psi\nn\\
\qs \psi,\Omega(e^{ikx})\qd&=&\Omega((e^{-i\lambda\partial_t}-1)e^{ikx})\psi=(e^{\lambda E}-1)\Omega(e^{ikx})\psi\nn\\
\qs \psi',\Omega(e^{ikx})\qd&=&\Omega((e^{i\lambda\partial_t}-1)e^{ikx})\psi'+i2\lambda \Omega(\partial_{x}e^{ikx})dx-\lambda^2 \Omega(e^{-i\lambda\partial_t}\nabla^2e^{ikx})\psi\nn\\
&&=\Omega(e^{ikx})[(e^{-\lambda E}-1)\psi'+2\lambda k dx+\lambda^2 e^{\lambda E}k^2\psi]\label{commutatori}
\eea

The exterior derivative operator (\ref{EDO1}) can be equivalently
expressed in terms of the forms $\psi,dx,\psi'$: \be
d\Omega(e^{ikx})=
\Omega(e^{ikx})[\E_{\psi}(k)\psi+\E_x(k)dx+\E_{\psi'}\psi']\label{EDO2}
\ee

Using the Leibnitz rule (that must be satisfied by the exterior
derivative operator): \bea
d(\Omega(e^{ikx})\Omega(e^{ipx}))&=&d\Omega(e^{i(k\oplus p)x})\\
&=&(d\Omega(e^{ikx}))\Omega(e^{ipx})+\Omega(e^{ikx})d\Omega(e^{ipx})
\eea
and making use of the (\ref{EDO2}), we find the following conditions:
\bea
1.&& \E_{\psi}(k\oplus p)=e^{\lambda p^0}\E_{\psi}(k)+\E_{\psi}(p)+\lambda e^{\lambda p^0}p \E_x(k)+\lambda^{2}e^{\lambda p^0}p^2\E_{\psi'}(k)\\
2.&& \E_x(k\oplus p)=\E_x(k)+\E_x(p)+2\lambda p \E_{\psi'}(k) \\
3.&&\E_{\psi'}(k\oplus p)=\E_{\psi'}(p)+e^{-\lambda p^0}\E_{\psi'}(k)
\eea
where $k\oplus p=(k^0+p^0,k+e^{-\lambda k^0}p)$.

It is easy to find that the solution of these equations is (see \cite{Sitarz}):
\bea
\E_{\psi'}(k)&=&a(1-e^{-\lambda k^0})\nn\\
\E_x(k)&=&-2a\lambda k\nn\\
\E_{\psi}(k)&=&-c(1-e^{\lambda k^0})-a\lambda^2e^{\lambda k^0}k^2\nn
\eea
with $a,c\in R$ that will be determined imposing the commutative limit $\lambda=0$.

Thus:
\bea
d\Omega(e^{ikx})&=&[-c(1-e^{\lambda k^0})-a\lambda^2e^{\lambda k^0}k^2]\psi+\nn\\
&&-2a\lambda k dx+\nn\\
&&+a(1-e^{-\lambda k^0})\psi'\nn\\
&=&[-c(1-e^{\lambda k^0})-a\lambda^2e^{\lambda k^0}k^2+a(1-e^{-\lambda k^0})]\Omega(e^{ikx})dt+\nn\\
&&-2a\lambda k \Omega(e^{ikx})dx+\nn\\
&&[-ai(1-e^{-\lambda k^0})-ic(1-e^{\lambda
k^0})-ai\lambda^2e^{\lambda k^0}k^2]\Omega(e^{ikx})\phi \eea
Imposing that in the limit $\lambda\to 0$ the exterior
differential operator be the standard one: \be d\Omega(e^{ikx})\to
d e^{ikx}=ik^0e^{ikx}dt-i k e^{ikx}dx \ee we obtain that
$a=c=\frac{i}{2\lambda}$.

Finally: \bea d\Omega(e^{ik\x})&=&\Omega(e^{ikx})\ts
\frac{i}{\lambda}[\sinh(\lambda k^0)-\frac{\lambda^2}{2}e^{\lambda
k^0}k^2]dt,\,-i k dx,\,\frac{1}{\lambda}[1-\cosh(\lambda
k^0)+\frac{\lambda^2}{2}e^{\lambda k^0}k^2]\phi\td\nn \eea

The result obtained can be straightforwardly extended to the case of $4$D \kM\ spacetime, in which the differential calculus turns out to be $5$D $(i,j=1,2,3)$:
\bea
\qs dx^0,x^0\qd=-\lambda \phi,&&[dx^0,x^i]=i\lambda dx^i\nonumber\\
\qs dx^i,x^0\qd=0,&&[dx^i,x^j]=i\delta^{ij}\lambda(dx^0-i\phi)\nonumber\\
\qs \phi,x^0\qd=\lambda dx^0,&&[\phi,x^i]=\lambda
dx^i\label{diffcalc} \eea where $(dx^0,dx^1,dx^2,dx^3,\phi)$ are
the five one-form generators. The exterior derivative operator is:
\bea
d\Omega(e^{ikx})&=&\Omega(e^{ikx})\E_a(k)\tau^a\\
\eta_{ab}&\equiv&(+1,-1-1-1,+1)\\
\tau^a&\equiv&(dx^0,dx^1,dx^2,dx^3,\phi)\\
e_a(k)&=&\ts\frac{i}{\lambda}[\sinh(\lambda
k^0)-\frac{\lambda^2}{2}e^{\lambda
k^0}k^2],ik_1,ik_2,ik_3,\frac{1}{\lambda}[\cosh(\lambda
k^0)-1-\frac{\lambda^2}{2}e^{\lambda k^0}k^2]\td\nn \eea
and it can be cast in the following form, in terms of $\mP_\mu$:
\bea
d\Omega(f(x))&=&\E_a(P)\Omega(f(x))\; \tau^a\nn\\
\E_a(P)&=&\ts\frac{i}{\lambda}[\sinh(\lambda
P_0)-\frac{\lambda^2}{2}e^{\lambda
P_0}P^2],iP_i,\frac{1}{\lambda}[\cosh(\lambda
P_0)-1-\frac{\lambda^2}{2}e^{\lambda P_0}P^2]\td\label{result}
\eea

\bigskip
\bigskip
\bigskip

One could write the exterior derivative operator in an alternative form (through the fields $\D$):
\bea
d\Omega(f(x))&=&\tau^a\,\D_a(P)\Omega(f(x))\nn\\
\eea
To determine the relations between $\E_a$ and $\D_a$ one has to compare the expressions:
\bea
\E_a(k)\Omega(e^{ikx})\tau^a&=&\tau^a\Omega(e^{ikx})\D_a(k)\label{compare}
\eea
From the relations (\ref{commutatori}) one obtains:
\bea
\Omega(e^{ikx})dx&=&[dx-\lambda kdt+i\lambda k\phi]\Omega(e^{ikx})\nn\\
\Omega(e^{ikx})dt&=&[(\cosh(\lambda E)+\frac{\lambda^2}{2}k^2 e^{\lambda E})dt+i(\sinh(\lambda E)+\frac{\lambda^2}{2}k^2 e^{\lambda E})\phi-\lambda ke^{\lambda E}dx]\Omega(e^{ikx})\nn\\
\Omega(e^{ikx})\phi&=&[-i(\sinh(\lambda E)-\frac{\lambda^2}{2}k^2e^{\lambda E})dt+(\cosh(\lambda E)-\frac{\lambda^2}{2}k^2e^{\lambda E})\phi+i\lambda ke^{\lambda E}dx]\Omega(e^{ikx})\nn
\eea
Thus the left side of (\ref{compare}) becomes:
\bea
\E_a(k)\Omega (e^{ikx})\tau^a&=&\E_0(k)[(\cosh(\lambda E)+\frac{\lambda^2}{2}k^2 e^{\lambda E})dt-i(-\sinh(\lambda E)+\frac{\lambda^2}{2}k^2 e^{\lambda E})\phi-\lambda ke^{\lambda E}dx]\Omega(e^{ikx})\nn\\
&&+\E_x(k)[dx+\lambda kdt+i\lambda k \phi]\Omega(e^{ikx})\nn\\
&&+\E_{\phi}(k)[i(-\sinh(\lambda E)-\frac{\lambda^2}{2}k^2e^{\lambda E})dt+(\cosh(\lambda E)-\frac{\lambda^2}{2}k^2e^{\lambda E})\phi+i\lambda ke^{\lambda E}dx]\Omega(e^{ikx})\nn\\
&&\label{left} \eea While the right sided of (\ref{compare}) is:
\bea
\tau^a\Omega(e^{ikx})\D_a(k)=[\D_t(k)dt+\D_x(k)dx+\D_{\phi}(k)d\phi]\Omega(e^{ikx})\label{rightdc}
\eea then, comparing (\ref{left}) and (\ref{rightdc}), we obtain:
\bea
\D_t&=&(\cosh(\lambda E)+\frac{\lambda^2}{2}k^2e^{\lambda E})\E_0-\lambda k\D_x-i(\sinh(\lambda E)+\frac{\lambda^2}{2}k^2e^{\lambda E})\E_{\phi}\nn\\
\D_x&=&-\lambda ke^{\lambda E}\E_0-\lambda k\D_x+i\lambda ke^{\lambda E}\E_{\phi}(k)\nn\\
\D_{\phi}&=&i(\sinh(\lambda E)+\frac{\lambda^2}{2}k^2 e^{\lambda E})\E_0+i\lambda k\D_x+(\cosh(\lambda E)-\frac{\lambda^2}{2}k^2e^{\lambda E})\E_{\phi}\nn
\eea
Substitute the expressions of the $\E(k)$ (\ref{result}, we find:
\bea
\D_t(k)&=&\frac{i}{\lambda}(\sinh(\lambda E)+\frac{\lambda^2}{2}k^2 e^{\lambda E})\nn\\
\D_x(k)&=&-\frac{i}{\lambda}ke^{\lambda E}\nn\\
\D_{\phi}(k)&=&-\frac{1}{\lambda}(1-\cosh(\lambda E)+\frac{\lambda^2}{2}k^2 e^{\lambda E})\nn
\eea

Finally:
\bea
d\Omega(f(x))&=&\tau^a\,\D_a(P)\Omega(f(x))\nn\\
\D_a(P)&=&\ts\frac{i}{\lambda}[\sinh(\lambda
P^0)+\frac{\lambda^2}{2}e^{\lambda P^0}P^2],iP_ie^{\lambda
P_0},-\frac{1}{\lambda}(1-\cosh(\lambda E)+\frac{\lambda^2}{2}k^2
e^{\lambda E})\td \label{result1}\eea This formula is just the
(\ref{5Ddiffcalc1}) used in Chapter 4.

\appendix{Determination of the spinorial representation of the \kkP\ algebra \label{appspinrepr}} In this appendix we
construct a spinorial representation of \kkP\ and determine the
possible $\beta$ matrices appearing in the (\ref{jodirac1}) such
that the Dirac equation is invariant under the \kkP\ action.

 We consider a generic form of the boost $\mN_j^t$ in terms of the differential representation
 of the \kkP\ boosts $\mN_j$ (\ref{bicrossbasis}) and the finite-dimensional
 representation of the Poinca\'e boosts $n_j$ (\ref{n}): \be
{\mN}^T_j=A(P)_{jk}{\mN}_k+B(P)_{jk}M_k+C(P)_{jk}n_k+E(P)_{jk}m_k\rightarrow
N^j_T=N^j+n^j
 \ee

Imposing that \be
[{\mN}^T_j,\mP_0]=i\mP_j,\;\;\;[{\mN}^T_j,\mP_k]=i\delta_{jk}[\frac{1-e^{-2\lambda
\mP_0}}{2\lambda}+\frac{\lambda}{2}\mP^2]-i\lambda \mP_j\mP_k \ee
we find: \be A(\mP)_{jk}=\delta_{jk},\;\;\;B(\mP)_{jk}=0 \ee then:
\be {\mN}^T_j={\mN}_j+C(\mP)_{jk}n_k+E(\mP)_{jk}m_k
 \ee
Now we impose the invariance condition: \be
[{\mN}^T_j,\Dirac_{\lambda}]=0 \label{eq0}\ee where the expression
of the dirac operator is given by:
 \be
\Dirac_{\lambda}=i\D_0(\mP)\gamma^0+i\D_i(\mP)\gamma^i-M[\frac{\lambda
\D_4(\mP)}{1-\cosh(\lambda
m)}+1]\gamma^0\bt+\frac{\D_4(\mP)\sinh(\lambda m)}{1-\cosh(\lambda
m)}I \ee
Thus, the (\ref{eq0}) can be written in the following way: \bea
0&=&i[{\mN}_j,\D_0(\mP)]\gamma^0+i[{\mN}_j,\D_i(\mP)]\gamma^i+\nn\\
&+&i\D_0(\mP)C(\mP)_{jk}[n_k,\gamma^0]+i\D_i(\mP)C(\mP)_{jk}[n_k,\gamma^i]+\nn\\
&-&M[\frac{\lambda \D_4}{1-\cosh(\lambda m)}+1]C(\mP)_{jk}[n_k,\gamma^0\bt]+\nn\\
&+&i\D_i(\mP)E(\mP)_{jk}[m_k,\gamma^i]\nn \eea where we have taken
into account that $[{\mN}_j,\D_4]=[m_k,\gamma^0]=[m_k,\bt]=0$.

The commutators present in the previous equation are given by:
\bea
[{\mN}_j,\D_0(\mP)]=i\D_j(\mP)&&[{\mN}_j,\D_i(\mP)]=i\D_0(\mP)\delta_{ji}\nn\\
\qs
n_k,\gamma^0\qd=i\gamma_k&&[n_k,\gamma^i]=i\eta_{ki}\gamma^0\;\;\;[m_k,\gamma^i]=-i\epsilon_{kil}\gamma_l
\eea and using them, we find: \bea
0&=&-\D_j\gamma^0-\D_0\gamma^j-\D_0C_{jk}\gamma_k+\D_kC_{jk}\gamma^0+\nn\\
&-&M[\frac{\lambda \D_4}{1-\cosh(\lambda m)}+1]C_{jk}(i\gamma_k\bt+\gamma^0[n_k,\bt])+\nn\\
&+&\D_iE_{jk}\epsilon_{kil}\gamma_l\nn\\
&=&(-\D_j+\D_kC_{jk})\gamma^0-\D_0\gamma^j-\D_0C_{jk}\gamma_k+\nn\\
&-&M[\frac{\lambda \D_4}{1-\cosh(\lambda m)}+1]C_{jk}(i\gamma_k\bt+\gamma^0[n_k,\bt])+\nn\\
&+&\D_i(P)E_{jk}\epsilon_{kil}\gamma_l\nn \eea Since
$\gamma^{\mu}$ form a basis, $\gamma^0$ cannot be expressed as
linear combination of $\gamma^k$, then the coefficient of
$\gamma^0$ must be zero and $C_{jk}=\delta_{jk}$. Moreover it must
be:\bea 0=-M[\frac{\lambda \D_4(\mP)}{1-\cosh(\lambda
m)}+1](i\gamma_j\bt+\gamma^0[n_j,\bt])+\D_i(\mP)E_{jk}\epsilon_{kil}\gamma_l\nn
\label{eqapp}\eea In order to determine $\bt$ let us write a
general explicit expression for $E_{jk}$: \be
E_{jk}(P)=\delta_{jk}E_1+P_jP_kE_2+\epsilon_{jkl}P_lE_3 \ee \be
\D_iE_{jk}\epsilon_{kil}\gamma_l=\D_i\epsilon_{jil}E_1\gamma_l+\epsilon_{jkr}P_rE_3\D_i\epsilon_{kil}\gamma_l
\ee \be
\D_iE_{jk}\epsilon_{kil}\gamma_l=E_1\D_i\epsilon_{jil}\gamma_l+(P_i\D_i\gamma_j-P_l\D_j\gamma_l)E_3
\ee In terms of $E_1,E_2,E_3$ the (\ref{eqapp}) is:
\bea
0&=&-M[\frac{\lambda \D_4}{1-\cosh(\lambda m)}+1](i\gamma_j\bt+\gamma^0[n_j,\bt])+P_i\D_i\gamma_jE_3\nn\\
&&+E_1\D_i\epsilon_{jil}\gamma_l-P_l\D_j\gamma_lE_3\nn \eea from
which follows $E_1=E_3=0$, and $i\gamma_j\bt+\gamma^0[n_j,\bt]=0$.
While $E_2$ can be determined by imposing the commutator among
boost generators: \bea
[\mN_j^T,\mN_s^T]&=&[\mN_j,\mN_s]+[\mN_j,E_{sl}]m_l\nn\\
&+&[n_j,n_s]+E_{sl}[n_j,m_l]\nn\\
&+&[E_{jk},\mN_s]m_k+E_{jk}[m_k,n_s]+E_{jk}E_{sl}[m_k,m_l] \eea
where we have taken into account that $C_jk=\delta_jk$. Thus \bea
[\mN_j^T,\mN_s^T]&=&-i\epsilon_{jst}M_t+([\mN_j,E_{sl}]-[E_{jl},\mN_s])m_l+
i\epsilon_{klt}E_{jk}E_{sl}m_t\nn\\
&&i(\epsilon_{kst}E_{jk}-\epsilon_{kjt}E_{sk})n_t \eea Thus:  \bea
&&([N_j,E_{sl}]-[N_s,E_{jl}])m_l+i\epsilon_{klt}E_{jk}E_{sl}m_t=0\nn\\
&&i(\epsilon_{kst}E_{jk}-\epsilon_{kjt}E_{sk})n_t=0\nn \eea and
considering $E_1=E_3=0$ the last equation states that $E_2=0$.

The equation $i\gamma_j\bt+\gamma^0[n_j,\bt]=0$ admits three
solutions for $\bt$, which are $\bt=0,\gamma^0,\gamma^0\gamma^5$.

In this way we have obtained that the form of $\mN^T_j$ and the
compatible $\beta$ matrices are given by: \bea
&&{\mN}_j^T={\mN}_j+n_j\nn\\
&&\bt=0\;\;\mbox{or}\;\;\gamma^0\;\;\mbox{or}\;\;\gamma^0\gamma^5
\eea

\bigskip
\bigskip
\bigskip

We proceed exactly in the same way in order to determine the
spinorial representation of the \kkP that leaves invariant  the
Dirac operator constructed with the other deformed derivatives
$\E_a$: \be
\Dirac_{\lambda}=i\E_0(\mP)\gamma^0+i\E_i(\mP)\gamma^i+M[\frac{\lambda
\E_4(\mP)}{1-\cosh(\lambda
m)}-1]\gamma^0\bt+\frac{\E_4(\mP)\sinh(\lambda m)}{1-\cosh(\lambda
m)}I \ee The solution in this case is the following: \bea
{\mN}^T_j&=&{\mN}_j+e^{-\lambda P_0}n_j-\lambda
\epsilon_{jkl}P_km_l\nn\\
\bt&=&0,\,\gamma^0,\,\gamma^0\gamma^5 \eea


\end{document}